\documentclass[]{aa}  

\usepackage{graphicx}
\usepackage{color}
\usepackage{txfonts}
\definecolor{darkred}{rgb}{0.7, 0.0, 0.0}
\definecolor{carnelian}{rgb}{0.7, 0.11, 0.11}
\definecolor{ao(english)}{rgb}{0.0, 0.5, 0.0}

\newcommand{\logne}[1]{{log$(n_\mathrm{e}\,[\mathrm{cm}^{-3}])${#1}}}

\usepackage{amstext}

\begin{document}

   \title{Non-equilibrium ionization by a periodic electron beam}
   \subtitle{II. Synthetic Si IV and O IV transition region spectra}

   \author{Elena Dzif\v{c}\'akov\'a
           \and Jaroslav Dud\'ik
          }

   \institute{Astronomical Institute of the Czech Academy of Sciences, Fri\v{c}ova 298, 251 65 Ond\v{r}ejov, Czech Republic
         \email{elena.dzifcakova\,@\,asu.cas.cz}
                }

   \date{Received ; accepted }

  \abstract
        {Transition region (TR) spectra typically show the \ion{Si}{IV} 1402.8\,\AA~line to be enhanced by a factor of 5 or more compared to the neighboring \ion{O}{IV} 1401.2\,\AA, contrary to predictions of ionization equilibrium models and the Maxwellian distribution of particle energies. Non-equilibrium effects in TR spectra are therefore expected.}
        {To investigate the combination of non-equilibrium ionization and high-energy particles, we apply the model of the periodic electron beam, represented by a $\kappa$-distribution that recurs at periods of several seconds, to plasma at chromospheric temperatures of 10$^4$ K. This simple model can approximate a burst of energy release involving accelerated particles.}
        {Instantaneous time-dependent charge states of silicon and oxygen were calculated and used to synthesize the instantaneous and period-averaged spectra of \ion{Si}{IV} and \ion{O}{IV}.}
        {The electron beam drives the plasma out of equilibrium. At electron densities of $N_\mathrm{e}$\,=\,10$^{10}$\,cm$^{-3}$, the plasma is out of ionization equilibrium at all times in all cases we considered, while for a higher density of $N_\mathrm{e}$\,=\,10$^{11}$\,cm$^{-3}$, ionization equilibrium can be reached toward the end of each period, depending on the conditions. In turn, the character of the period-averaged synthetic spectra also depends on the properties of the beam. While the case of $\kappa$\,=\,2 results in spectra with strong or even dominant \ion{O}{IV}, higher values of $\kappa$ can approximate a range of observed TR spectra. Spectra similar to typically observed spectra, with the \ion{Si}{IV} 1402.8\,\AA~line about a factor 5 higher than \ion{O}{IV} 1401.2\,\AA, are obtained for $\kappa$\,=\,3. An even higher value of $\kappa$\,=\,5 results in spectra that are exclusively dominated by \ion{Si}{IV}, with negligible \ion{O}{IV} emission. This is a possible interpretation of the TR spectra of UV (Ellerman) bursts, although an interpretation that requires a density that is 1--3 orders of magnitude lower than for equilibrium estimates.}
        {}

   \keywords{Sun: UV radiation -- Sun: transition region -- Radiation mechanisms: non-thermal}
   
   \titlerunning{Non-equilibrium ionization by a periodic electron beam. II. \ion{Si}{IV} and \ion{O}{IV} TR spectra}
   \authorrunning{Dzif\v{c}\'{a}kov\'{a} \& Dud\'{i}k}
   \maketitle
%
\section{Introduction}
\label{Sect:1}

The solar transition region (hereafter, TR) is an inhomogeneous thin interface between the solar chromosphere and the overlying hot corona. Its typical temperatures roughly span the region between a few times 10$^4$\,K to about 10$^6$\,K. TR observations show peculiar line intensities, with the lines from the Li and Na-like isoelectronic sequences being enhanced by a factor of several compared to other lines that are formed at similar temperatures \citep[e.g.,][]{DelZanna02}. A particular example is the \ion{Si}{IV} doublet at 1393.8\,and 1402.8\,\AA, whose peak formation temperature should lie at log($T_\mathrm{max}$\,[K])\,=\,4.9 under equilibrium conditions that are characterized by a Maxwellian distribution \citep{Dudik14b}. However, their intensities are usually enhanced by a factor of five or more compared to the neighboring intercombination \ion{O}{IV} lines that are formed at similar temperatures of log($T_\mathrm{max}$\,[K])\,=\,5.15 \citep[e.g.,][]{Doyle84,Judge95,Curdt01,Yan15,Polito16b,Doschek16,Dudik17b}. Furthermore, the TR is known to be highly dynamic, with intensity changes on the order of several seconds or tens of seconds \citep[e.g.,][]{Testa13,Regnier14,Hansteen14,Peter14Sci,Tian14,Tian16,Vissers15,Tajfirouze16,Martinez16,Hou16,Warren16,Samanta17} as identified in recent observations performed by the Interface Region Imaging Spectrograph \citep[IRIS,][]{DePontieu14} as well as by the Hi-C \citep{Kobayashi14} rocket-borne instrument. 

This dynamics represents a modeling challenge, since the intensity changes can be shorter than the change in ionization equilibration timescales \citep[see Fig. 1 of][]{Smith10}. 1D hydrodynamic and 3D magnetohydrodynamic simulations have both shown that the TR ions can be out of ionization equilibrium as a result of rapid heating, cooling, advection (i.e., flows), or a combination of these processes \citep[e.g.,][]{Raymond78,Dupree79,Noci89,Raymond90,Hansteen93,Spadaro94,Bradshaw03a,Bradshaw03b,Bradshaw04,Doyle13,Olluri13b,Olluri15,Martinez16}. An overview of the modeling techniques and of the importance of non-equilibrium ionization can be found in \citet{Dudik17a}.

Furthermore, dynamical events involving turbulence, magnetic reconnection, or wave-particle interaction can lead to departures from the equilibrium Maxwellian distribution, for instance, by particle acceleration \citep{Hasegawa85,Laming07,Gontikakis13,Gordovskyy13,Gordovskyy14,Bian14,Che14,Vocks08,Vocks16}. Generally, any plasma where the collisional mean free-path exceeds
by about 10$^{-2}$  the local pressure or density scale-height can show departures from Maxwellians, usually in the form of high-energy tails of the distribution \citep{Roussel-Dupre80a,Shoub83,Ljepojevic88a,Scudder13}. \citet{Scudder13} argued that situations like this can occur as low as in the solar transition region, and they are expected
to be common in solar and stellar coronae.

High-energy tails can be conveniently modeled by non-Maxwellian $\kappa$-distributions, which exhibit strong high-energy tails that are described by one additional parameter $\kappa$ \citep{Olbert68,Vasyliunas68a,Vasyliunas68b,Livadiotis15b}. Non-Maxwellian distributions have also been detected in the solar corona by \citet{Dudik15}. In relation to the TR, the $\kappa$-distributions have previously been considered by \citet{Dzifcakova11Si}. These authors used ratios of the observed \ion{Si}{III} line intensities to show that $\kappa$-distributions are likely to be present in TR. Some of the results were revised using newer atomic data by \citet{DelZanna15a}. Indications of the possible presence of $\kappa$-distributions in TR have been obtained by fitting the TR line profiles \citep{Dudik17b} and the \ion{O}{IV} intensity ratios relative to \ion{Si}{IV} one \citep{Dudik14b,Dudik17b}. The \ion{Si}{IV}\,/\,\ion{O}{IV} ratios are also strongly sensitive to the presence of non-equilibrium ionization, however,
which also decrease the \ion{O}{IV} lines relative to \ion{Si}{IV} \citep{Doyle13,Martinez16}.

\citet[][hereafter, Paper~I]{Dzifcakova16a} provided a first simple model incorporating both the non-equilibrium ionization and non-Maxwellian distributions. This was done by considering a periodic passage of an electron beam through plasma with coronal temperatures. The beam was modeled through $\kappa$-distributions. It was shown that the occurrence of the electron beam drives the plasma out of ionization equilibrium, and that the plasma is out of equilibrium at all times. This had significant consequences for the analysis of the plasma under the assumption of equilibrium. In particular, the plasma looked multithermal when it was interpreted using ionization equilibrium and a Maxwellian distribution. 

In this paper, we apply the periodic electron beam model presented in Paper~I to TR plasma (Sect. \ref{Sect:2}), in particular, to chromospheric plasma at 10$^4$\,K that is heated by the beam. Synthetic period-averaged and instantaneous spectra are presented in Sect. \ref{Sect:3}. Results are discussed and summarized in Sect. \ref{Sect:4}.

%
\begin{figure*}[!ht]
        \centering 
        \includegraphics[width=5.5cm]{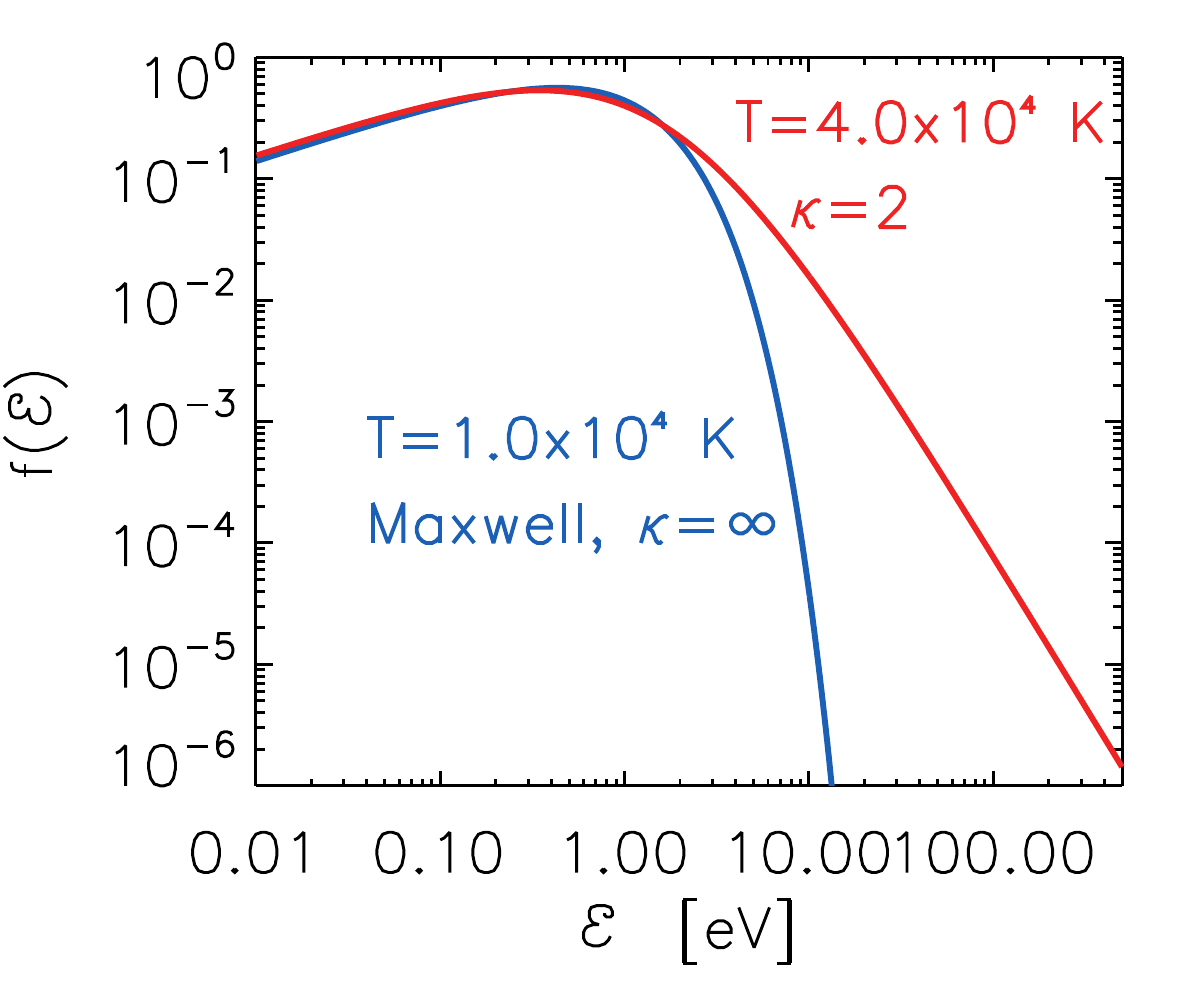}
        \includegraphics[width=5.5cm]{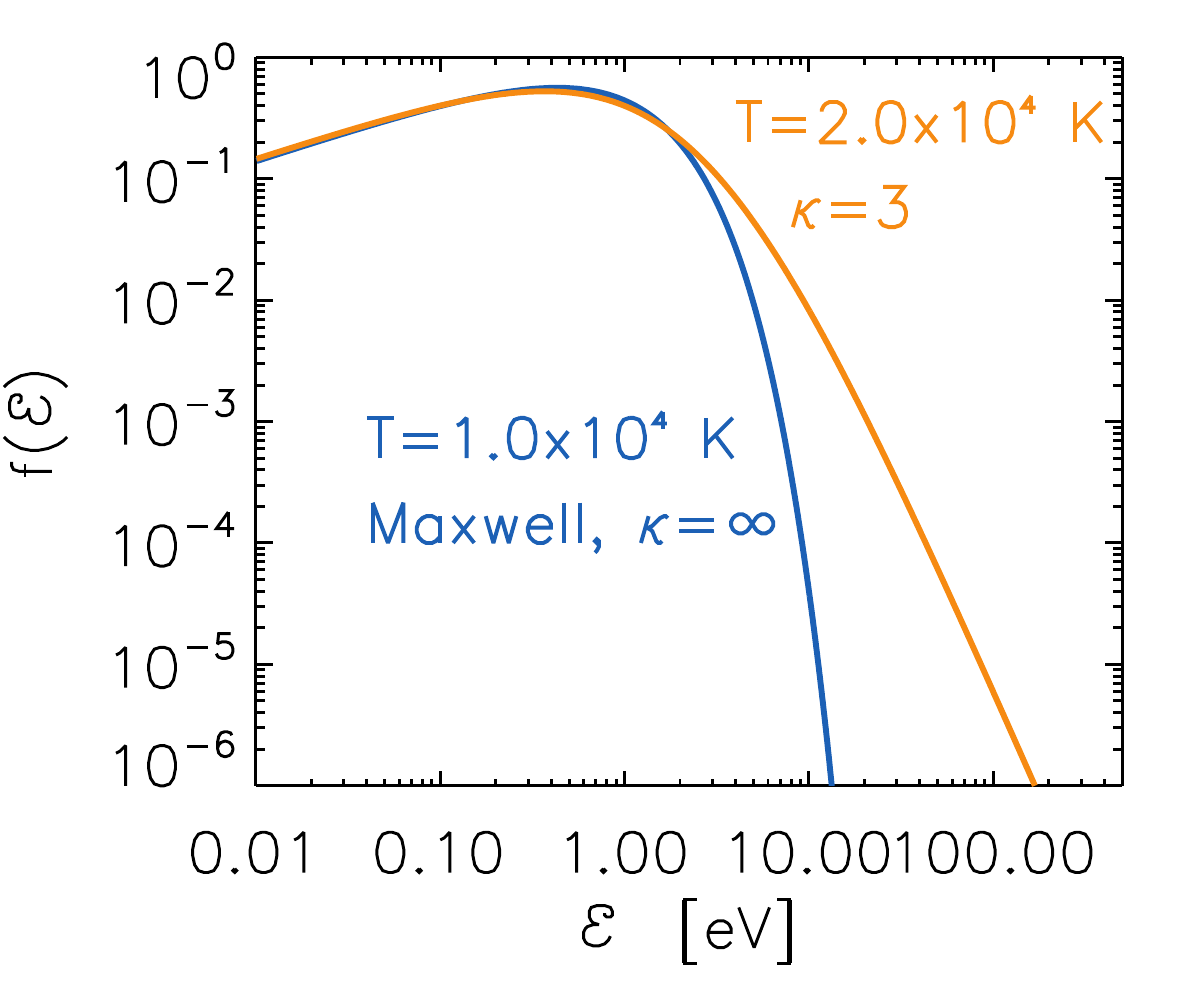}
        \includegraphics[width=5.5cm]{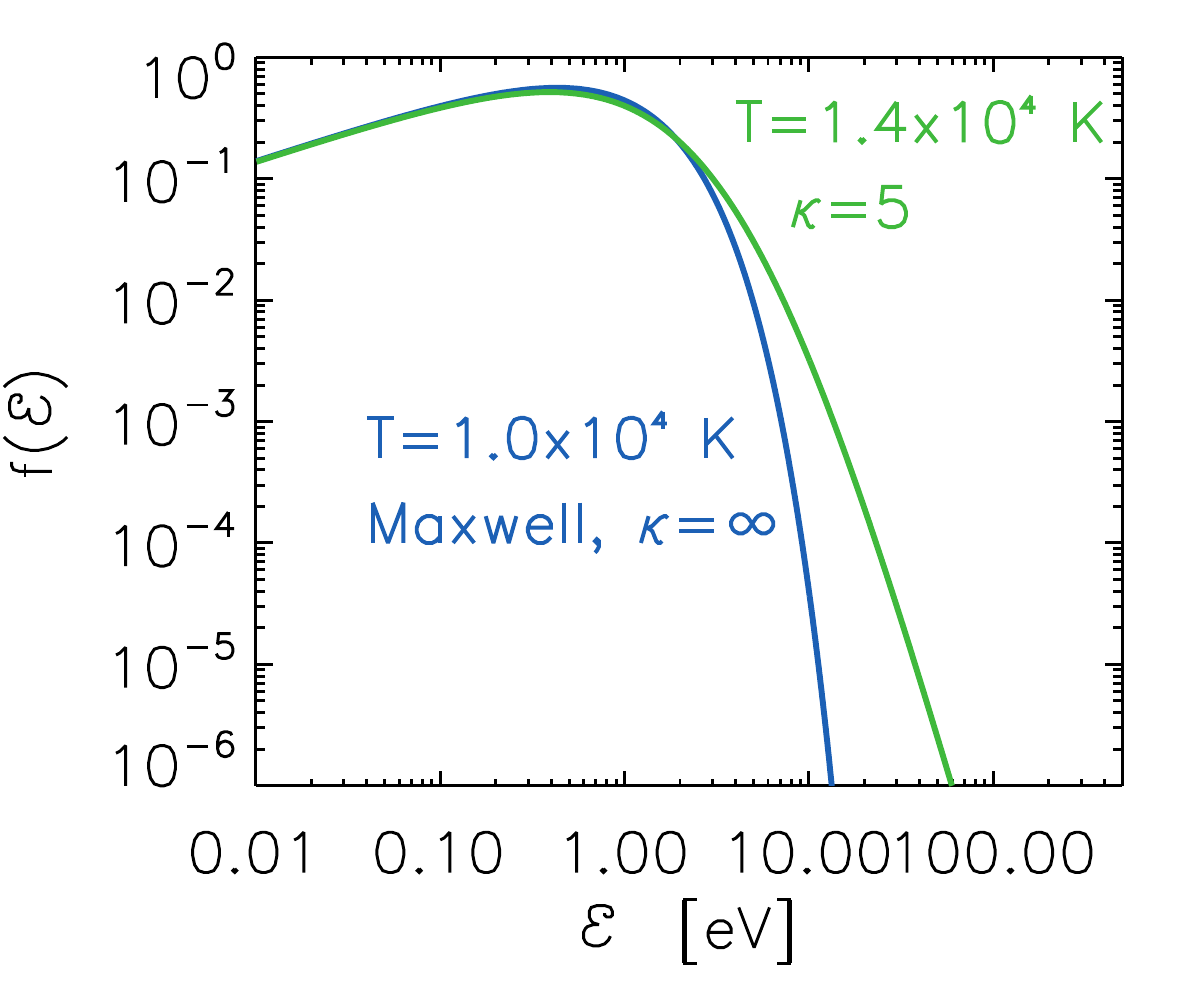}
\caption{Electron distributions assumed in our model Maxwellian distribution with temperature $10^4$ K ({\it blue line}) shown together with the $\kappa$-distribution with $\kappa=2$ for temperatures
of $4\times10^4$ K ({\it left, red line}), $\kappa=3$, $T=2\times10^4$ K ({\it middle, orange line}), and $\kappa=5$,   $T=1.4\times10^4$ K ({\it right, green line}). }
\label{Fig:distr}
\end{figure*}
%
%
\begin{figure*}[!t]
        \centering  
        \includegraphics[width=7.5cm]{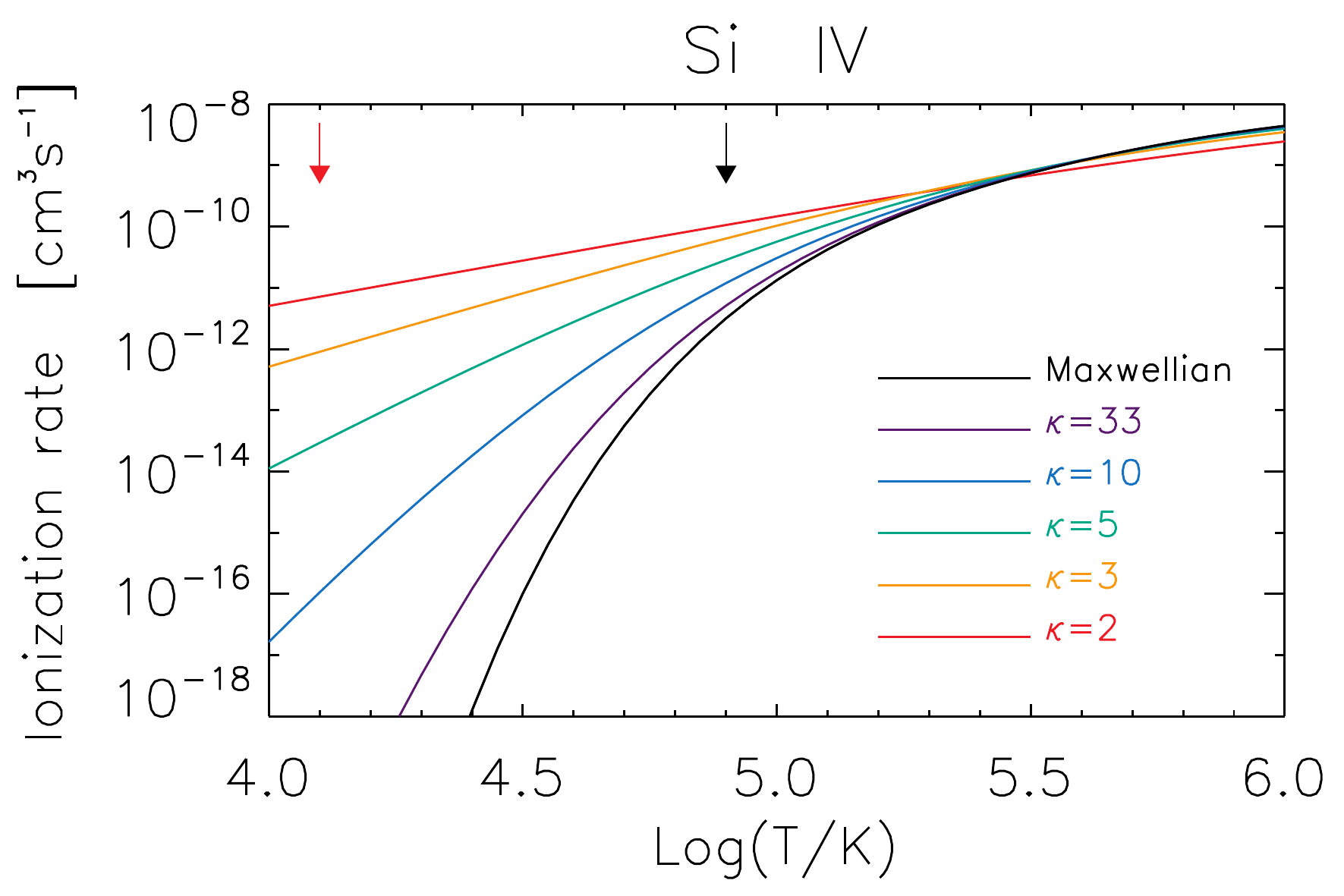}
        \includegraphics[width=7.5cm]{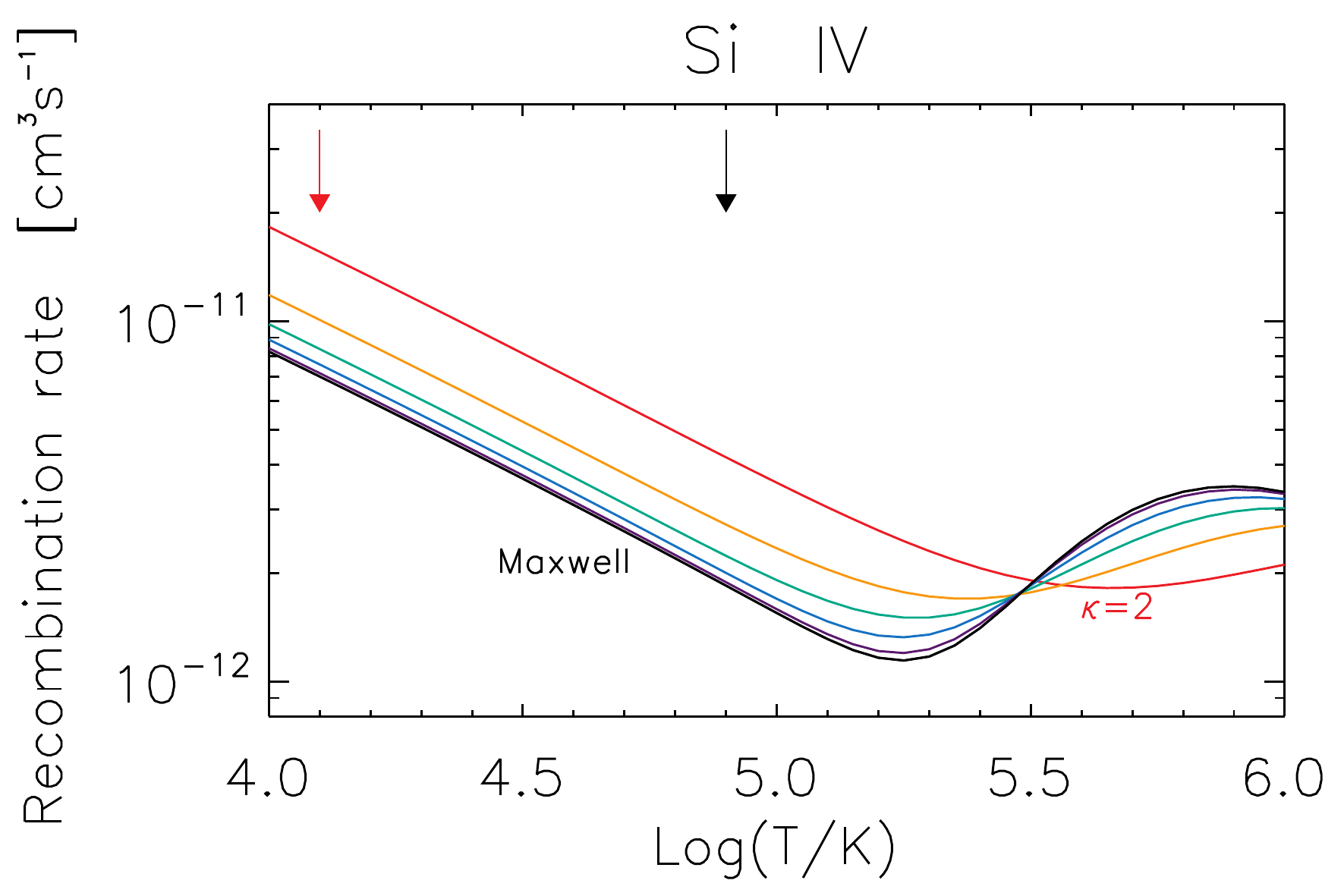}
        \includegraphics[width=7.5cm]{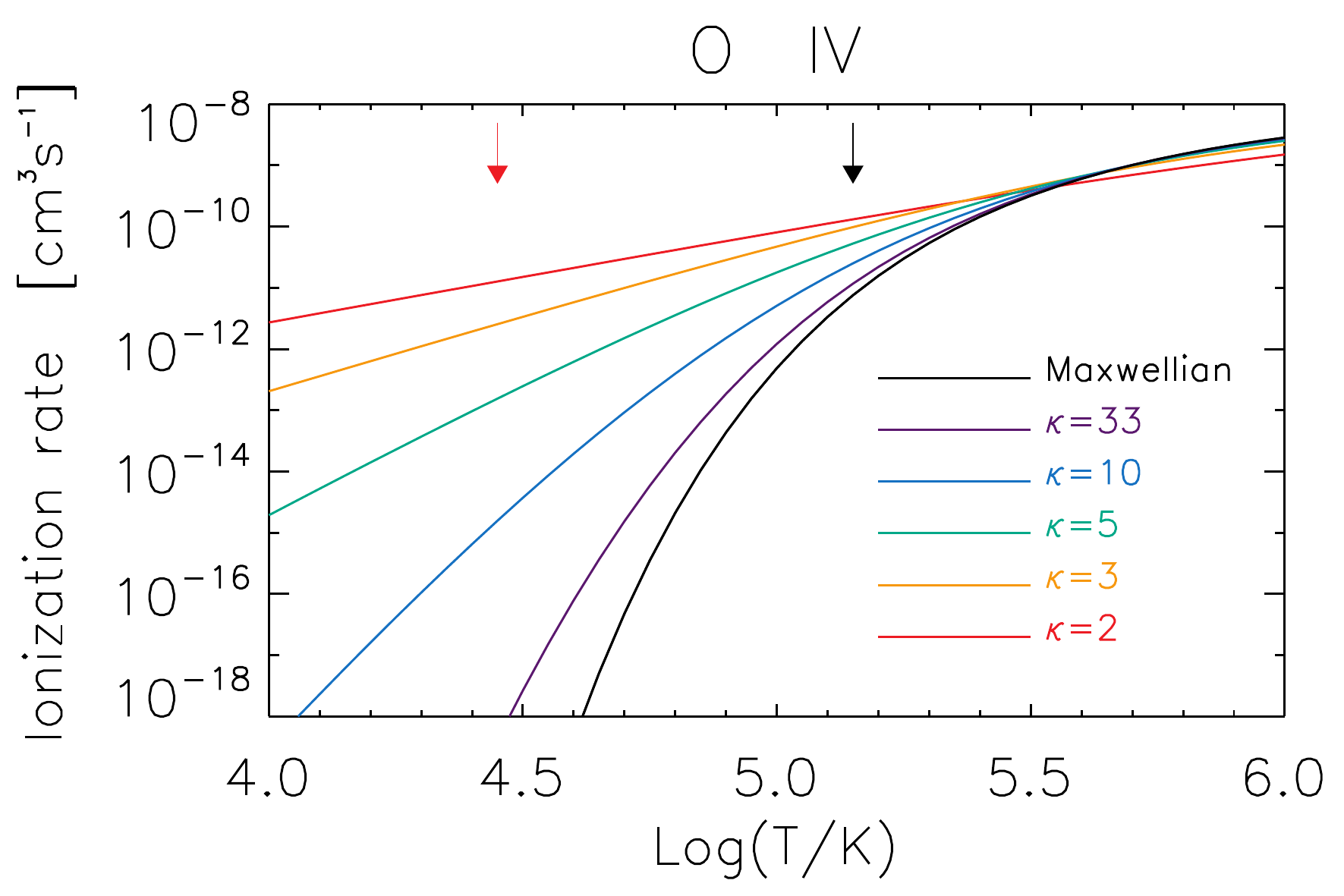}
        \includegraphics[width=7.5cm]{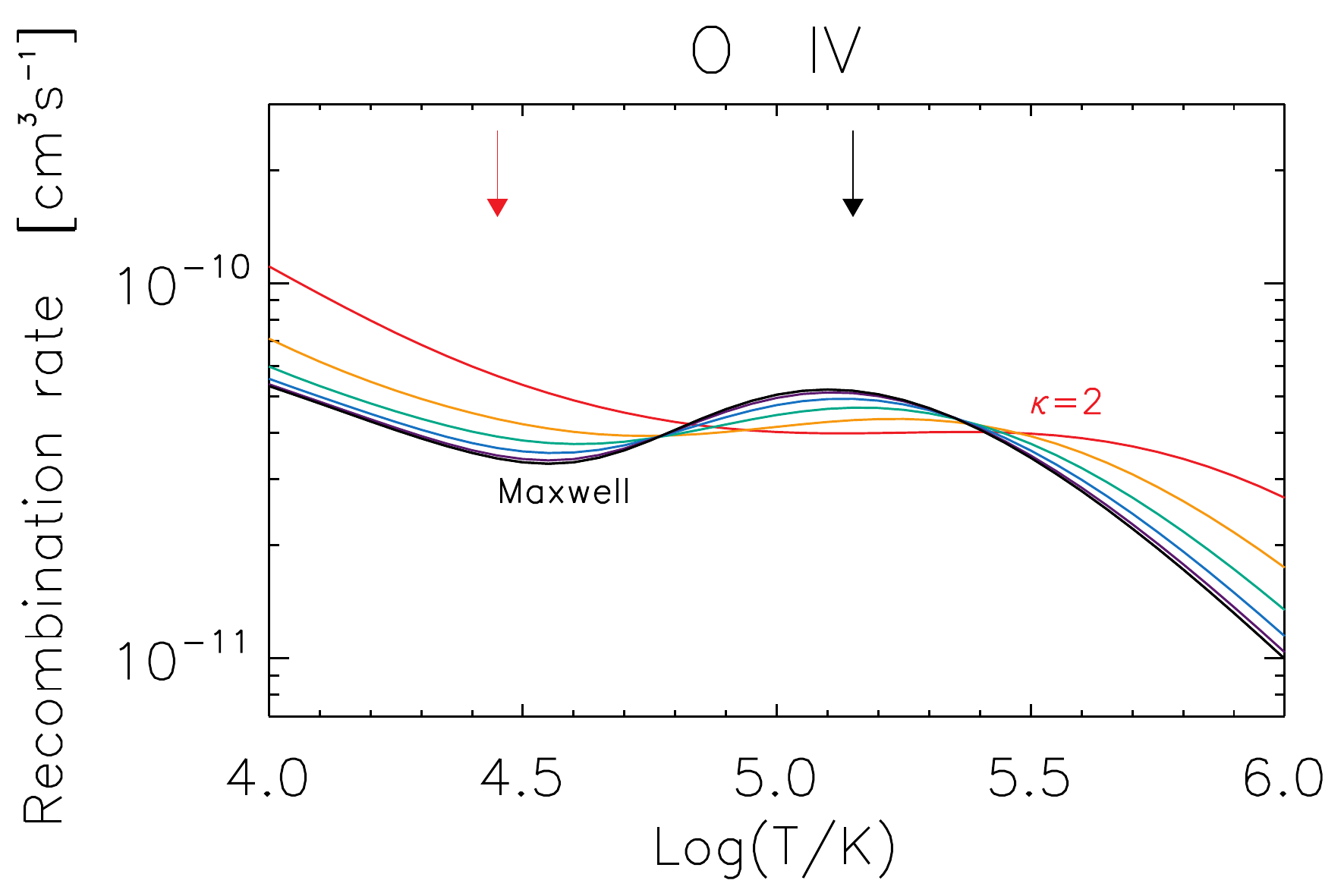}
\caption{Ionization and recombination rates for \ion{Si}{IV} (\textit{top}) and \ion{O}{IV} (\textit{bottom}) for  a Maxwellian and a $\kappa$-distribution with $\kappa=2$, 3, 5, 10, and 33. Colors denote individual $\kappa$ values. Black and red arrows denote temperatures of the maximum relative ion abundance for the Maxwellian and $\kappa$-distribution with $\kappa=2$.}
\label{Fig:rates}
\end{figure*}
%
%
\begin{figure}[!t]
        \centering
        \includegraphics[width=7.1cm,viewport= 0 55 396 275,clip]{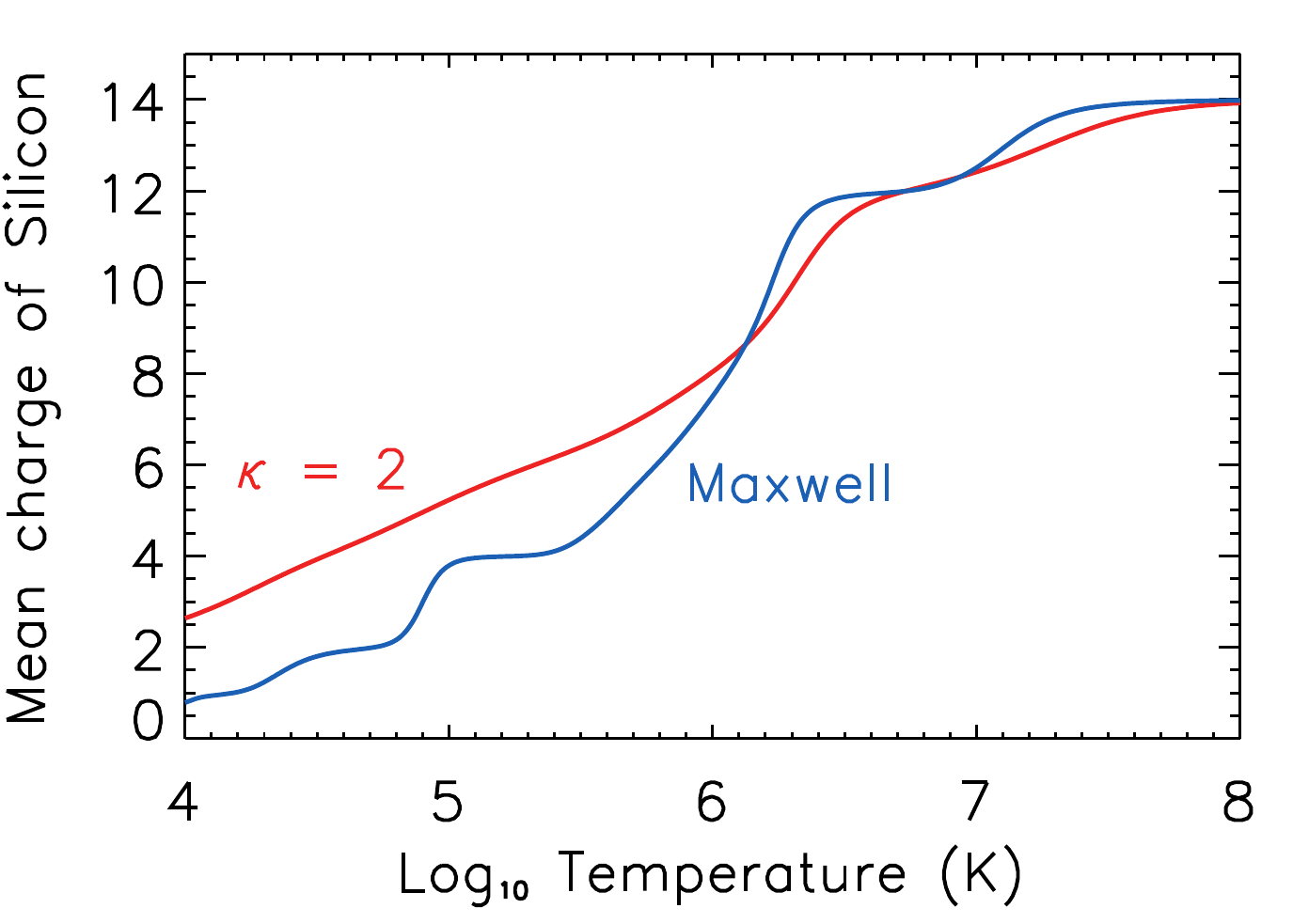}
        \includegraphics[width=7.1cm,viewport= 0  0 396 275,clip]{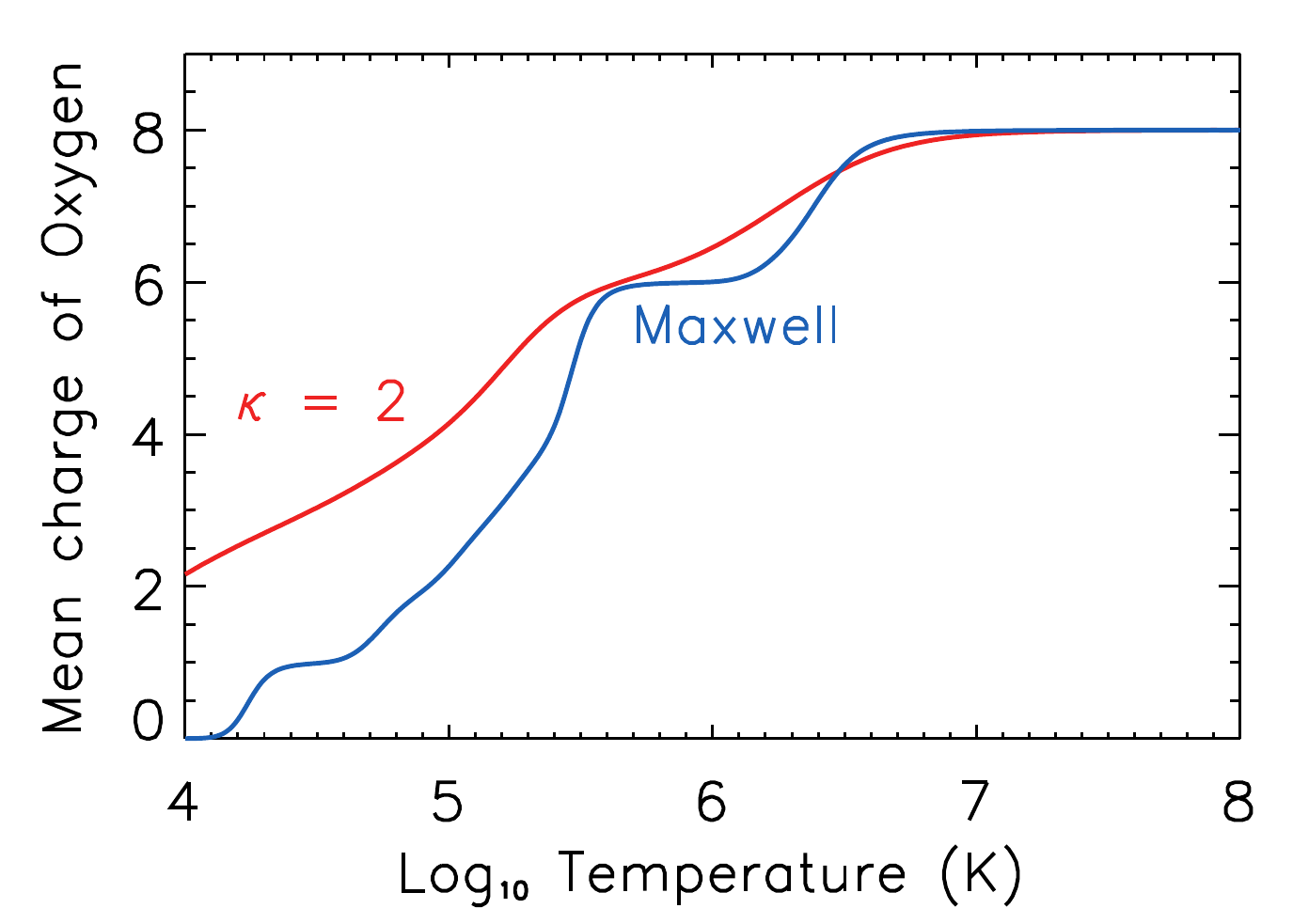}
\caption{Mean charge of silicon (\textit{top}) and oxygen (\textit{bottom}) for the Maxwellian distribution and the $\kappa$-distribution with $\kappa=2$ (blue and red lines, respectively).}  
\label{Fig:mean_charge}
\end{figure}
%

%
%
\section{Method}
\label{Sect:2}

We used the same simple model setup as in Paper~I, but with a much lower temperature $T$, corresponding to the non-Maxwellian conditions in the TR. We briefly summarize the salient points here. The model consists of a periodic electron beam passing through plasma that is switched on and off with a period of $P/2$. Initially, the plasma is Maxwellian with a temperature $T$. At time $t=P/2$s, the electron beam is switched on until $t$\,=\,$P,$ when it is switched off for another $P/2$. The distribution of plasma with the electron beam is represented by a $\kappa$-distribution \citep[e.g.,][]{Olbert68,Vasyliunas68a,Vasyliunas68b,Owocki83,Livadiotis15b,Dzifcakova15,Dudik17b},
\begin{equation}
        f_\kappa(E)dE= A_\kappa  \frac{2}{\sqrt{\pi}\left(k_\mathrm{B}T_\kappa\right)^{3/2}} \frac{E^{1/2}dE}{\left(1+\frac{E}{(\kappa-3/2)k_\mathrm{B}T_\kappa}\right)^{\kappa+1}}\,,
 \label{Eq:Kappa}
\end{equation}
which is chosen since (i) it contains a power-law high-energy tail and (ii) its bulk can be approximated by a Maxwellian with temperature $T$ \citep[see][]{Oka13},
\begin{equation}
        T = \frac{\kappa -3/2}{\kappa}T_\kappa\,.
        \label{Eq:T_kappa}
\end{equation}
Examples of this approximation are shown in Fig. \ref{Fig:distr}. We note that in this manner, the switching on of the beam effectively adds particles that traverse the plasma. For $\kappa$\,=\,2, 3, 5 (Fig. \ref{Fig:distr}), $T_\kappa$ is 4, 2, and 1.4 times higher than the $T$ corresponding to the Maxwellian bulk, while the $\kappa$\,=\,distribution adds 56\%, 34\%, and 19\% particles, respectively.

It is assumed that except for the switch from the Maxwellian to a $\kappa$-distribution, the electron distribution function does not evolve any further, that is, there is no heating by thermalization of energetic particles. As we described in Paper~I, this is a crude approximation, meaning that the beam is generated elsewhere along the same magnetic structure, and it only traverses the region of interest for a duration of $P/2$ at each period. The switch on of the beam alters the ionization and recombination rates, however, and we only aim to show general non-equilibrium effects in TR line intensities of \ion{Si}{IV}, \ion{O}{IV}, and \ion{S}{IV}, without any additional assumptions on TR geometry, details of particle acceleration, or the (magneto-)hydrodynamic evolution of a particular TR structure element.

Changes in ionization rates $I_i$ and recombination rates $R_i$ for $\kappa$ were derived by \citet{Dzifcakova13a} and are shown in Fig. \ref{Fig:rates}. Generally, the $\kappa$-distributions produce changes in the recombination rates $R_i$ up to a factor of about $\approx$2, while the ionization rates are increased by orders of magnitude by the high-energy tails of the $\kappa$-distributions. For example, the Maxwellian ionization rate for  \ion{Si}{IV} is approximately $10^{-15}$ cm$^3$s$^{-1}$ at $T$\,=\,4\,$\times10^4$\,K, while for $\kappa$\,=\,2, it is about $10^{-11}$ cm$^3$s$^{-1}$,
or in other words, four orders higher at the same temperature. However, for \ion{O}{IV}, these ionization rates are at least ten times lower than the \ion{Si}{IV} rates at the same temperature (Fig. \ref{Fig:rates}). The corresponding ionization equilibration times for $\kappa$\,=\,2 are typically tens of seconds at the electron densities $10^{10}-10^{11}$ cm$^{-3}$ for \ion{Si}{IV} and ten times longer for \ion{O}{IV} at a given electron density \citep[see also][]{Smith10}. Therefore, any periods $P$ of the electron beam recurrence shorter than these timescales are expected to drive the plasma out of ionization equilibrium (cf. Paper~I). 

The instantaneous ionic composition is obtained by solving the first-order differential equation for the non-equilibrium ionization \cite[e.g.,][]{Bradshaw03b,Kaastra08},
\begin{equation}
 \frac{d Y_i}{d t} = N_\mathrm{e} \left( I_{i-1} Y_{i-1} + R_i Y_{i+1} - I_i Y_i - R_{i-1} Y_i\right)\,,
 \label{Eq:noneqion}
\end{equation}
through the Runge-Kutta method. In this equation, $Y_i$ represents the ion abundance of the $+i$ ion of the element $Y$, normalized to unity, and $N_\mathrm{e}$ stands for the electron density. We note that the ionization rates $I_i$ in Eq. \ref{Eq:noneqion} contain contributions from direct ionization and autoionization, while the recombination rates are the sum of the radiative and dielectronic recombination. The ionization and recombination rates for the $\kappa$-distributions were taken from \cite{Dzifcakova13a}.
There, the ionization rates were obtained by direct integration of the cross sections that are included in the CHIANTI database \citep{Dere97,Dere09,DelZanna15b} over the $\kappa$-distributions, while the recombination rates were obtained through the parametric approximate method \citep[as detailed in Sect. of][]{Dzifcakova13a}. This method allows calculating non-Maxwellian recombination rates if the Maxwellian rates (available in CHIANTI) are known.

Similarly to Paper~I, we specified the ionization state and its evolution through the effective ionization temperature $T_\mathrm{eff}$. This$\text{}_\mathrm{}$ is a quantity derived from the mean charge, assuming ionization equilibrium. The mean charge of an element increases with temperature. For TR ions, it is higher for the  $\kappa$-distributions than for the Maxwellian distribution (Fig. \ref{Fig:mean_charge}). This will have important consequences if the non-equilibrium charge state of the plasma (Sect. \ref{Sect:3}) is interpreted in terms of an equilibrium Maxwellian or a $\kappa$-distribution.

The time-dependent ion abundances obtained using Eq. (\ref{Eq:noneqion}) were used to calculate the synthetic spectra of \ion{Si}{IV}, \ion{O}{IV}, and \ion{S}{IV}  in the $1349-1410$\,\AA~wavelength range observed by the \textit{IRIS} instrument. Details of the spectral synthesis method are described in Sect. 2.3 of Paper~I. We note that the relative level population used in the spectral synthesis assumes instantaneous excitation equilibrium and is thus calculated from the respective excitation and de-excitation cross sections and radiative rates using the procedure outlined in Sect. 3 of \citet{Dudik14a}, and applied to TR lines in \citet{Dudik14b}. To do this, we used the atomic data of \citet{Liang09} for \ion{Si}{IV} and of \citet{Liang12} for \ion{O}{IV}. These atomic data are the same as in CHIANTI v8 \citep{Dere97,DelZanna15b}. For \ion{S}{IV}, we used the atomic data of \citet{DelZanna16}. This is a different set of atomic data than CHIANTI v8, which mainly relies on atomic data from \citet{Tayal99,Tayal00}. Further details on the atomic datasets for these ions can be found in Appendix A of \citet{Polito16b}.

We included photoexcitation by a photospheric blackbody with an effective temperature of 6000\,K in the calculation of the relative
level populations and assumed low TR heights above the photosphere. The importance of photoexcitation for TR line intensities for any $\kappa$ was shown by \cite{Dzifcakova11Si} for \ion{Si}{III} lines. \cite{Dudik14b} demonstrated that photoexcitation contributes significantly (several ten percents) to \ion{Si}{IV} intensities. For \ion{O}{IV,} however, the photoexcitation is negligible under typical TR conditions. Thus, including photoexcitation decreases the \ion{O}{IV}\,/\,\ion{Si}{IV} intensity ratios.
Finally, the results obtained here were derived for the electron densities $N_\mathrm{e}=10^{10}$ cm$^{-3}$ and $10^{11}$ cm$^{-3}$, several different temperatures $T$ of the initial Maxwellian plasma spanning log($T$/K)$=4.0-4.5$, and several values of $\kappa$\,=\,2, 3, and 5, representing a strong, a  medium, and a weak electron beam, respectively. 
In these calculations, the periods $P$ we considered are 5 and 10\,s. The values of $N_\mathrm{e}$ were chosen arbitrarily to represent typical TR densities while being different enough for non-equilibrium ionization to be expected for $N_\mathrm{e}$\,=\,10$^{10}$\,cm$^{-3}$, while the plasma is expected to be closer to and even reach ionization equilibrium at 10$^{11}$\,cm$^{-3}$ \citep[see Fig. 1 of][]{Smith10}. These combinations of parameters are sufficient to cover the phase space in which the plasma experiences non-equilibrium ionization as a result of the electron beam. We note that the choice of $T$\,=\,10$^4$\,K can be thought of as the electron beams heating plasma at chromospheric temperatures (but relatively low densities), such as in UV (Ellerman) bursts or in small arch filament systems undergoing magnetic reconnection \citep[cf.][]{Peter14Sci,Grubecka16,Zhao17}. Our simple model is not able to consider lower initial temperatures or higher densities, since in such conditions the plasma can become optically thick. In addition, this plasma would be too dense for non-equilibrium ionization and energetic particles to play a significant role.

\begin{figure*}[!ht]
        \centering
        \includegraphics[width=5.4cm]{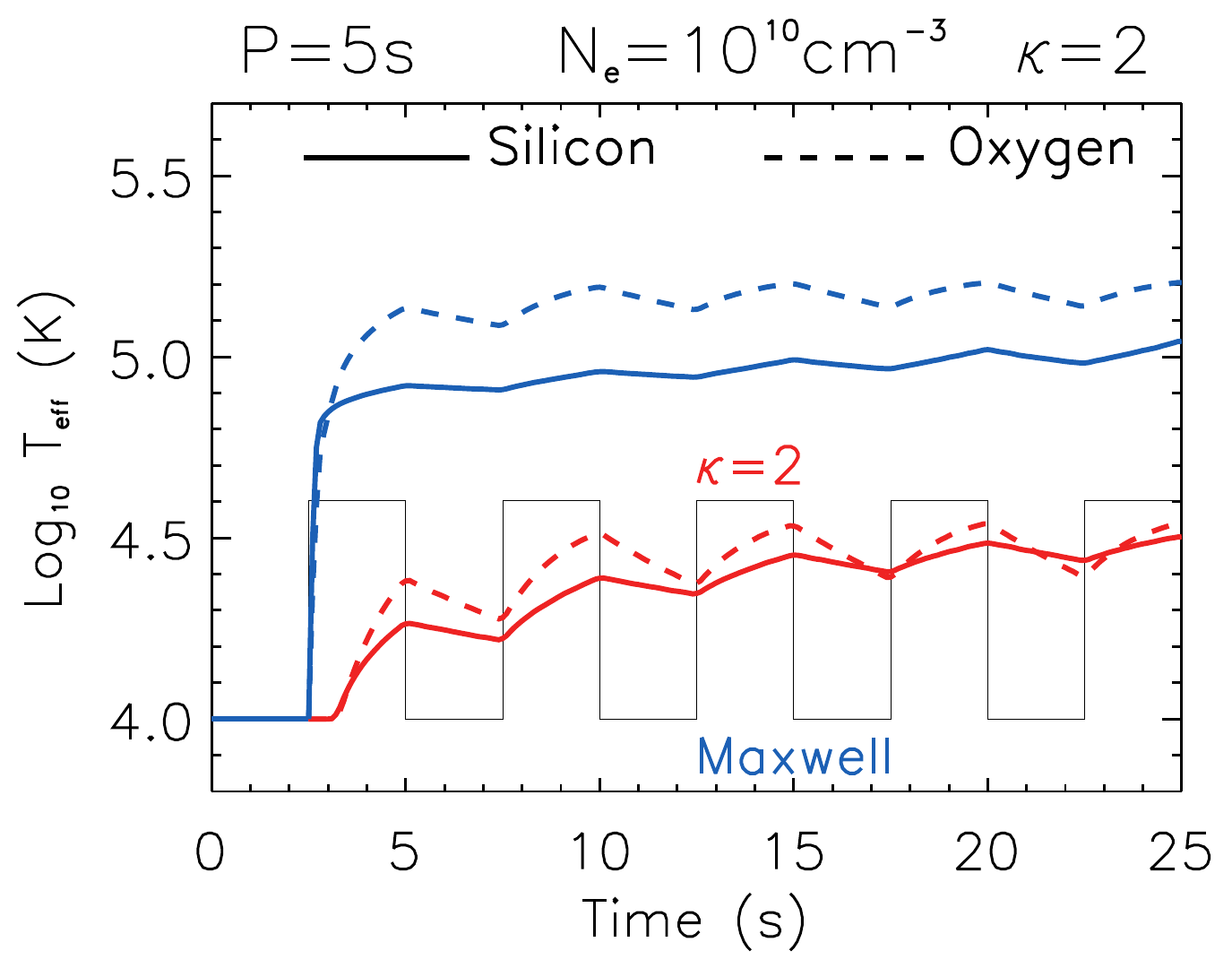}
        \includegraphics[width=5.4cm]{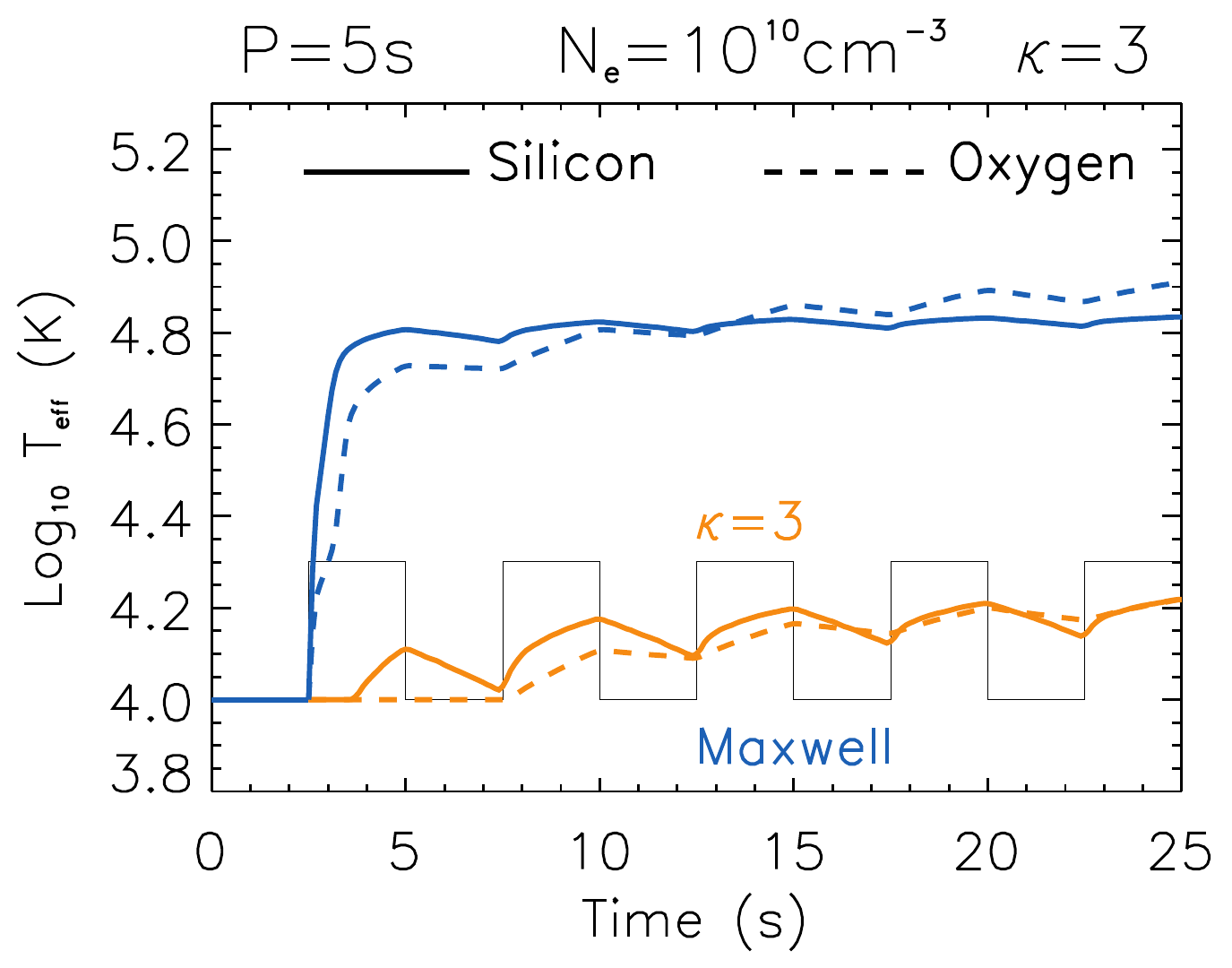}
        \includegraphics[width=5.4cm]{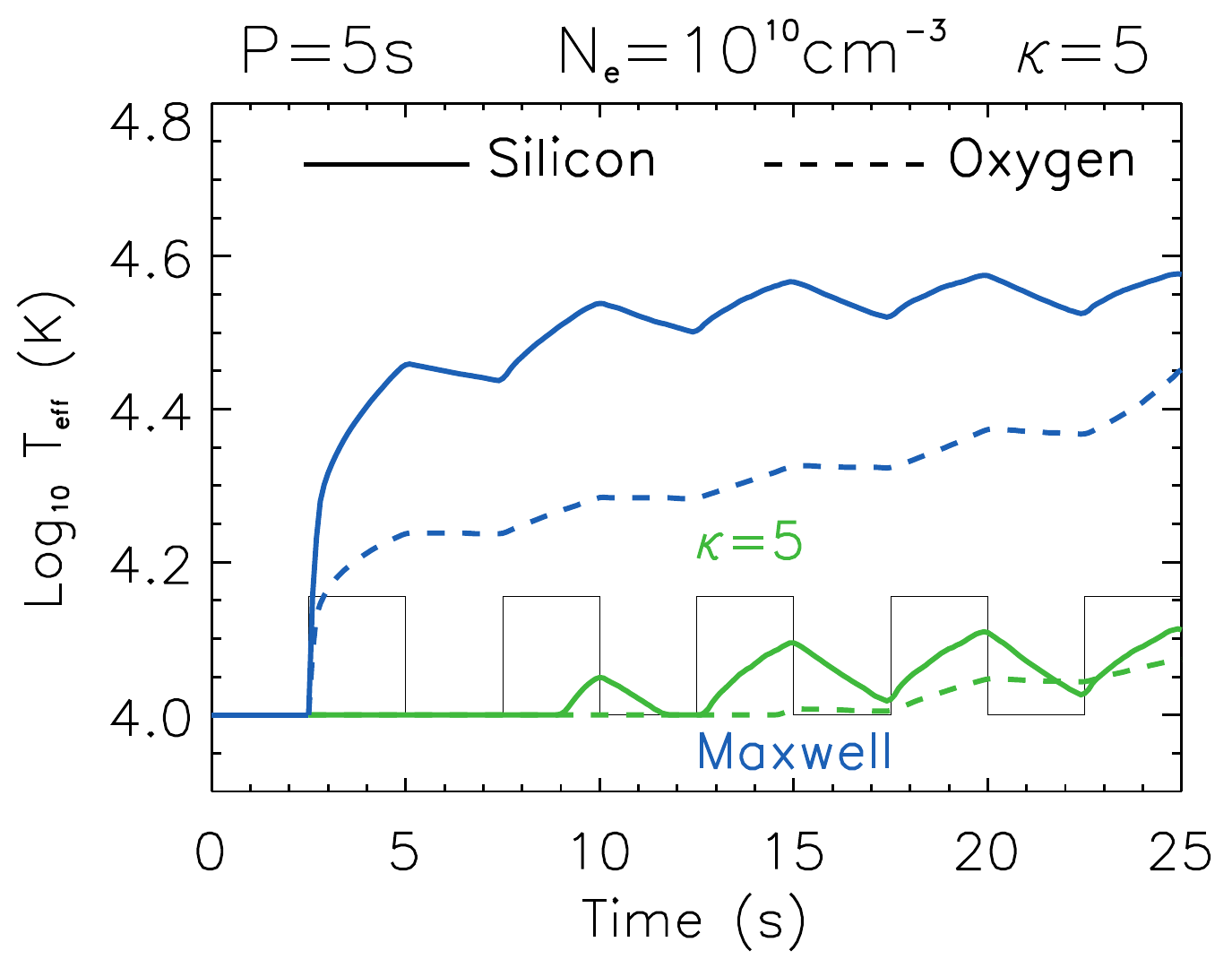}
        \includegraphics[width=5.4cm]{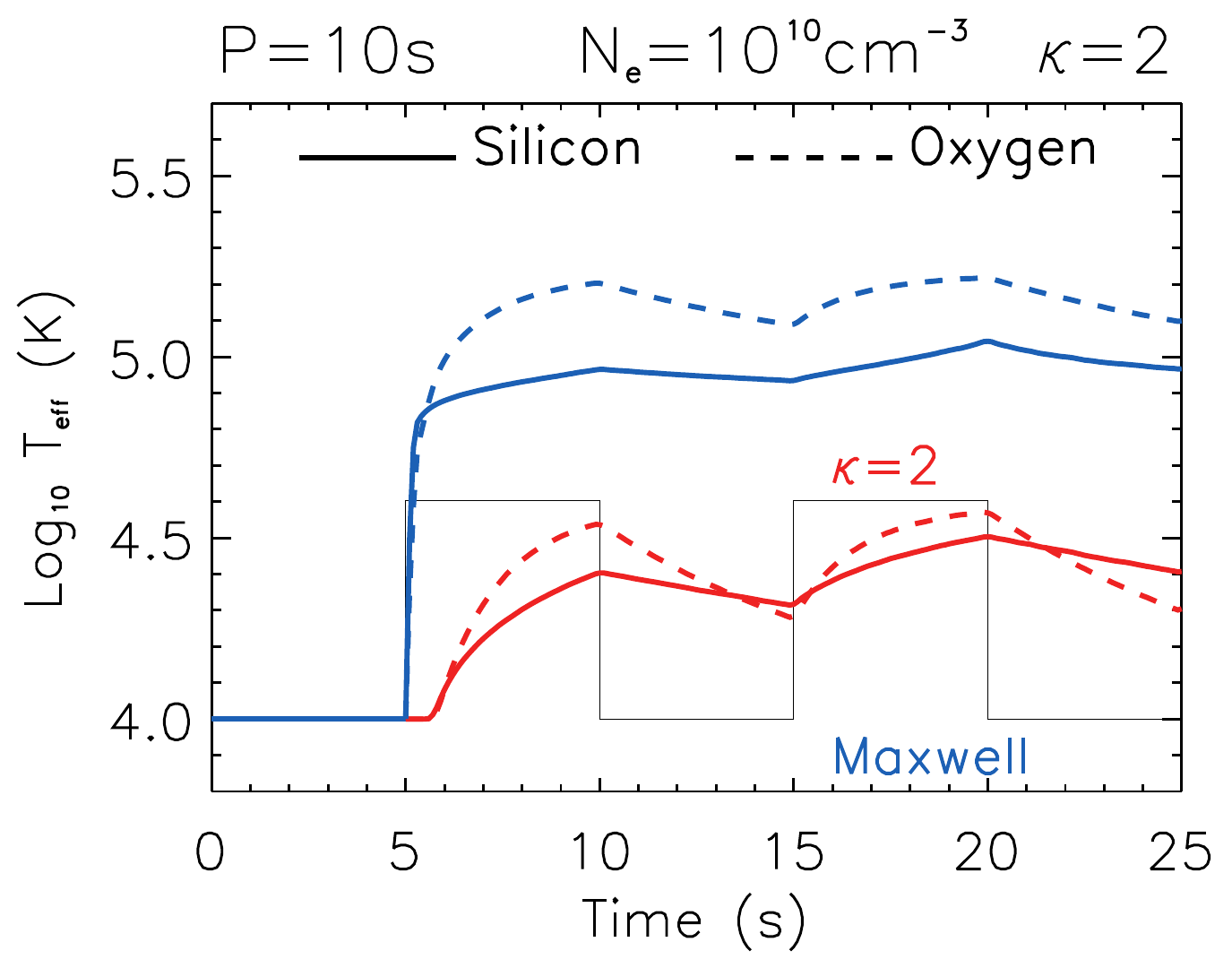}
        \includegraphics[width=5.4cm]{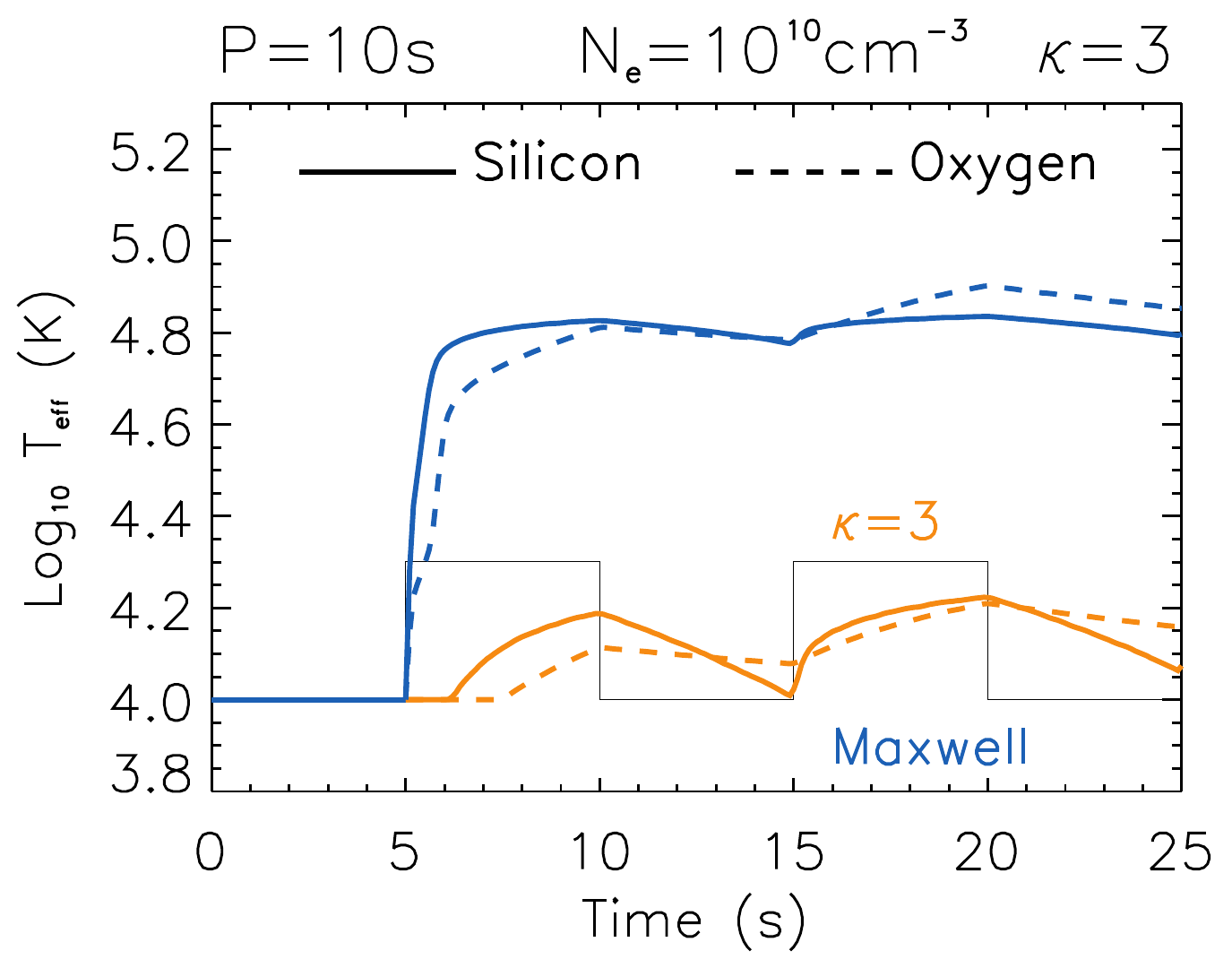}
        \includegraphics[width=5.4cm]{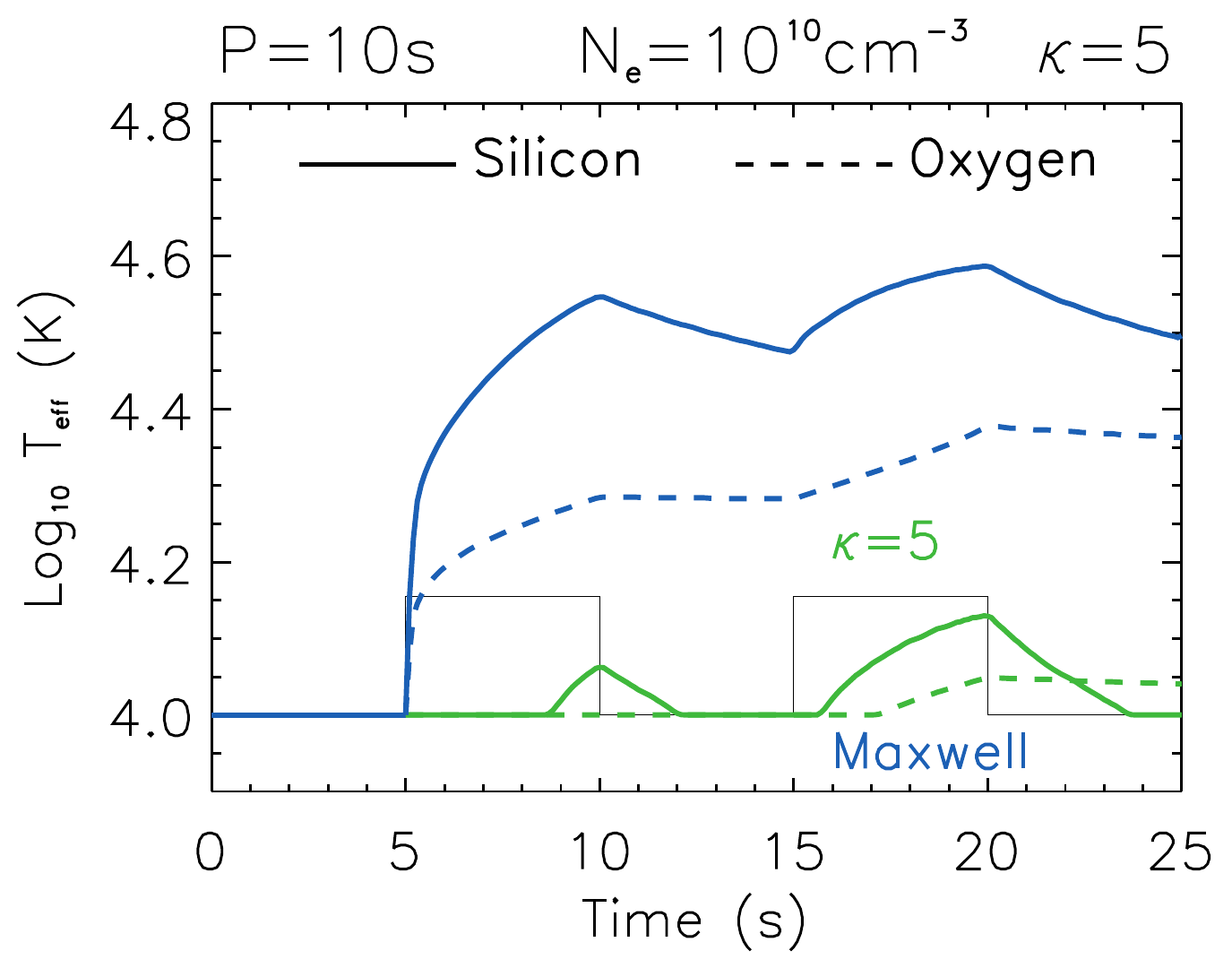}
        \includegraphics[width=5.4cm]{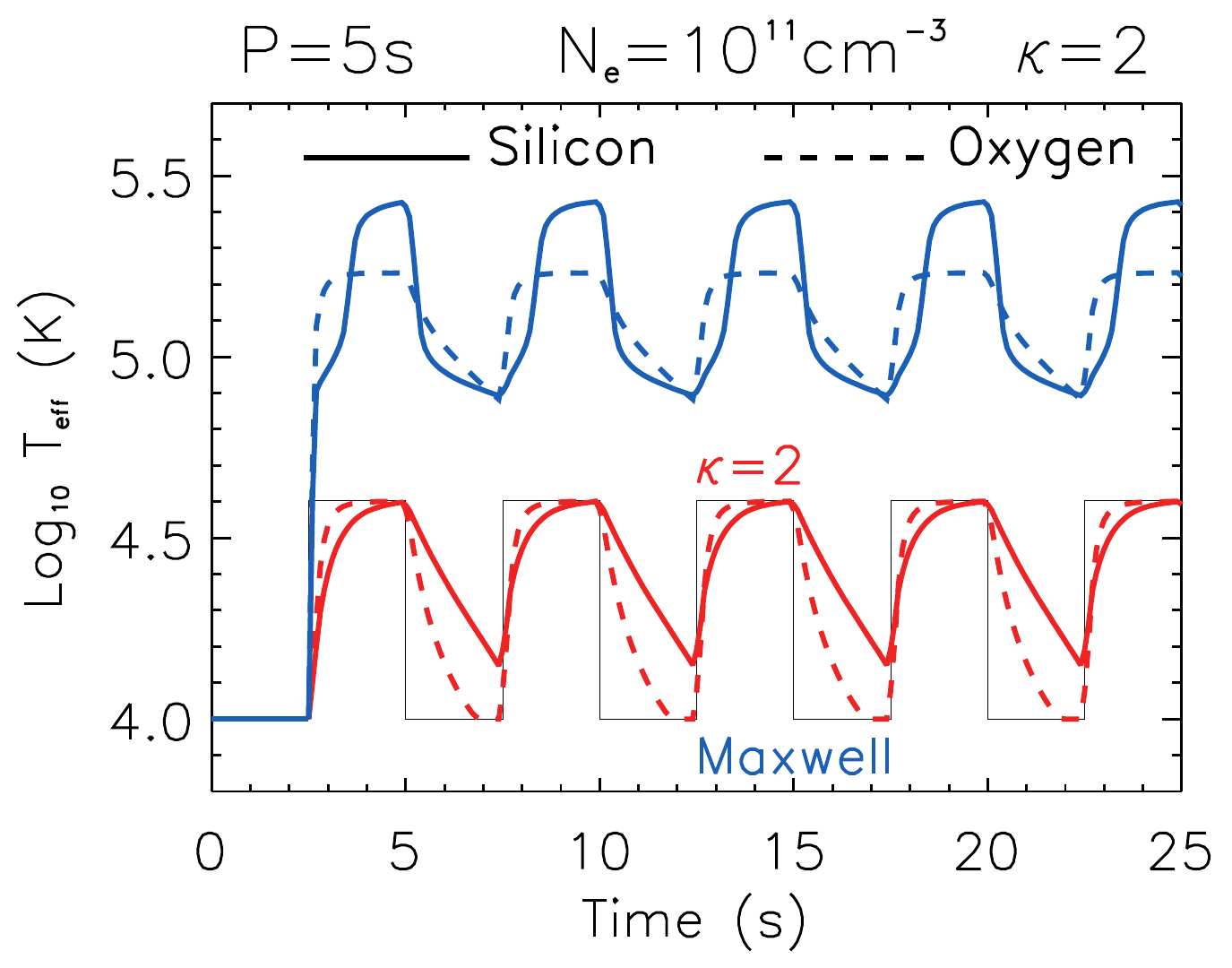}
        \includegraphics[width=5.4cm]{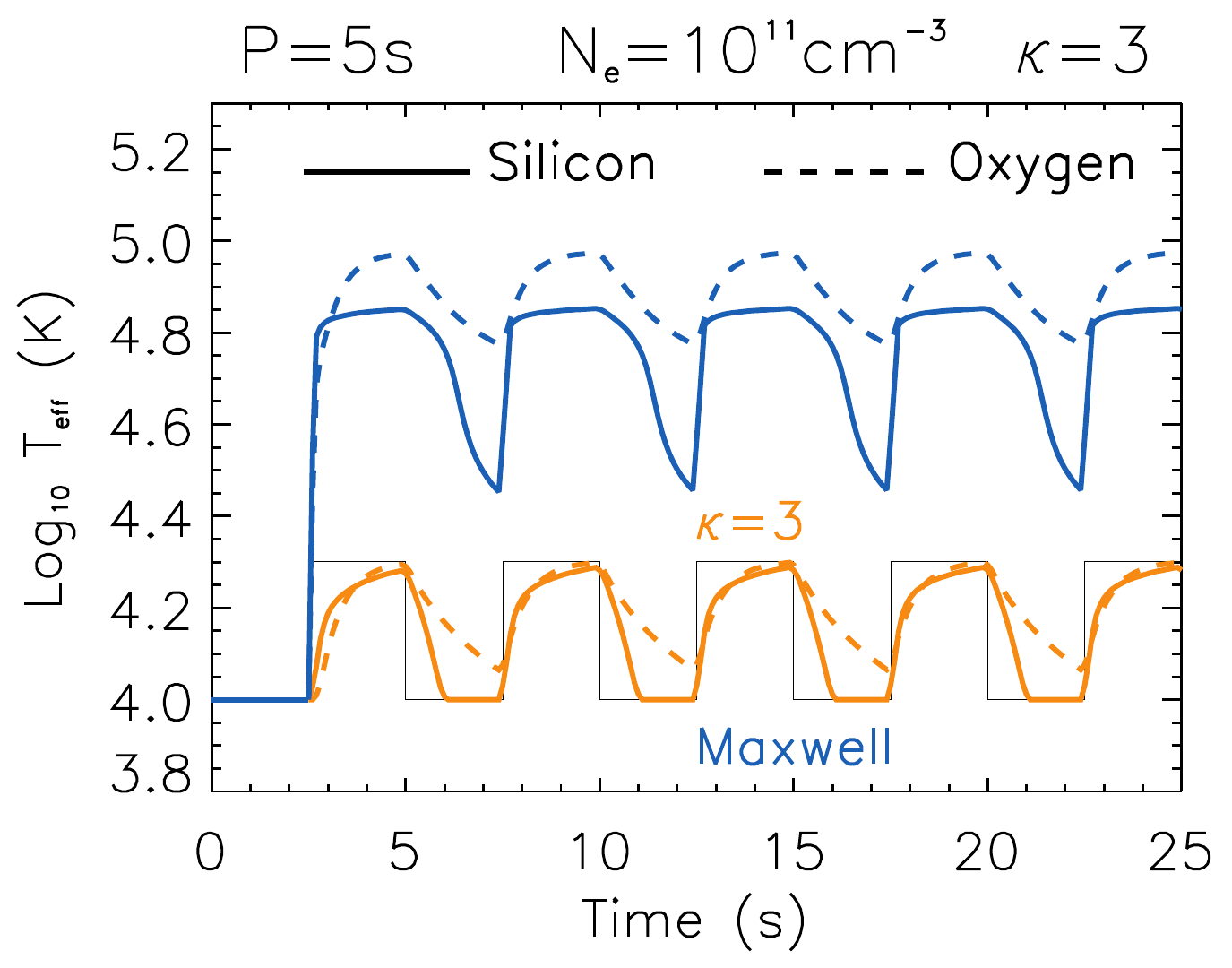}
        \includegraphics[width=5.4cm]{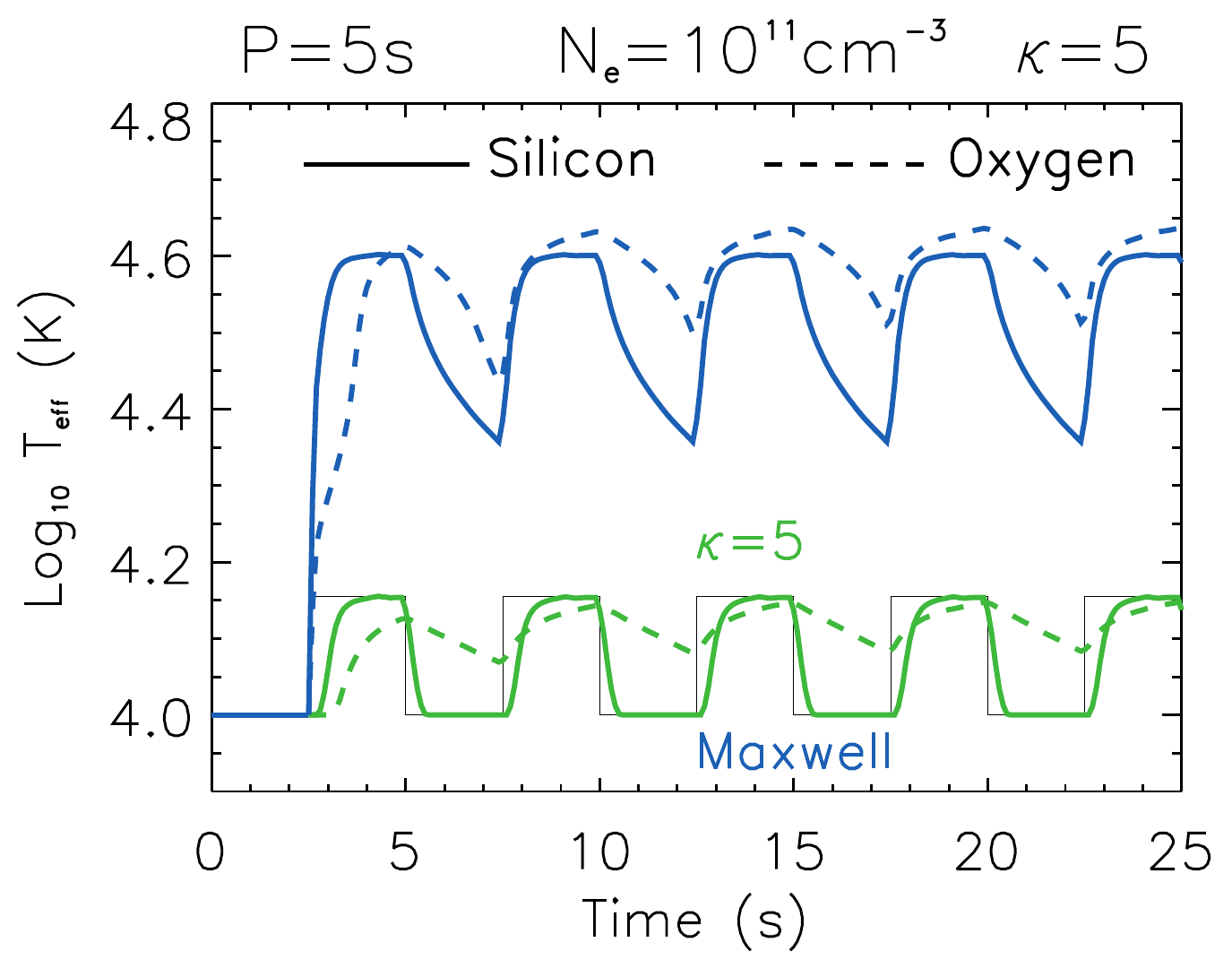}
        \includegraphics[width=5.4cm]{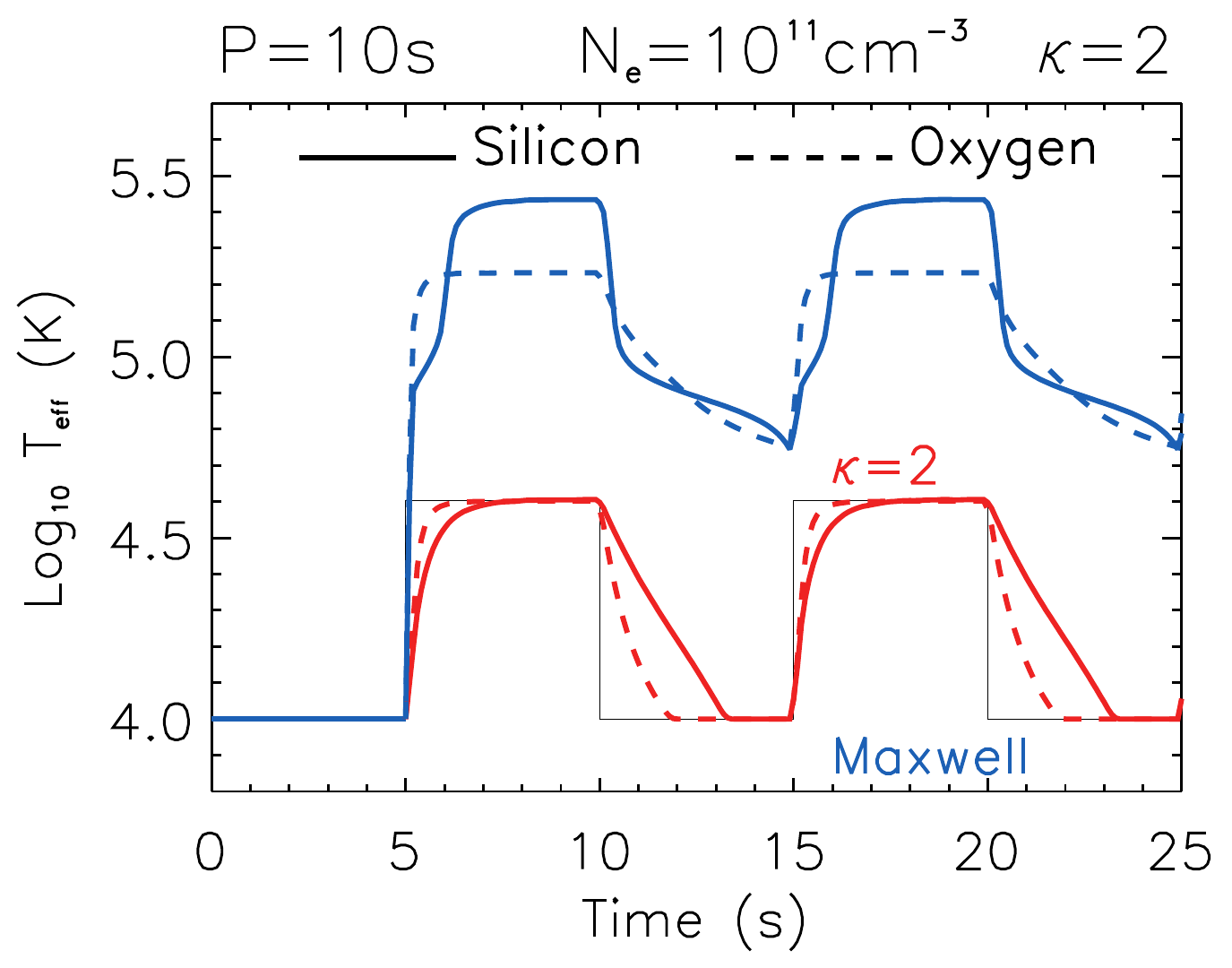}
        \includegraphics[width=5.4cm]{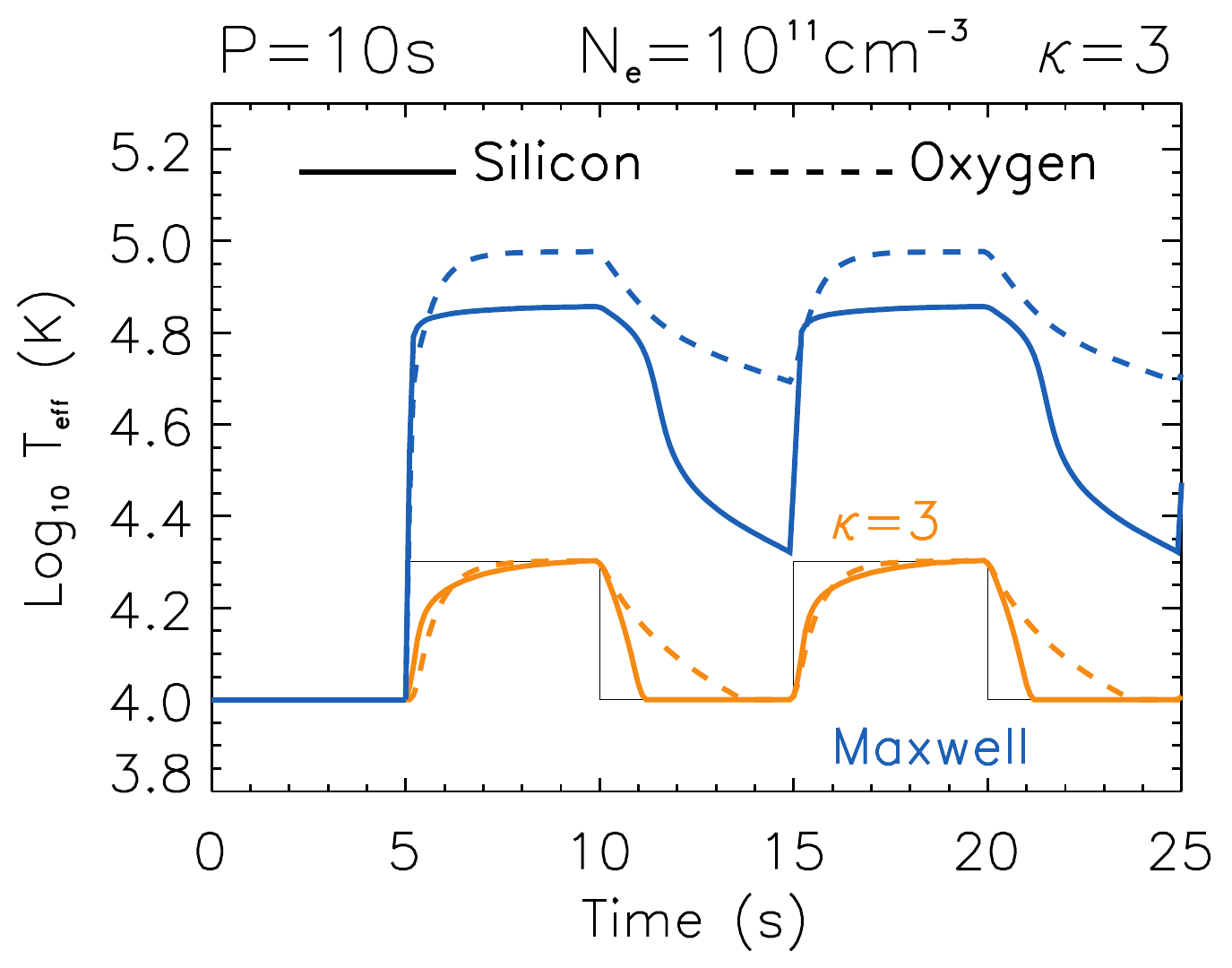}
        \includegraphics[width=5.4cm]{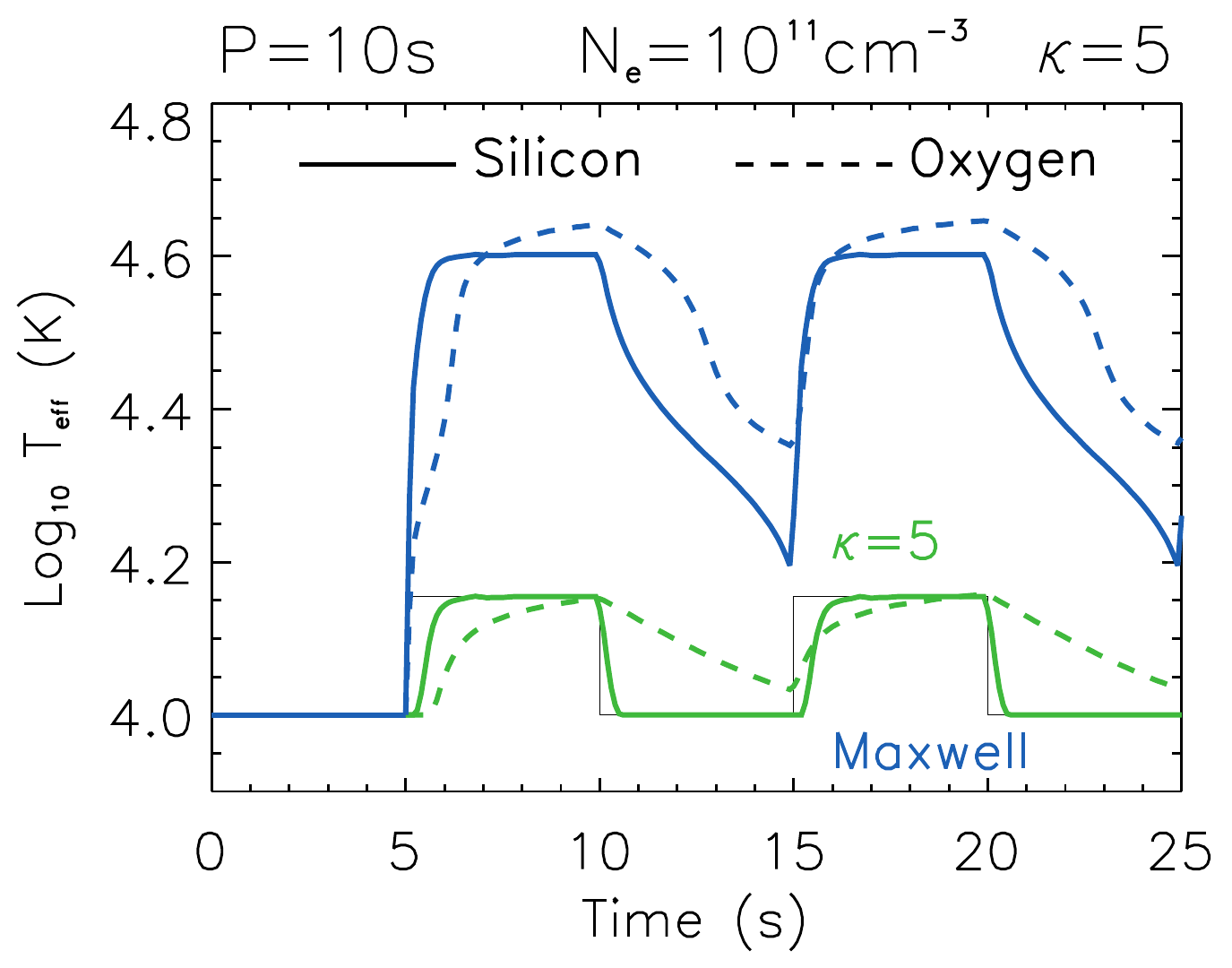}
\caption{Evolution of the effective temperature $T_\mathrm{eff}$ for Si (full lines) and O (dashed lines) under the assumption of the Maxwellian distribution (blue) and $\kappa$-distribution with $\kappa=2$, 3, and 5 (red, orange, and green, respectively). The model parameters are indicated in each panel. Rows 1--2 correspond to $N_\mathrm{e}=10^{10}$ cm$^{-3}$, while rows 3--4 stand for $N_\mathrm{e}$\,=\,10$^{11}$ cm$^{-3}$.}
\label{Fig:Teff_evol}
\end{figure*}
%
%
\begin{figure*}[!ht]
        \centering
        \includegraphics[width=5.4cm]{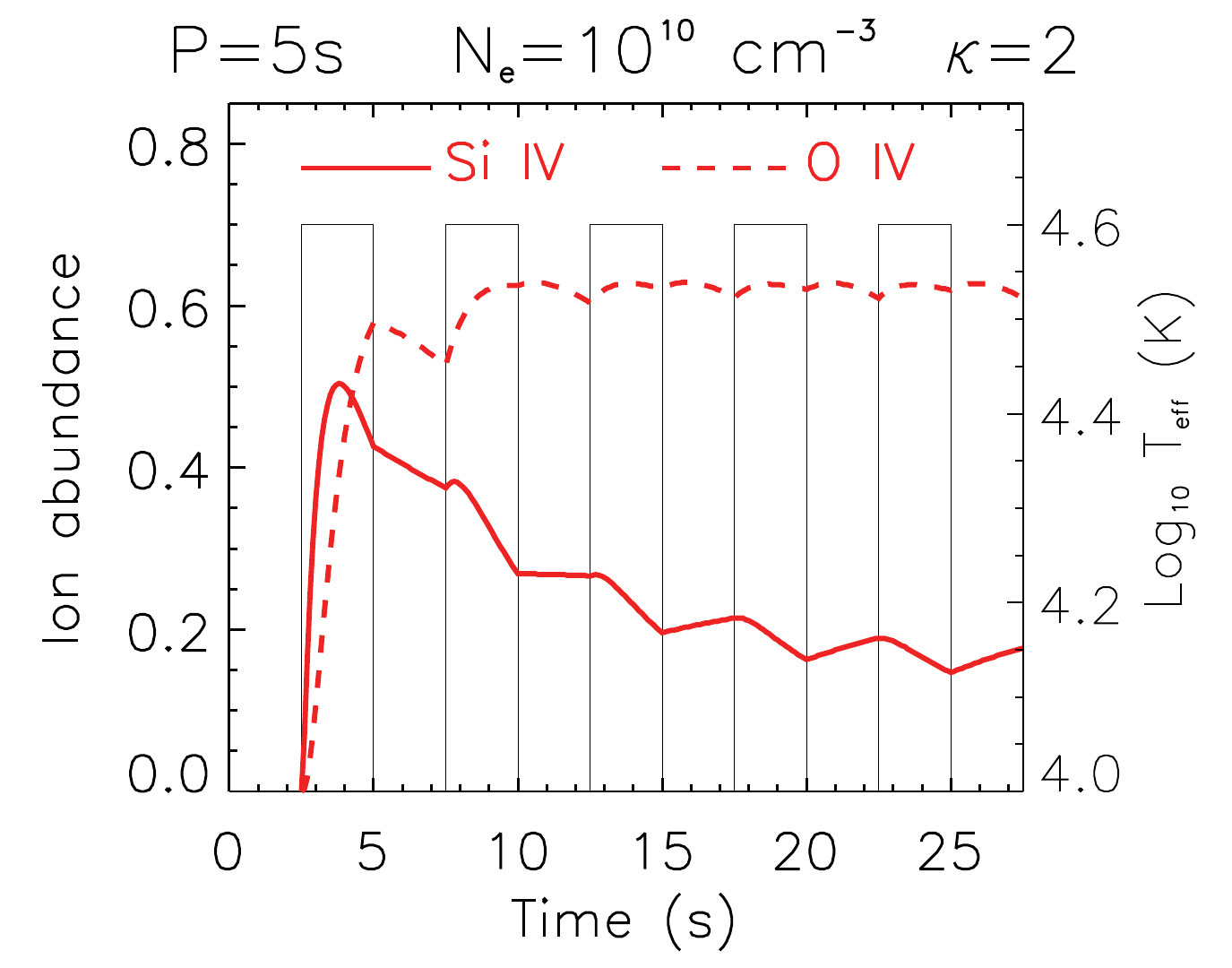}
        \includegraphics[width=5.4cm]{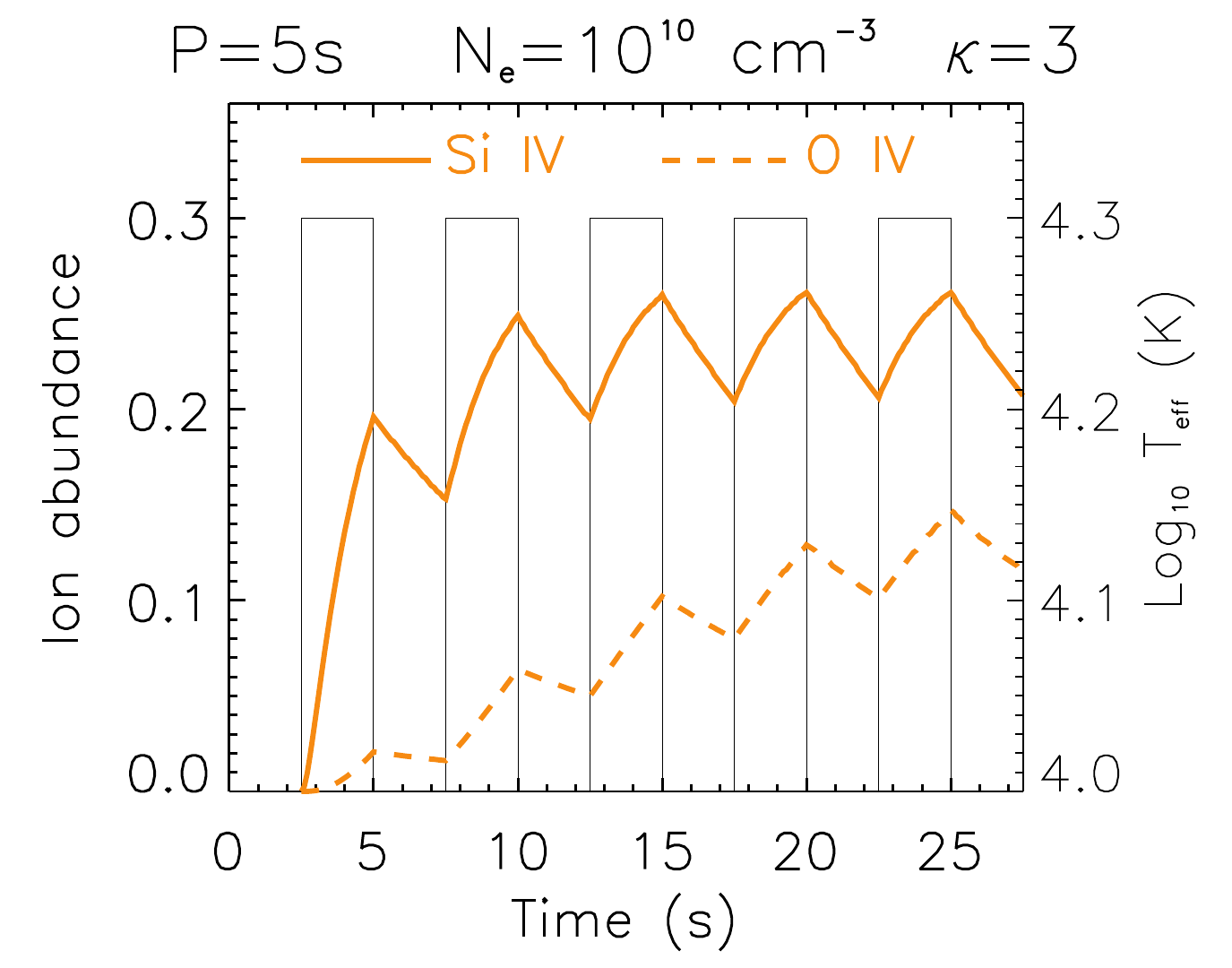}
        \includegraphics[width=5.4cm]{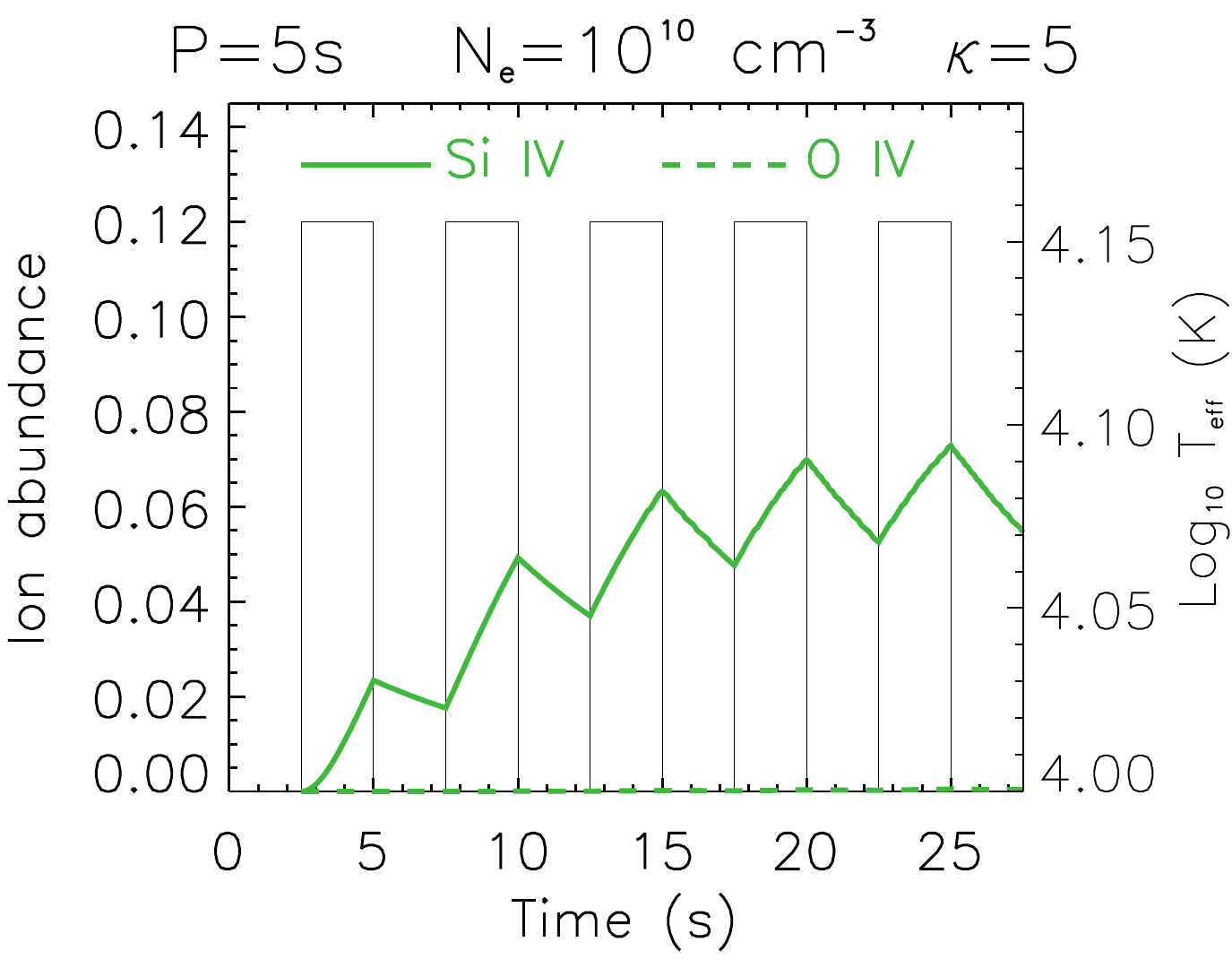}
        \includegraphics[width=5.4cm]{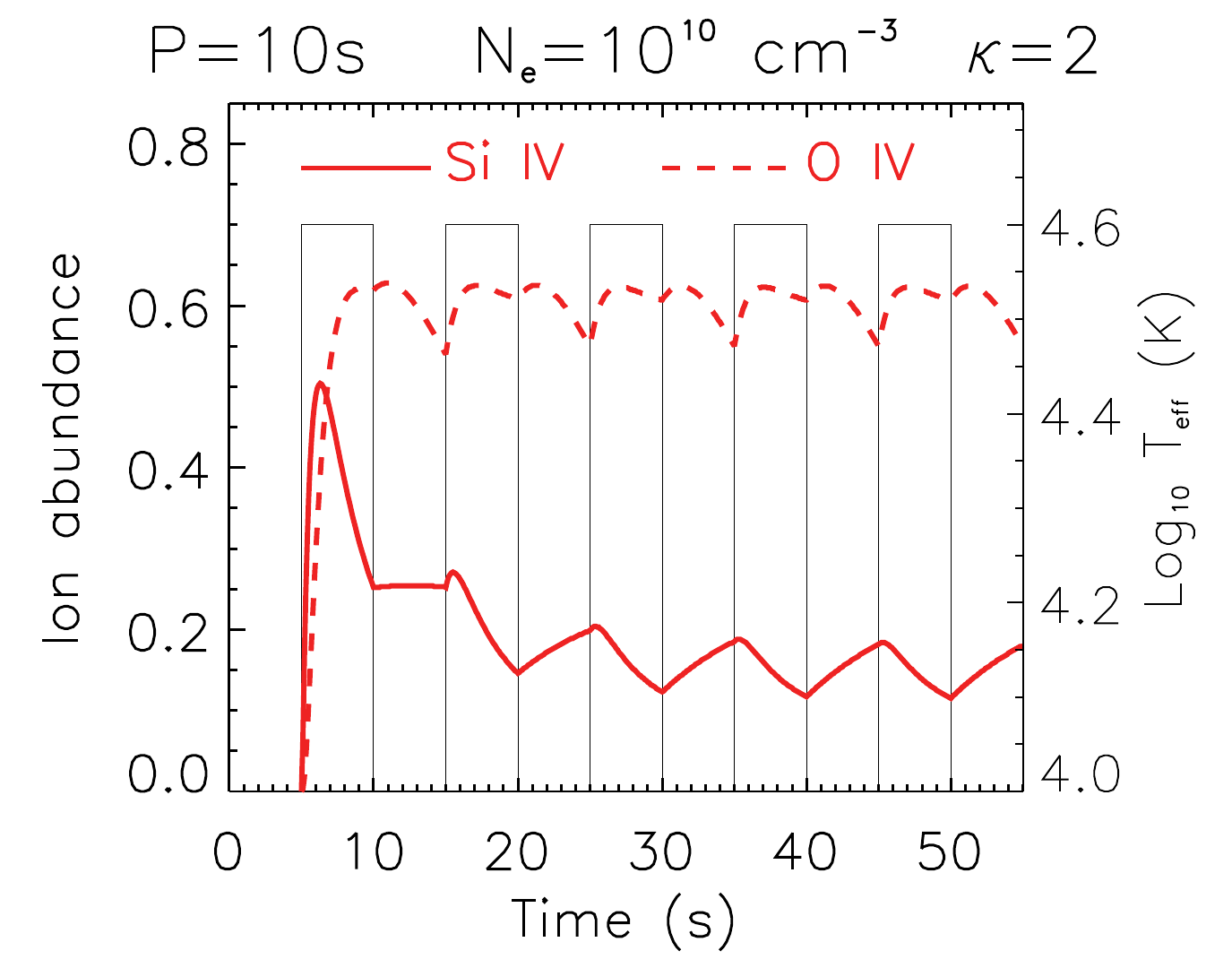}
        \includegraphics[width=5.4cm]{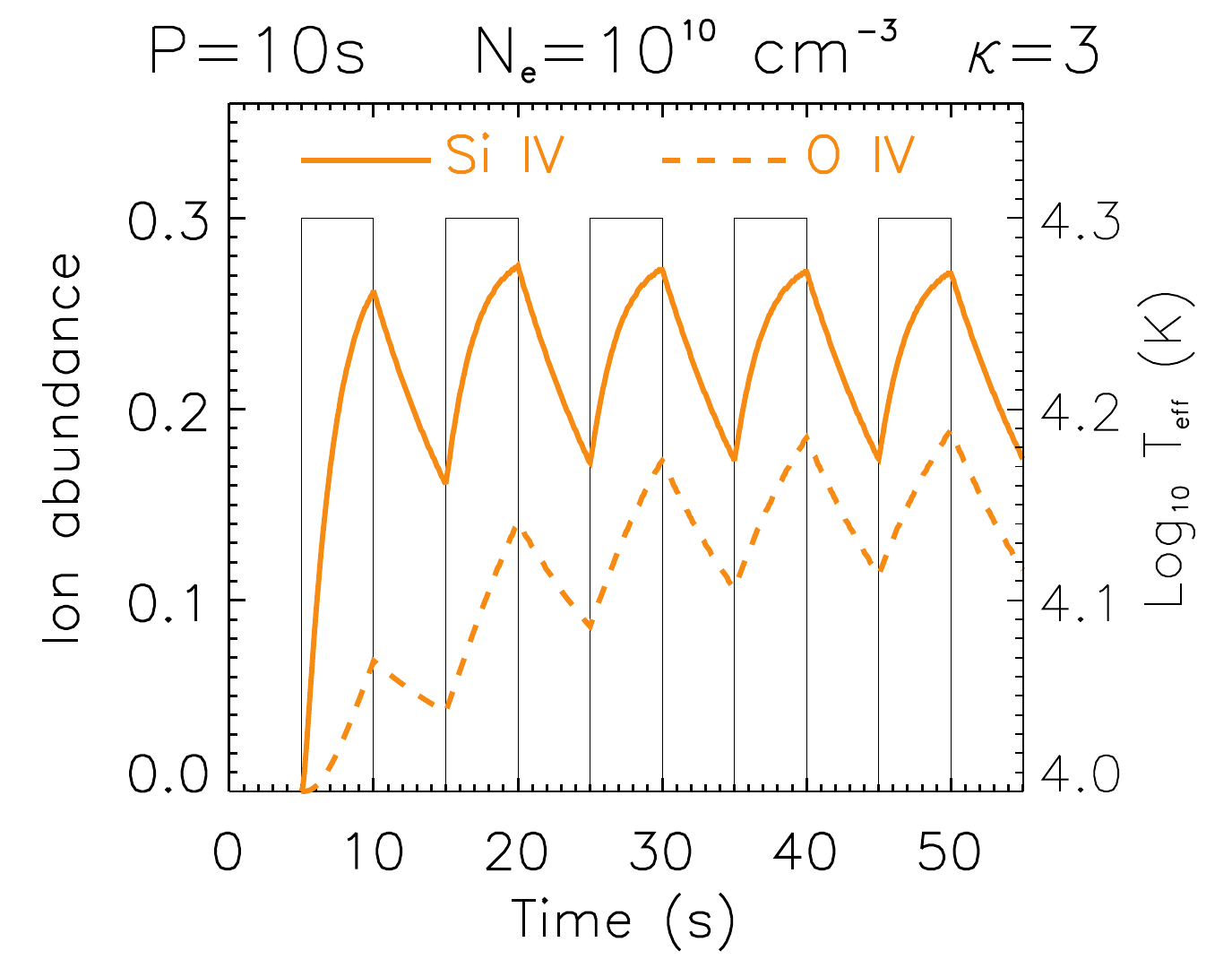}
        \includegraphics[width=5.4cm]{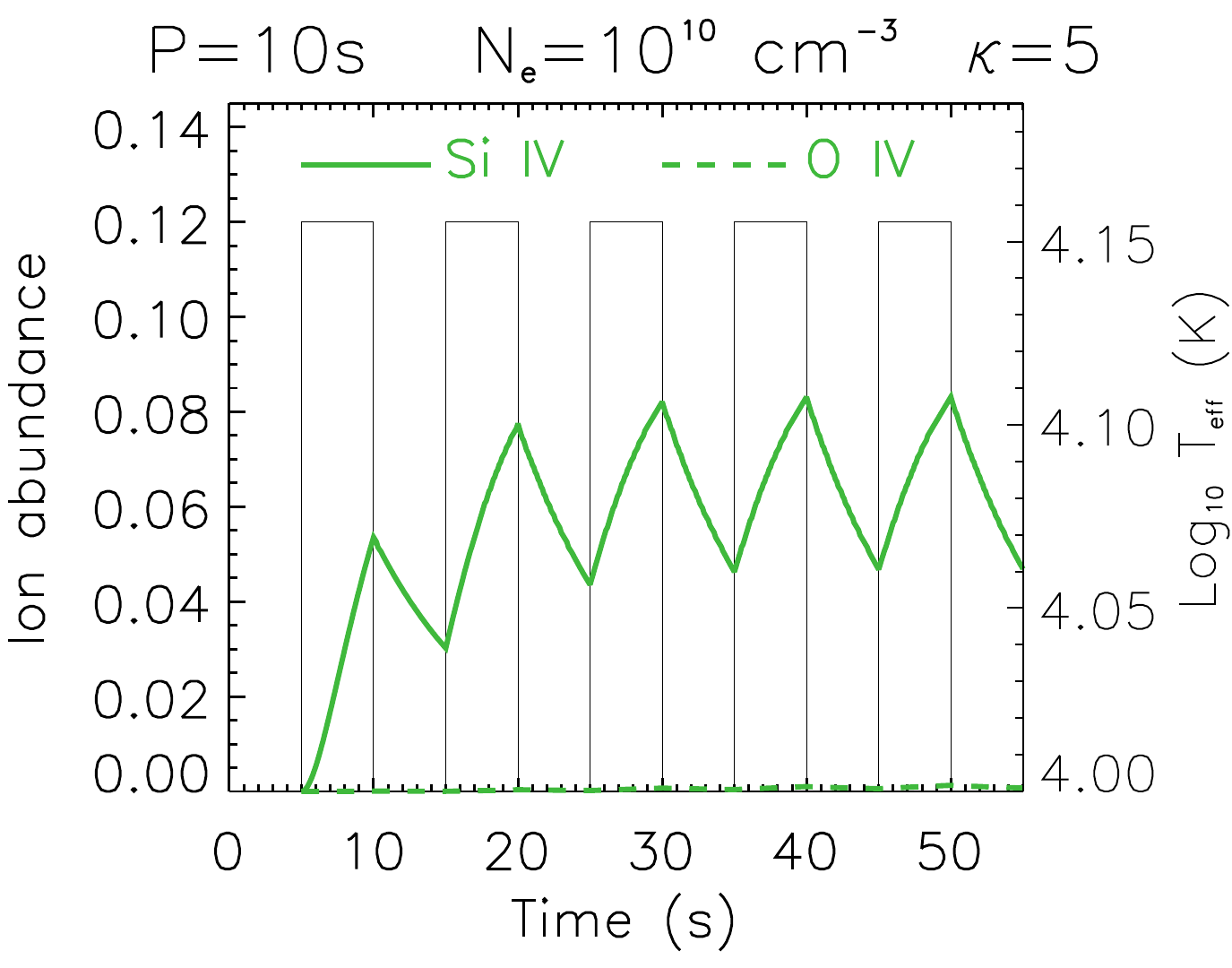}
        \includegraphics[width=5.4cm]{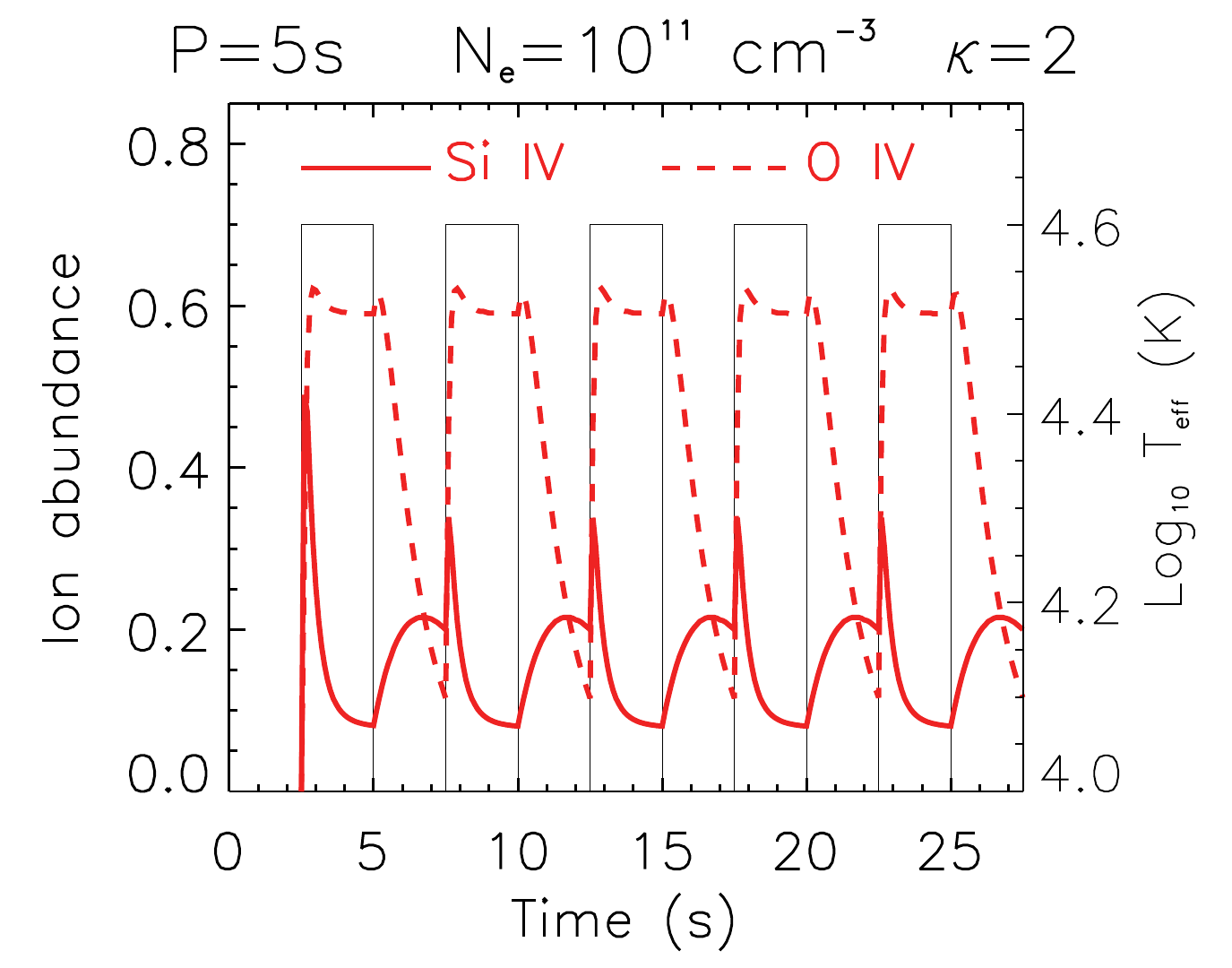}
        \includegraphics[width=5.4cm]{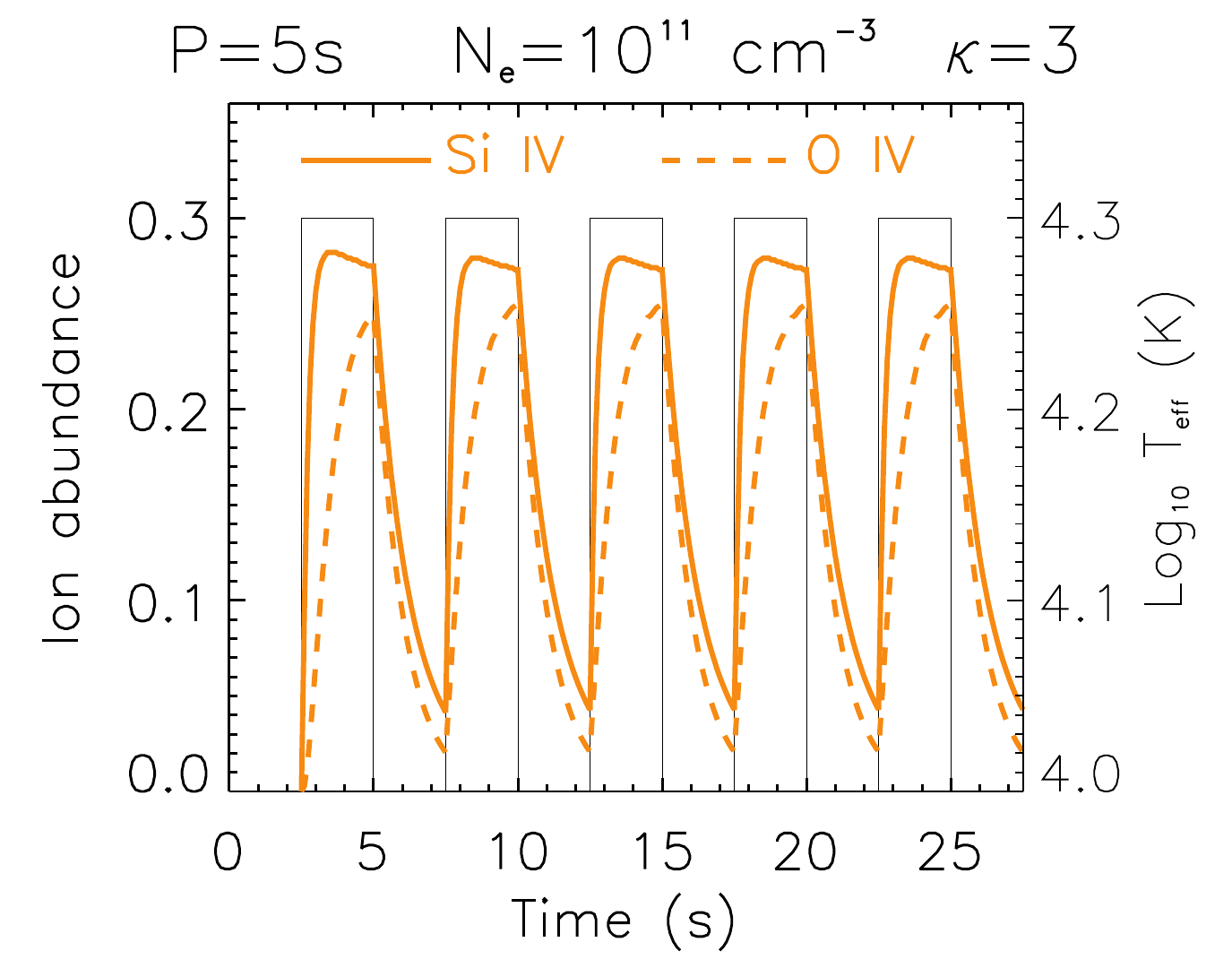}
        \includegraphics[width=5.4cm]{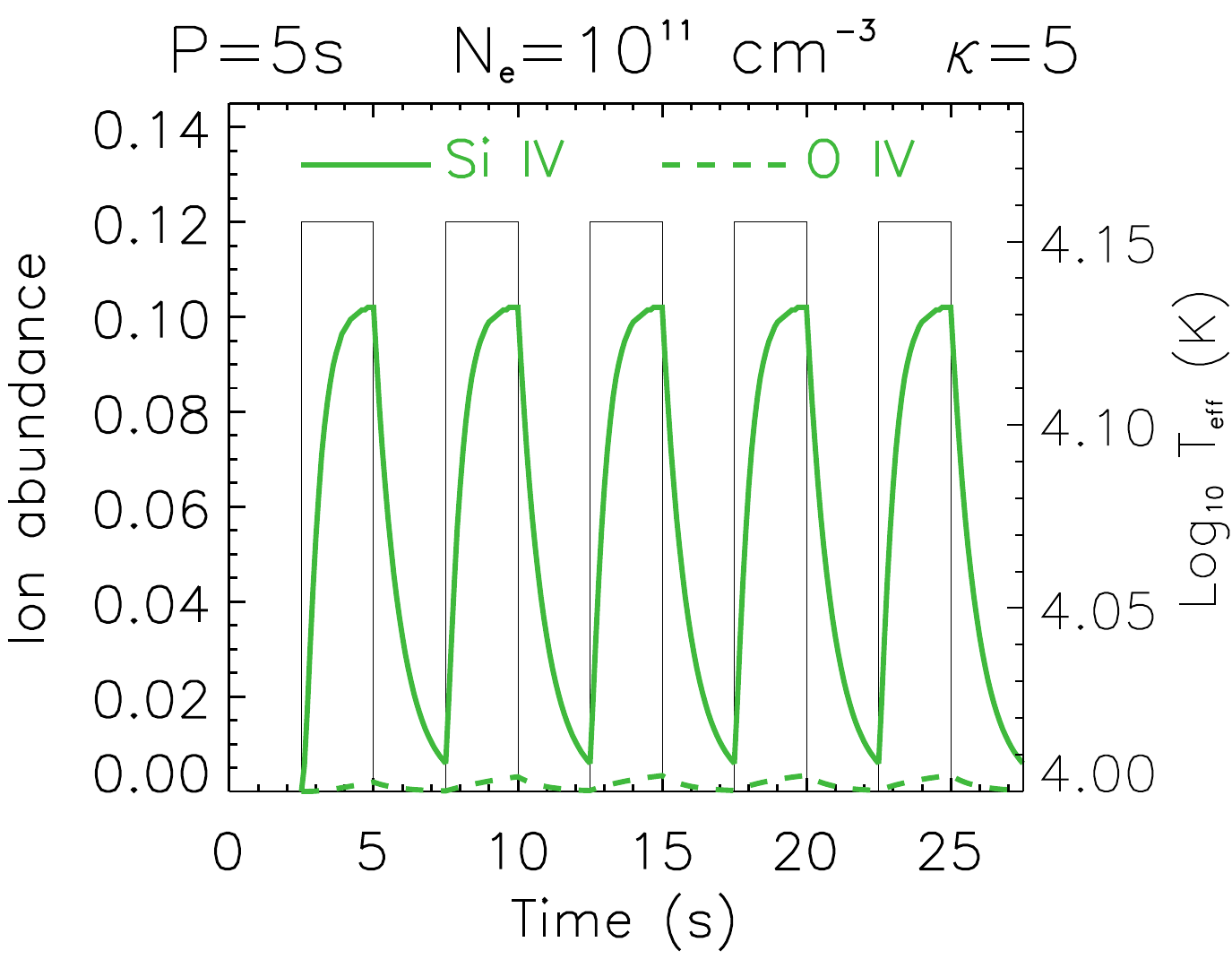}
        \includegraphics[width=5.4cm]{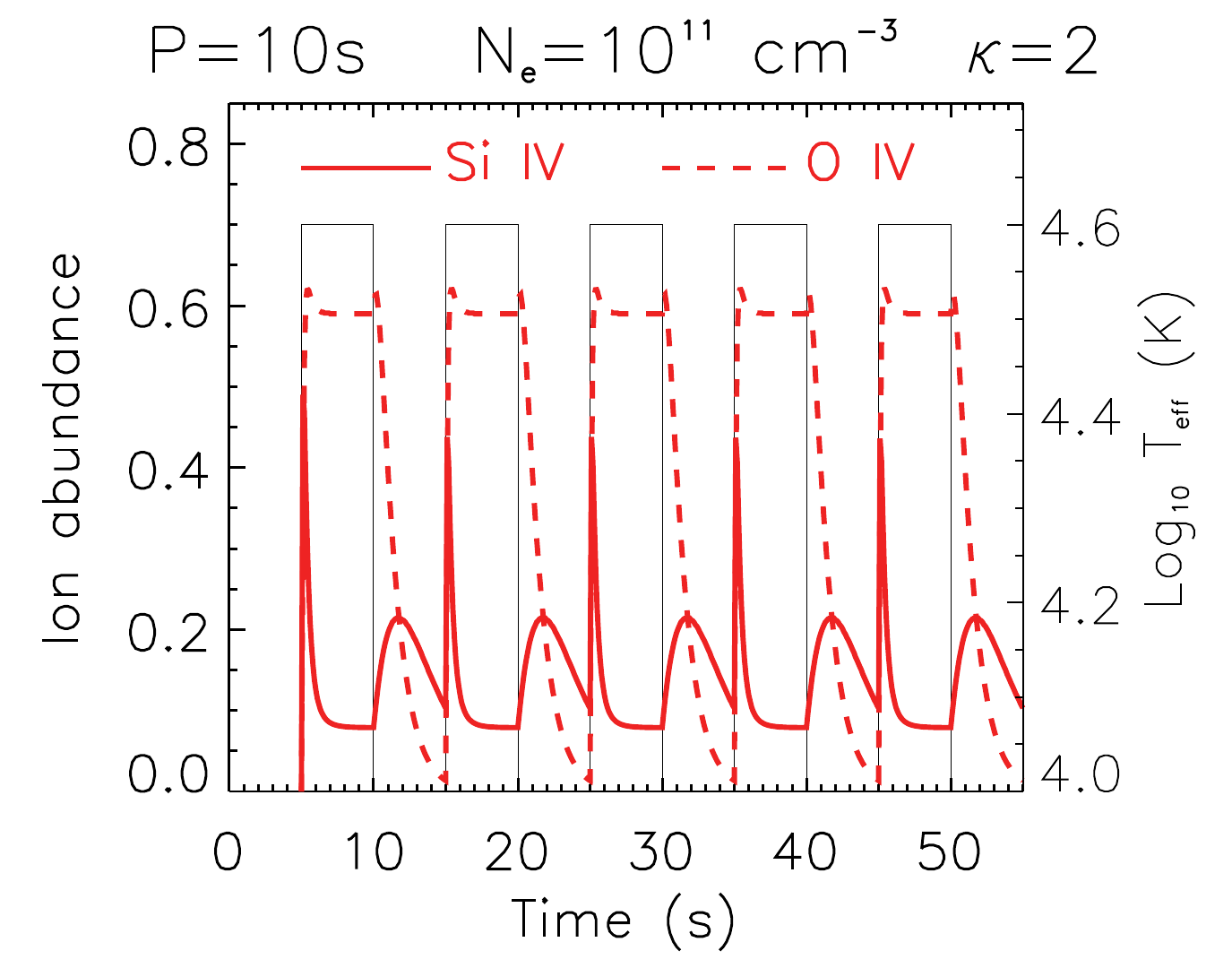}
        \includegraphics[width=5.4cm]{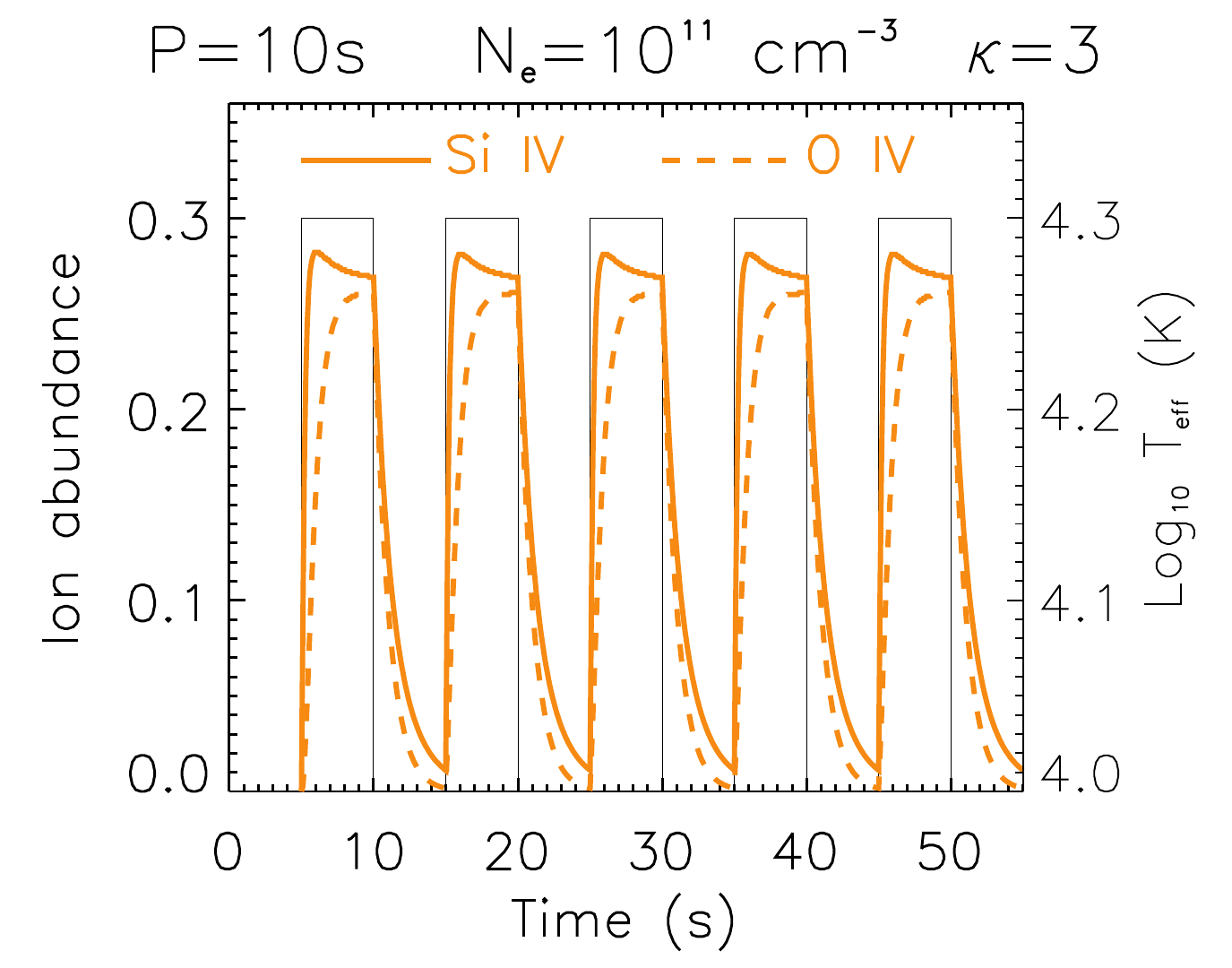}
        \includegraphics[width=5.4cm]{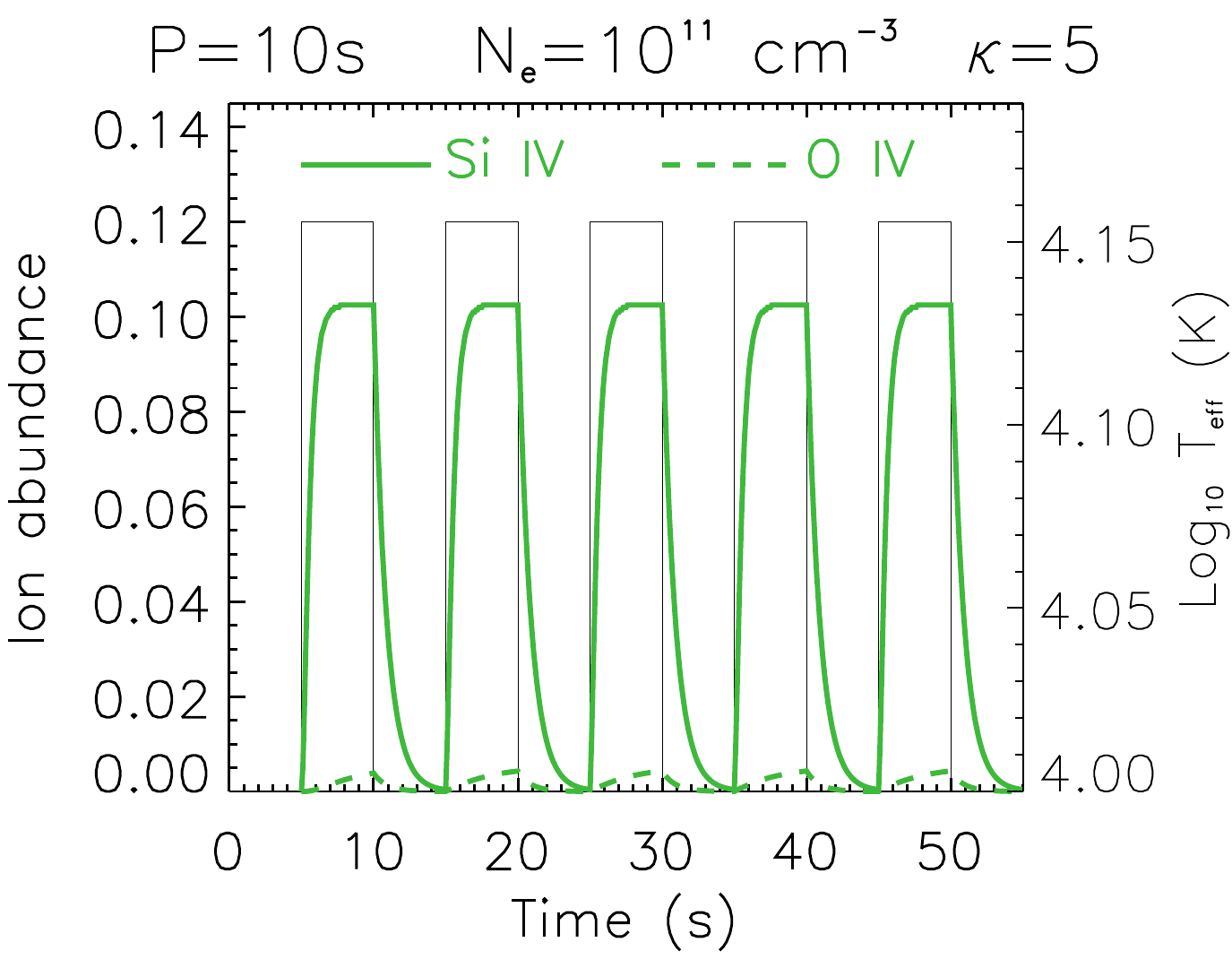}
\caption{Evolution of the relative abundances of \ion{Si}{IV} and \ion{O}{IV}. The layout corresponds to Fig. \ref{Fig:Teff_evol}. The thin black lines denote the evolution of $T$ during the periodic presence of the electron beam.}
\label{Fig:Si4_O4}
\end{figure*}
%
%

%
\begin{table}[!ht]
        \caption[]{Temperatures $T_\mathrm{max}$ of the relative ion abundance maxima.}
         \label{Table:1}
        $$ 
        \begin{tabular}{lcccc}
                \hline
                \noalign{\smallskip}
                                &  \multicolumn{4}{c}{log($T_\mathrm{max}$ [K])}           \\
                Ion             & $\kappa$ = 2  & $\kappa$ = 3       & $\kappa$ = 5  & Maxwellian    \\
                \noalign{\smallskip}
                \hline
                \noalign{\smallskip} 
                \ion{Si}{IV}    & 4.10          & 4.35          & 4.65            & 4.90          \\
                \ion{S}{IV}     & 4.20          & 4.50          & 4.75            & 5.00          \\
                \ion{O}{IV}     & 4.45          & 4.70          & 4.95            & 5.15          \\
                \noalign{\smallskip}
                \hline
        \end{tabular}
        $$ 
\end{table}
%
\begin{table}[!ht]
        \caption[]{Relative ion abundances in equilibrium for $\kappa$-distributions at temperatures $T$ during the second half-periods, when the distribution is a $\kappa$-distribution. Maxwellian ion abundances during the first half-period corresponding to $T$\,=\,10$^4$\,K are also shown in the last column.}
         \label{Table:2}
        $$ 
        \begin{tabular}{lccc}
                \hline
                \noalign{\smallskip}
                                &  \multicolumn{3}{c}{Relative ion abundances in equilibrium}                \\
                                & $\kappa$ = 2          & $\kappa$ = 3             & $\kappa$ = 5          \\
                Ion             & $T = 10^{4.6}$ K      & $T = 10^{4.3}$ K   & $T = 10^{4.15}$ K     \\
                \noalign{\smallskip}
                \hline
                \noalign{\smallskip} 
                \ion{Si}{IV}    & 0.079         & 0.267         & 0.019           \\
                \ion{S}{IV}     & 0.374         & 0.513         & 0.029           \\
                \ion{O}{IV}     & 0.591         & 0.260         & 0.001           \\
                \noalign{\smallskip}
                \hline
        \end{tabular}
        $$ 
\end{table}
%
%

%
%
%
%
\section{Results}
\label{Sect:3}

%
\subsection{Non-equilibrium ionization}
\label{Sect:3.1}

The behavior of $T_\mathrm{eff}$ during the first five periods as derived from Si and O charge states is shown in Fig. \ref{Fig:Teff_evol}. This figure is supplemented by Fig. \ref{Fig:Si4_O4}, which shows the corresponding behavior of the relative ion abundances of \ion{Si}{IV} and \ion{O}{IV}.

In Fig. \ref{Fig:Teff_evol}, the behavior of the actual electron temperature $T$ is shown with thin gray lines. It is a step-wise function as the temperature changes from 10$^4$\,K during the Maxwellian half-period to the higher $T_\kappa$ values (Eq. \ref{Eq:T_kappa}) during the half-period with a $\kappa$-distribution. The behavior of the effective temperature $T_\mathrm{eff}$ derived from the mean charge is shown with thick blue lines for $T_\mathrm{eff}^\mathrm{Mxw}$, and red, orange, and green lines for the corresponding $T_\mathrm{eff}^\kappa$ with $\kappa$\,=\,2, 3, and 5, respectively. Full lines denote the effective temperature derived from silicon and dashed lines the temperature derived
from oxygen.

It is easily seen that in all cases, the $T_\mathrm{eff}^\mathrm{Mxw}$ derived under the assumption of an equilibrium Maxwellian distribution is much higher, in some cases by almost an order of magnitude, than the $T_\mathrm{eff}^\kappa$ derived by assuming a $\kappa$-distribution. This difference originates from the shifts of the relative ion abundances of TR ions with $\kappa$. The difference obtained here is much more pronounced than the differences found in Paper~I, where $T_\mathrm{eff}$ differed by less than 0.2 in log($T_\mathrm{eff}$\,[K]) even for the extreme case of $\kappa$\,=\,2.

Here, these differences between the $T_\mathrm{eff}$ derived by assuming either a Maxwellian or a $\kappa$-distribution have an interesting consequence: The apparent energy requirements for the plasma to reach a given charge state are much lower when the charge state is interpreted in terms of a $\kappa$-distribution than for a Maxwellian one. 

Figure \ref{Fig:Teff_evol} shows that Si never reaches ionization equilibrium if the electron density $N_\mathrm{e}=10^{10}$\,cm$^{-3}$ (rows 1--2 in Fig. \ref{Fig:Teff_evol}, full lines). This conclusion is confirmed by the evolution of the relative ion abundance of \ion{Si}{IV} (rows 1--2 in Fig. \ref{Fig:Si4_O4}) and holds true independently of the period or the value of $\kappa$. Similarly as in Paper~I, the periods chosen are too short for Si to reach an oscillatory state at this density. Thus, Si experiences ionization pumping. During the first period, log($T_\mathrm{eff}^\mathrm{Mxw}$\,[K]) spikes from 4.0 to 4.5--4.9  depending on $\kappa$. It then reaches higher maxima with each additional period. Overall, a weaker beam (higher $\kappa$) results in a lower maximum of the effective temperature and also its mean value. For some cases, the effect of ionization pumping can appear to be weak, for example, for $\kappa$\,=\,2--3, when the $T_\mathrm{eff}^\mathrm{Mxw}$ increases slowly, while the effect is more clearly visible for $T_\mathrm{eff}^\kappa$.

A higher electron density of $N_\mathrm{e}=10^{11}$\,cm$^{-3}$ (rows 3--4 of Fig. \ref{Fig:Teff_evol}) leads to more pronounced changes in $T_\mathrm{eff}$ during any given period, with no ionization pumping. The charge state reaches ionization equilibrium for a given value of $\kappa$ at the end of the first half-period. This is indicated by the equivalence of $T_\mathrm{eff}^\kappa$\,=\,$T_\kappa$ toward the end of the half-period characterized by a $\kappa$-distribution, as well as the relative abundance of \ion{Si}{IV} reaching its equilibrium value (see rows 3--4 of Fig. \ref{Fig:Si4_O4}). The duration of equilibrium state is longer for $P$\,=\,10\,s than for 5\,s, and $T_\mathrm{eff}^\kappa$ is lower for higher $\kappa$. During the second half-period, Si recombines with a corresponding fast decrease in $T_\mathrm{eff}^\mathrm{Mxw}$, which never reaches the initial ionization state that corresponds to the Maxwellian temperature of $10^4$\,K. However, $T_\mathrm{eff}^\kappa$ approximates the behavior of the true plasma $T$ much better in all instances (red, orange, and green lines in rows 3--4 of Fig. \ref{Fig:Teff_evol}).

The charge state of oxygen and its evolution (dashed lines in Fig. \ref{Fig:Teff_evol}) is generally similar to that of silicon. However, some deviations are worth mentioning. The main difference is that its response to a periodic electron beam is usually slower than for Si if $\kappa$\,=\,3 and 5. Because of its lower ionization rates at a given $T$ (Fig. \ref{Fig:rates}), oxygen needs more time to ionize when the electron density is lower and the beam is relatively weak. In an extreme example of $N_\mathrm{e}=10^{10}$\,cm$^{-3}$ and $\kappa=5$, the mean charge $T_\mathrm{eff}^\kappa$ is not significantly perturbed even after several periods. However, for a strong beam with $\kappa$\,=\,2, oxygen can react faster than silicon; compare the full and dashed lines in the left column of Fig. \ref{Fig:Teff_evol}.

At higher densities of $N_\mathrm{e}=10^{11}$\,cm$^{-3}$, oxygen is able to reach equilibrium ionization stages for the corresponding $\kappa$-distributions only if $\kappa$\,=\,2 or 3 (dashed lines in Fig. \ref{Fig:Teff_evol}, rows 3--4, left and middle). This is not true for a weaker electron beam described by $\kappa=5$. In this case, oxygen is always out of equilibrium for both periods if $\kappa$\,=\,5 (Fig. \ref{Fig:Teff_evol} rows 3--4, right). 

The recombination timescales for oxygen are also substantially longer than for silicon. The minima of $T_\mathrm{eff}^\mathrm{Mxw}$ during the second half-periods are very high compared to the initial ionization state that corresponds to log($T$\,[K])\,=\,4. This is also true for higher densities of $N_\mathrm{e}=10^{11}$\,cm$^{-3}$ and longer $P$\,=\,10\,s (Fig. \ref{Fig:Teff_evol}).

%
\begin{figure*}[!ht]
        \centering
        \includegraphics[width=6.0cm]{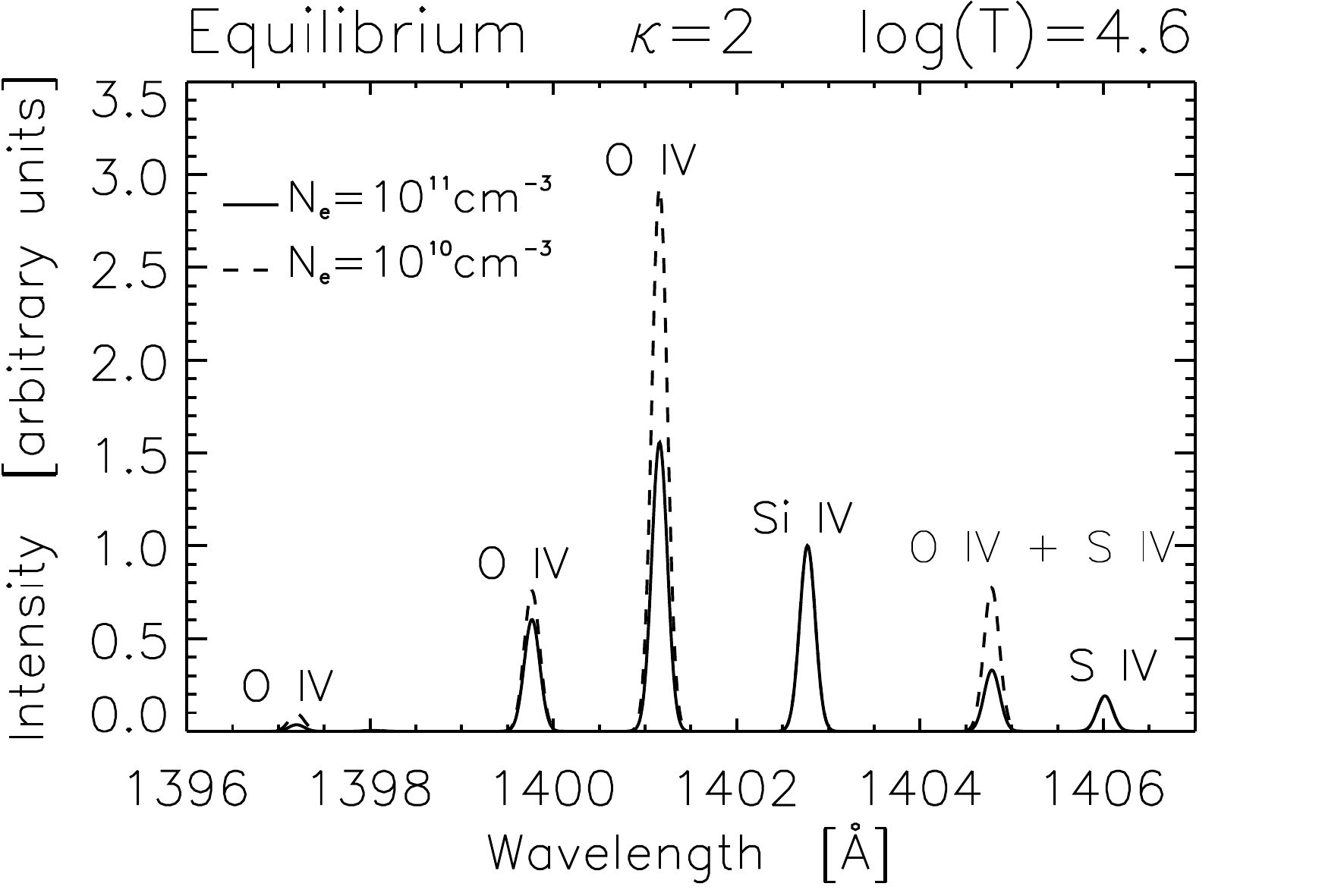}
        \includegraphics[width=6.0cm]{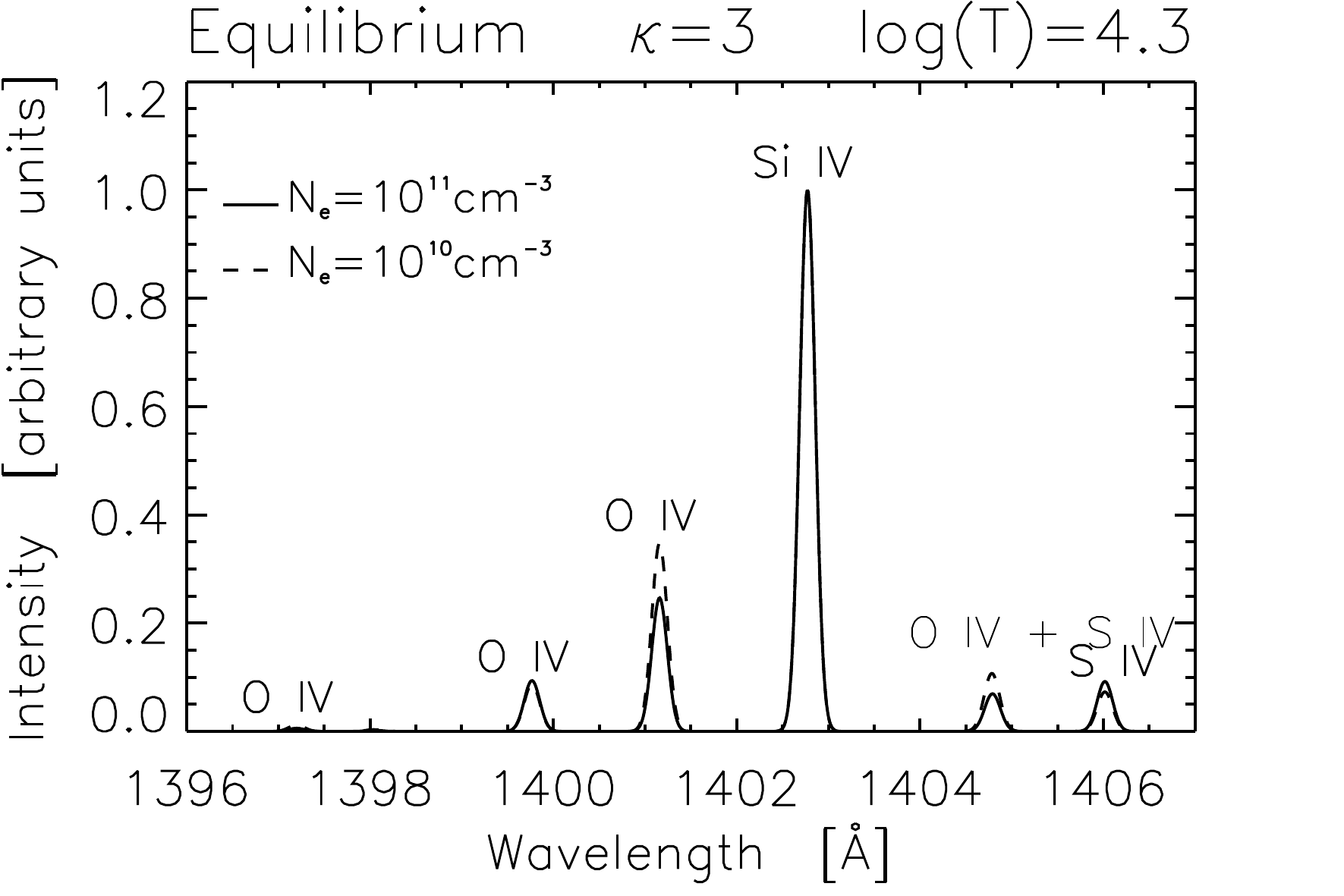}
        \includegraphics[width=6.0cm]{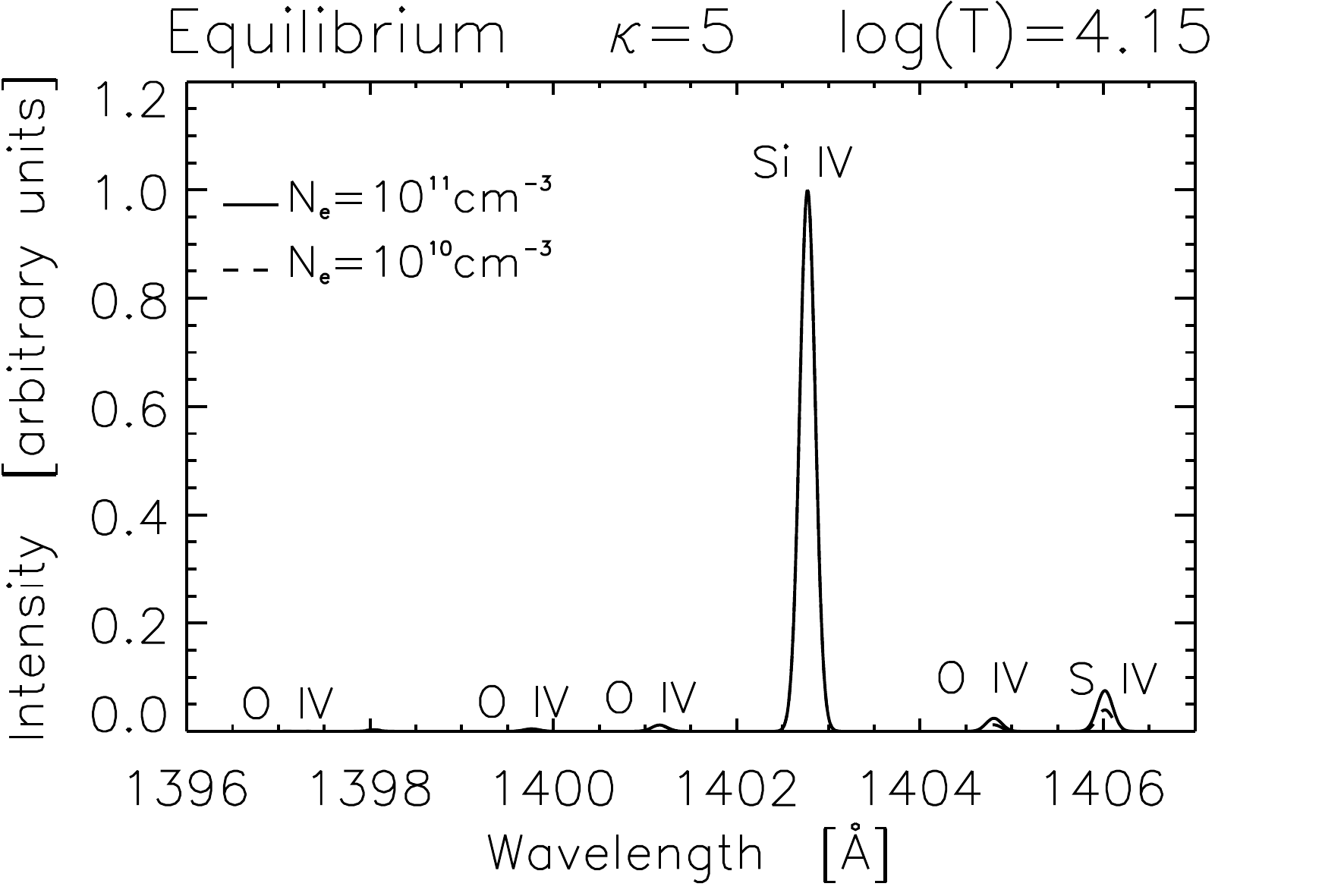}
\caption{Synthetic TR spectra in equilibrium for the IRIS FUV2 channel, calculated for $\kappa$\,=\,2 and $T_\kappa=10^{4.6}$ K (left), $\kappa$\,=\,3 and $T=10^{4.3}$ K (middle), and $\kappa=5$ with $T=10^{4.15}$ K (right). Dashed and full lines correspond to  $N_\mathrm{e}=10^{10}$ cm$^{-3}$ and 10$^{11}$ cm$^{-3}$, respectively.}
\label{Fig:sp_eq}
\end{figure*}
%

%
\begin{figure*}[!ht]
        \centering
        \includegraphics[width=6.0cm]{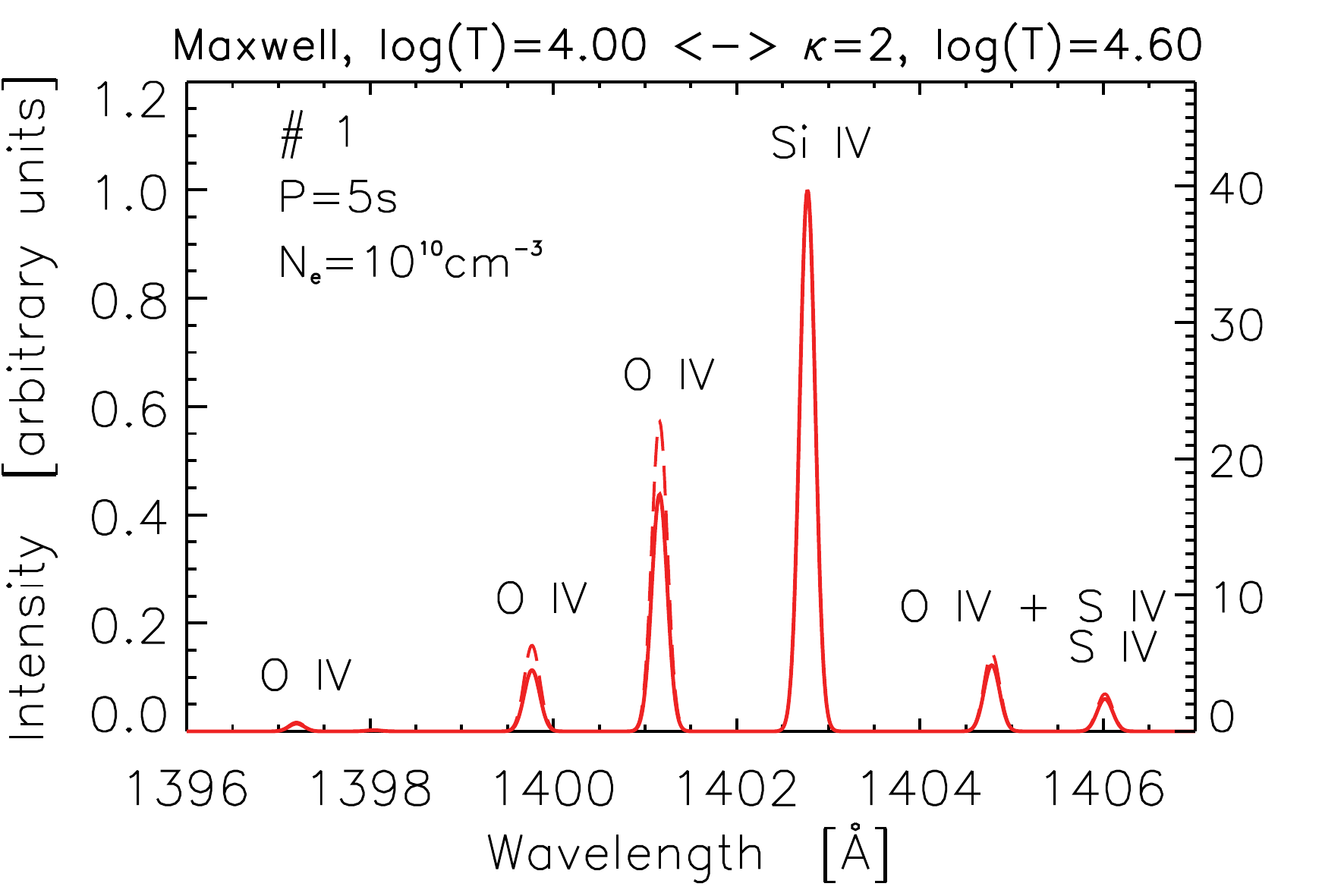}
        \includegraphics[width=6.0cm]{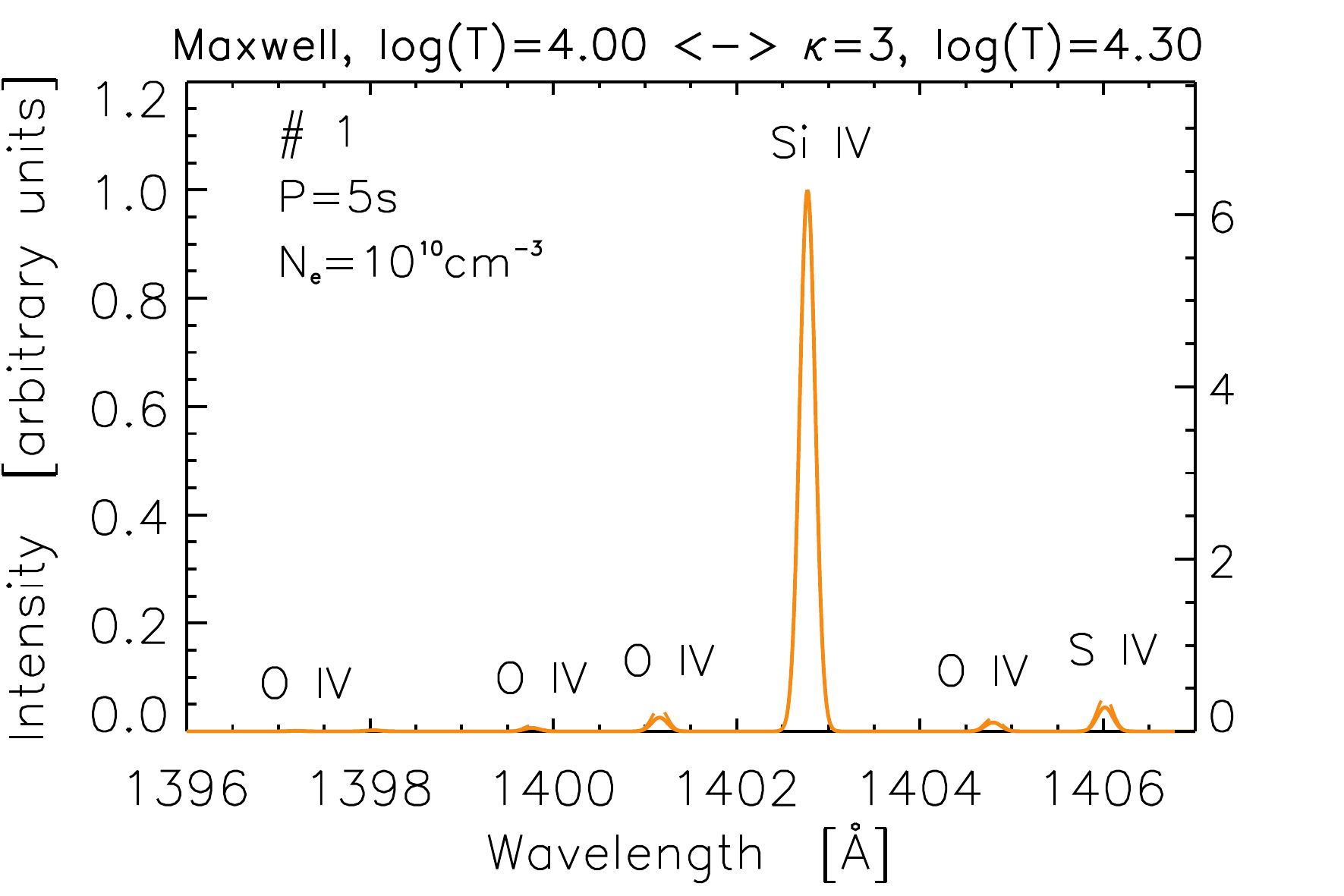}
        \includegraphics[width=6.0cm]{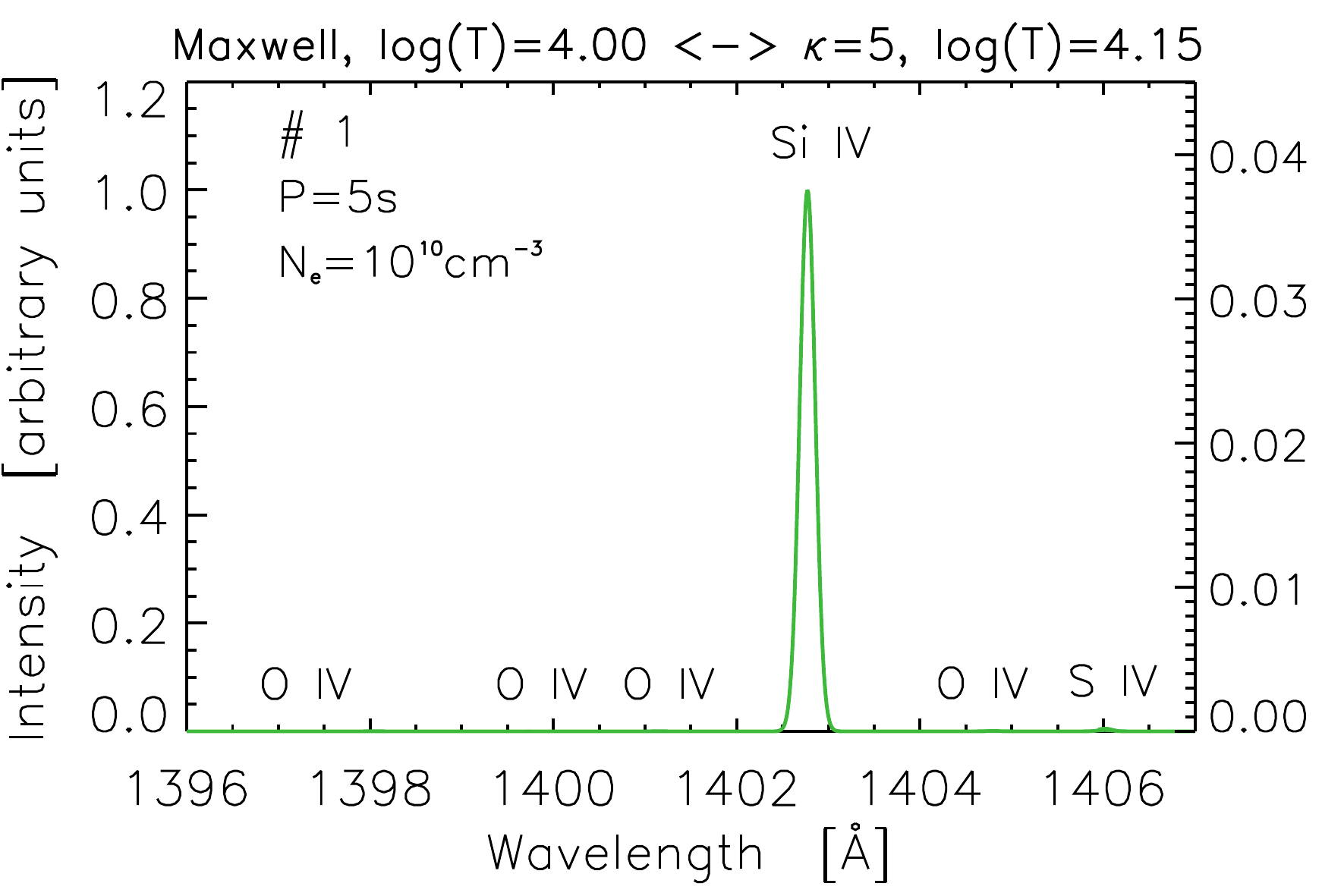}
        \includegraphics[width=6.0cm]{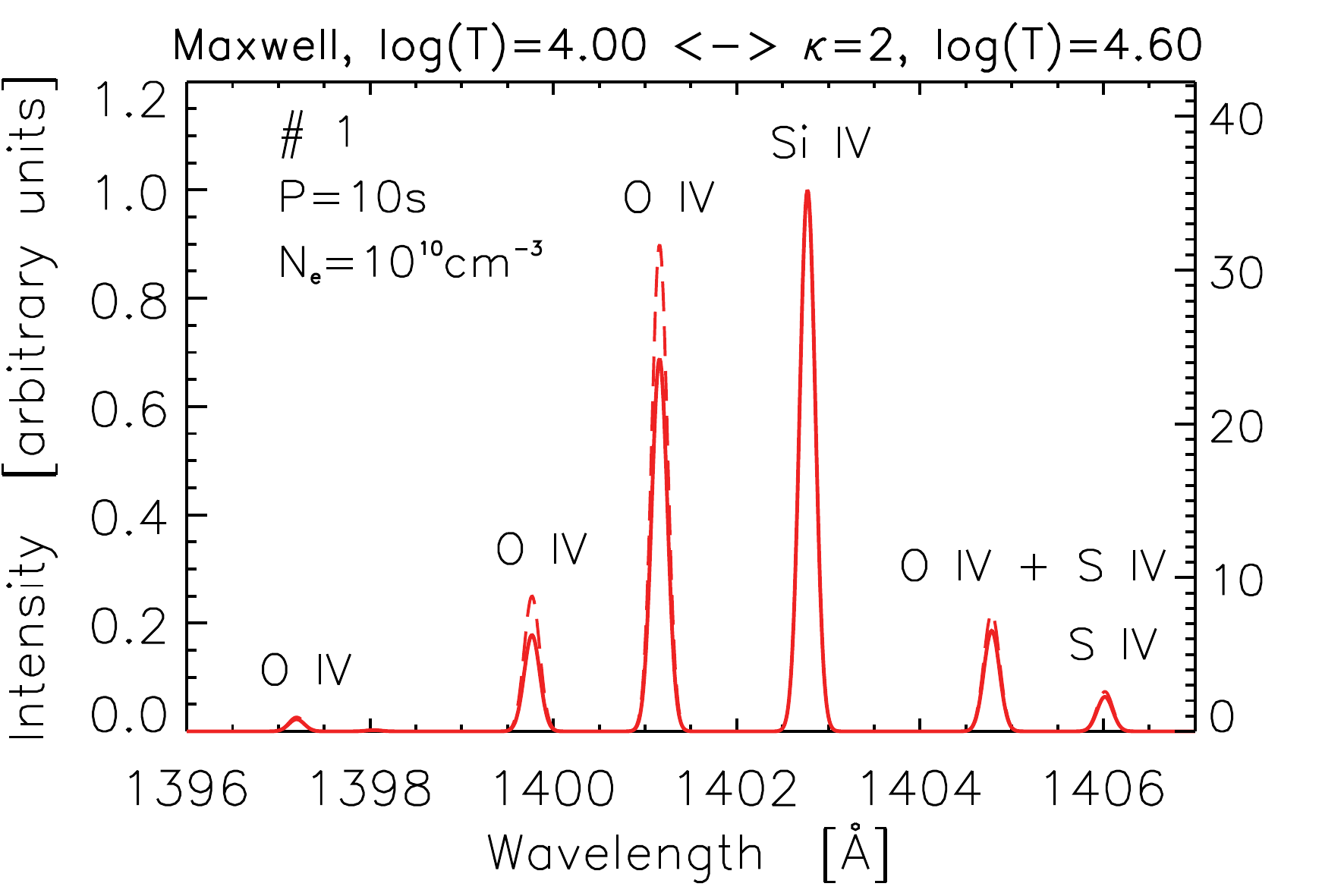}
        \includegraphics[width=6.0cm]{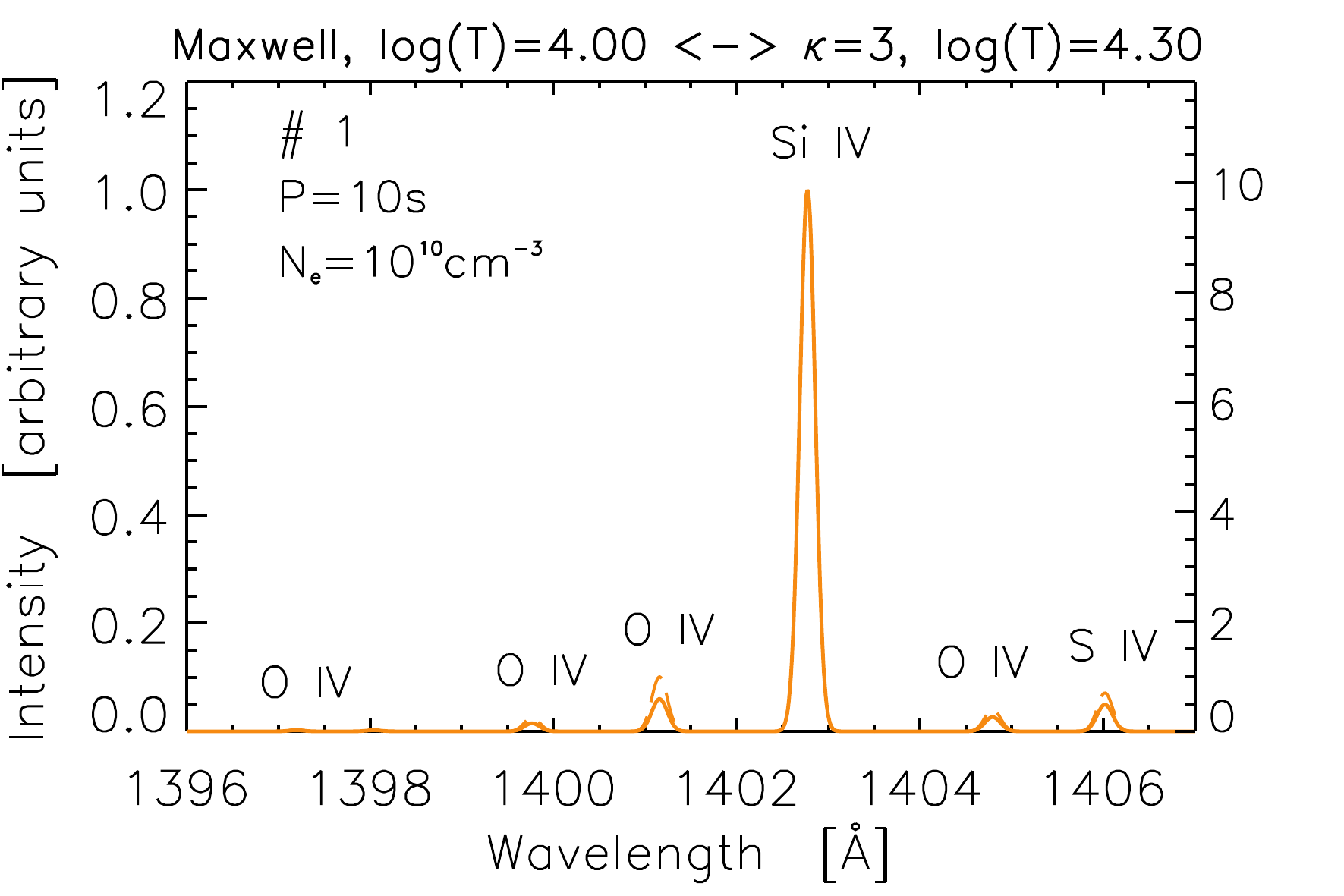}
        \includegraphics[width=6.0cm]{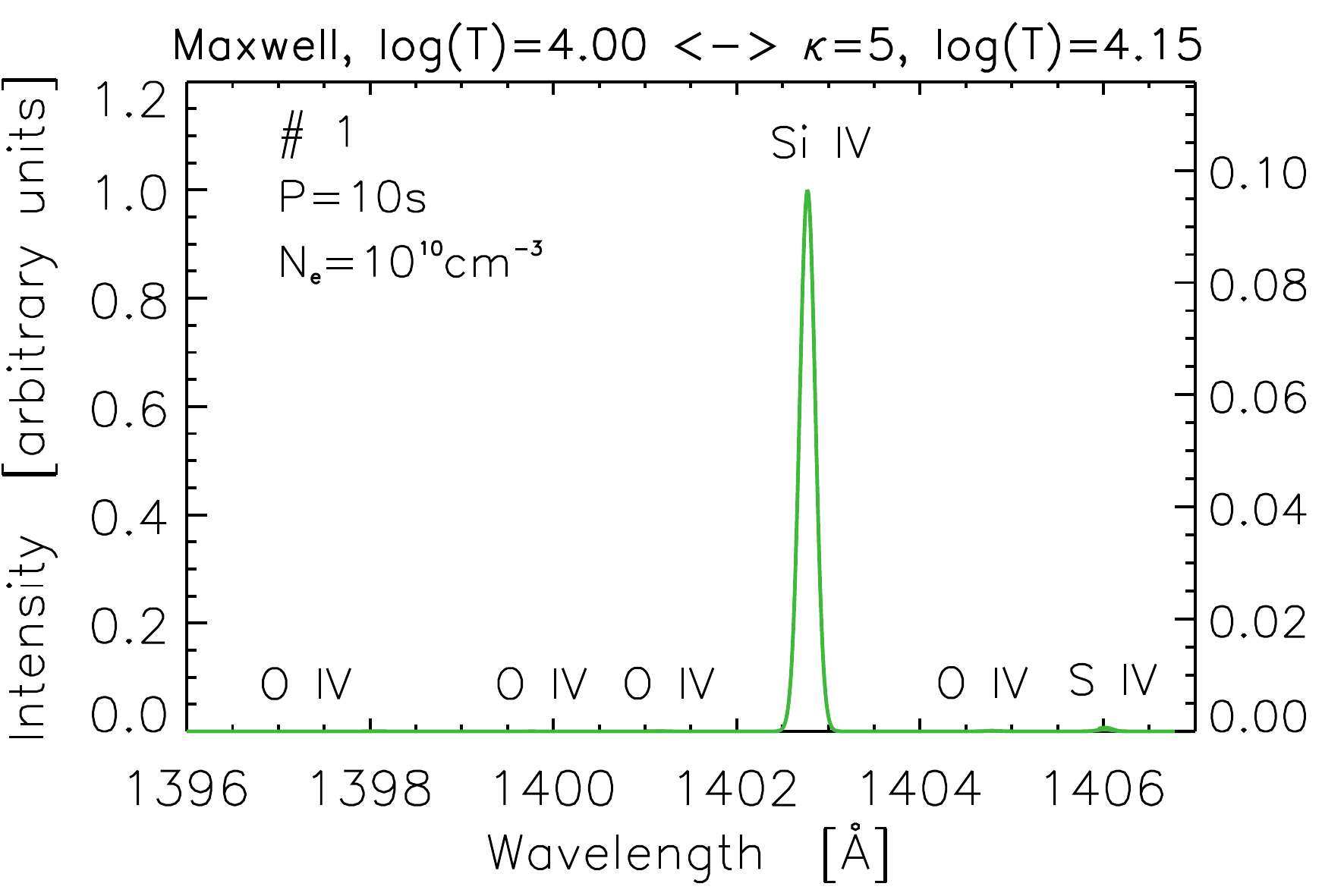}
        \includegraphics[width=6.0cm]{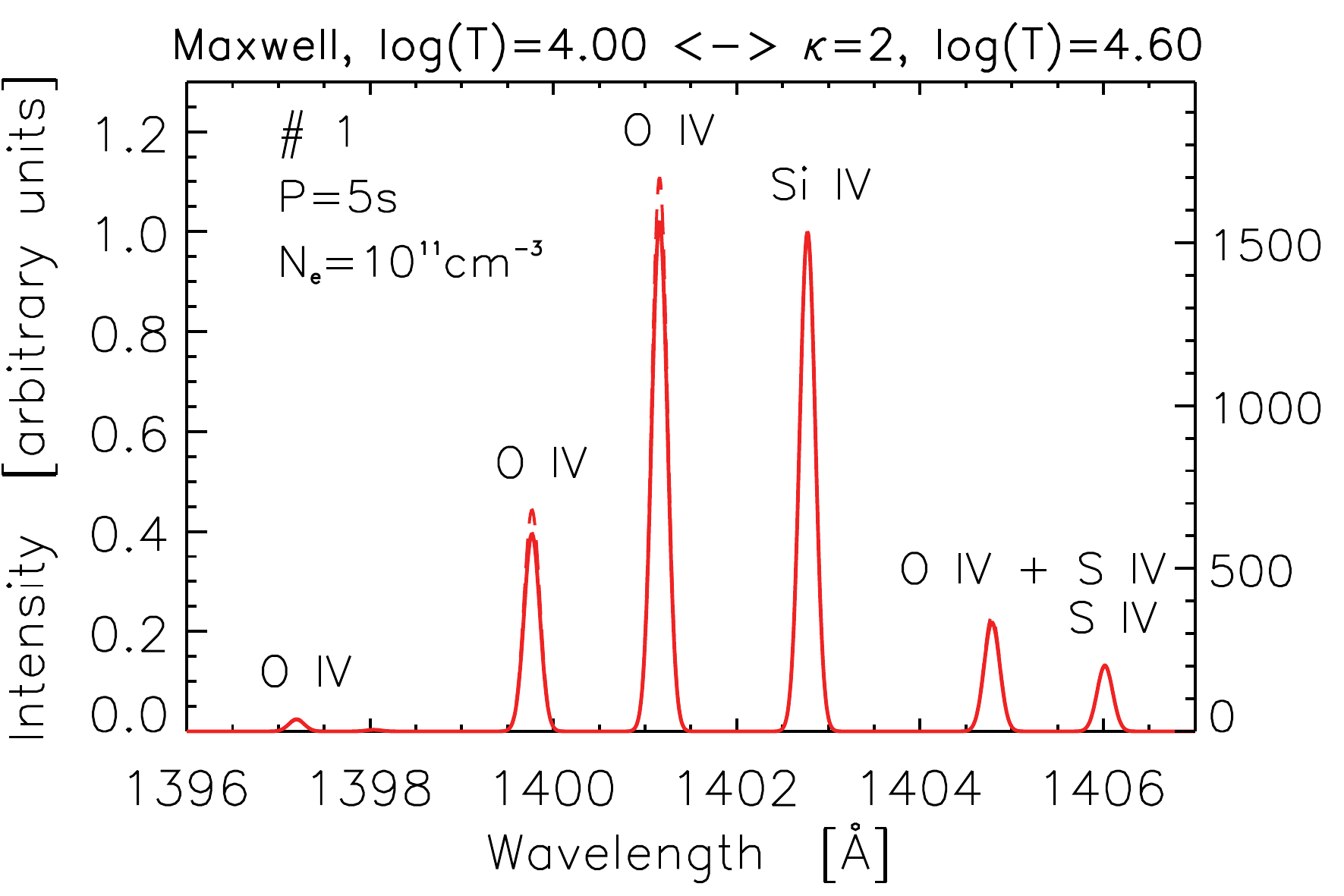}
        \includegraphics[width=6.0cm]{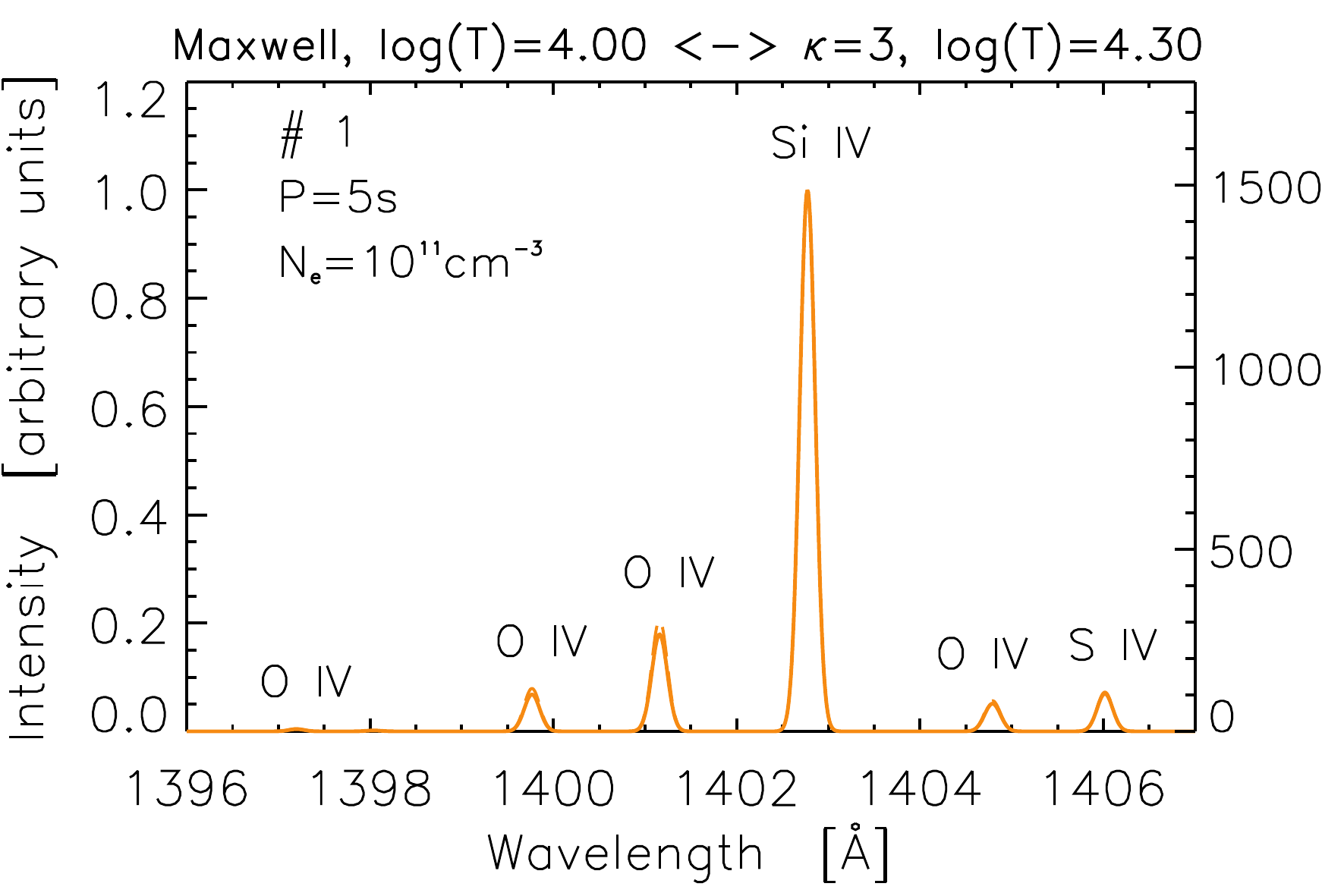}
        \includegraphics[width=6.0cm]{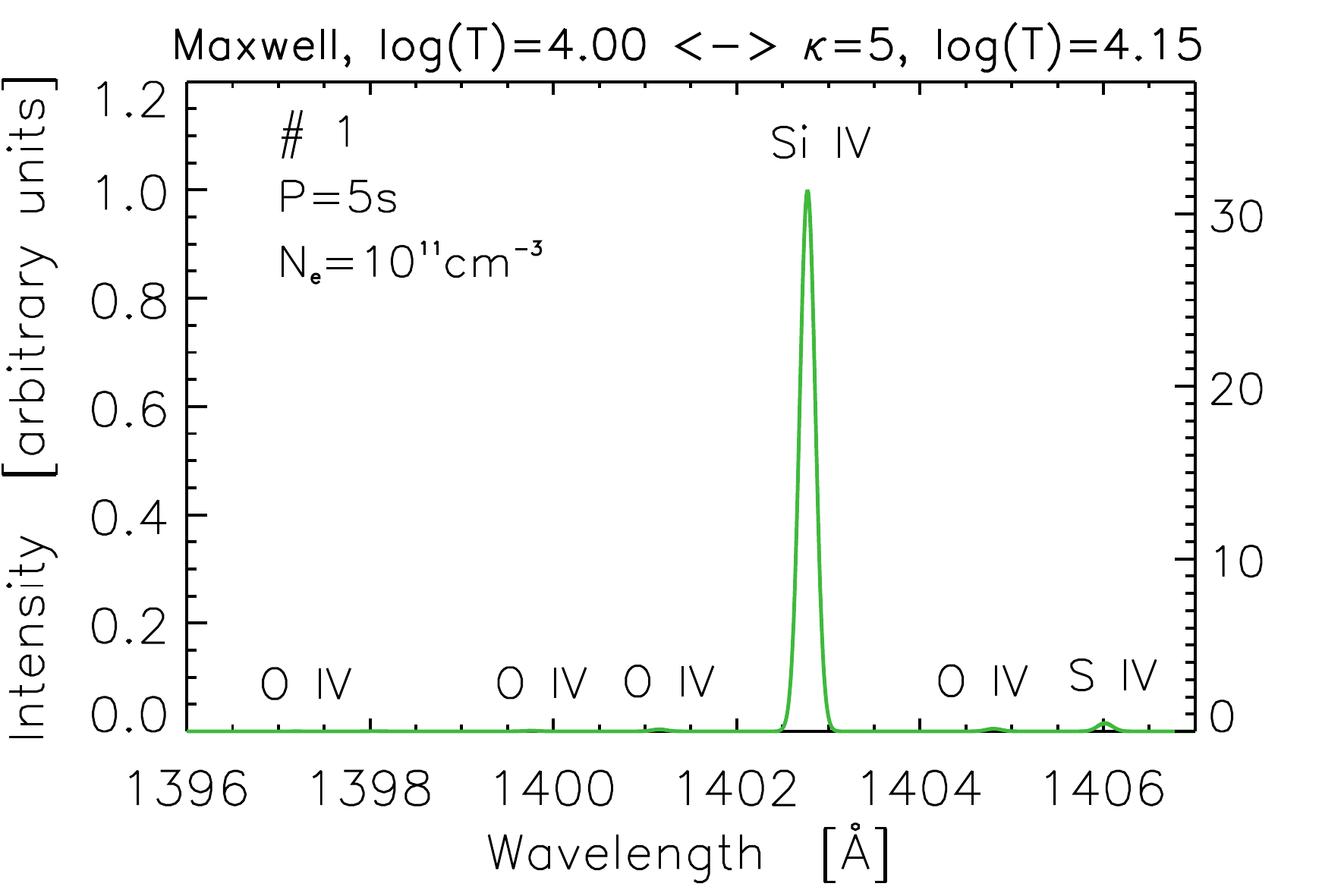}
        \includegraphics[width=6.0cm]{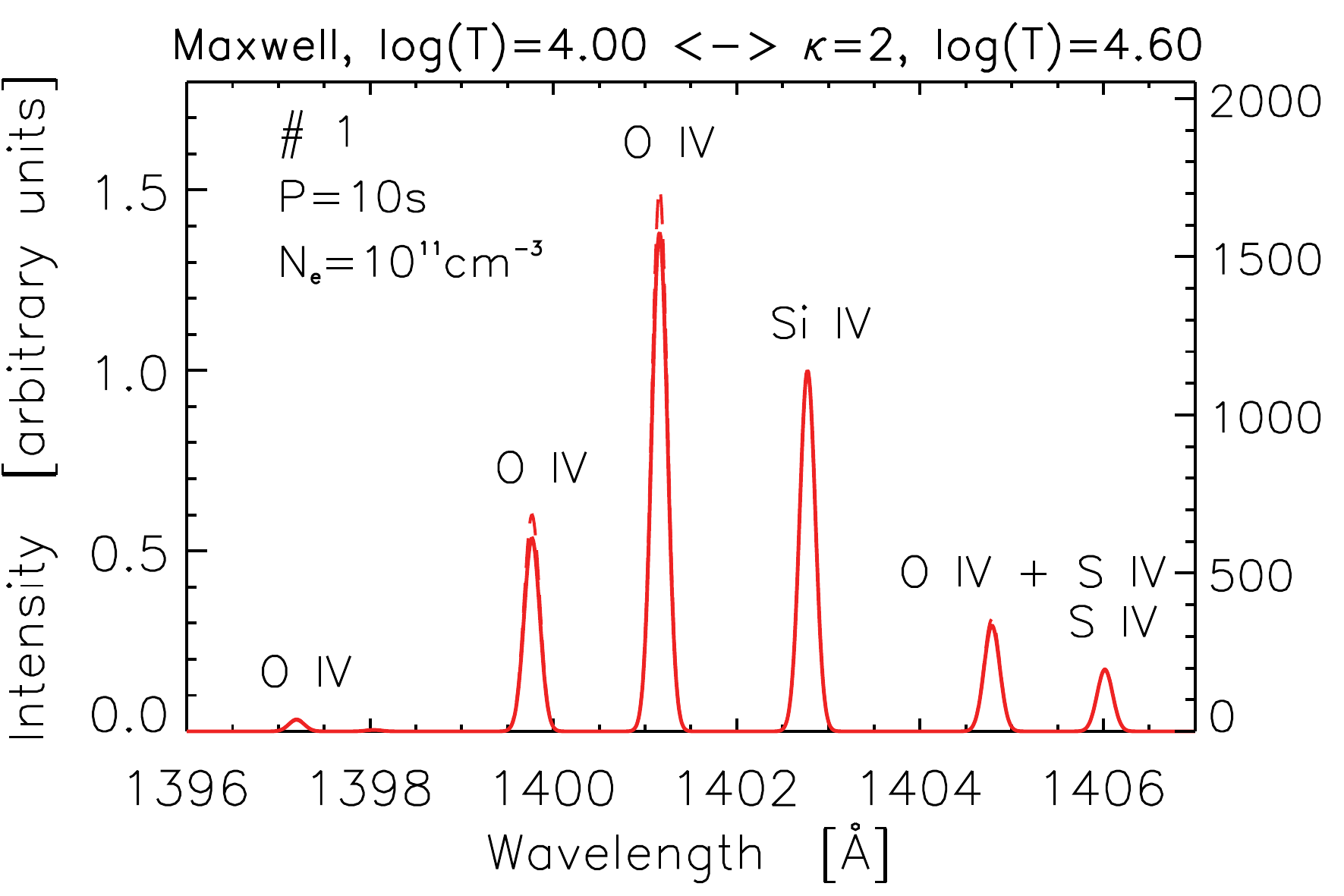}
        \includegraphics[width=6.0cm]{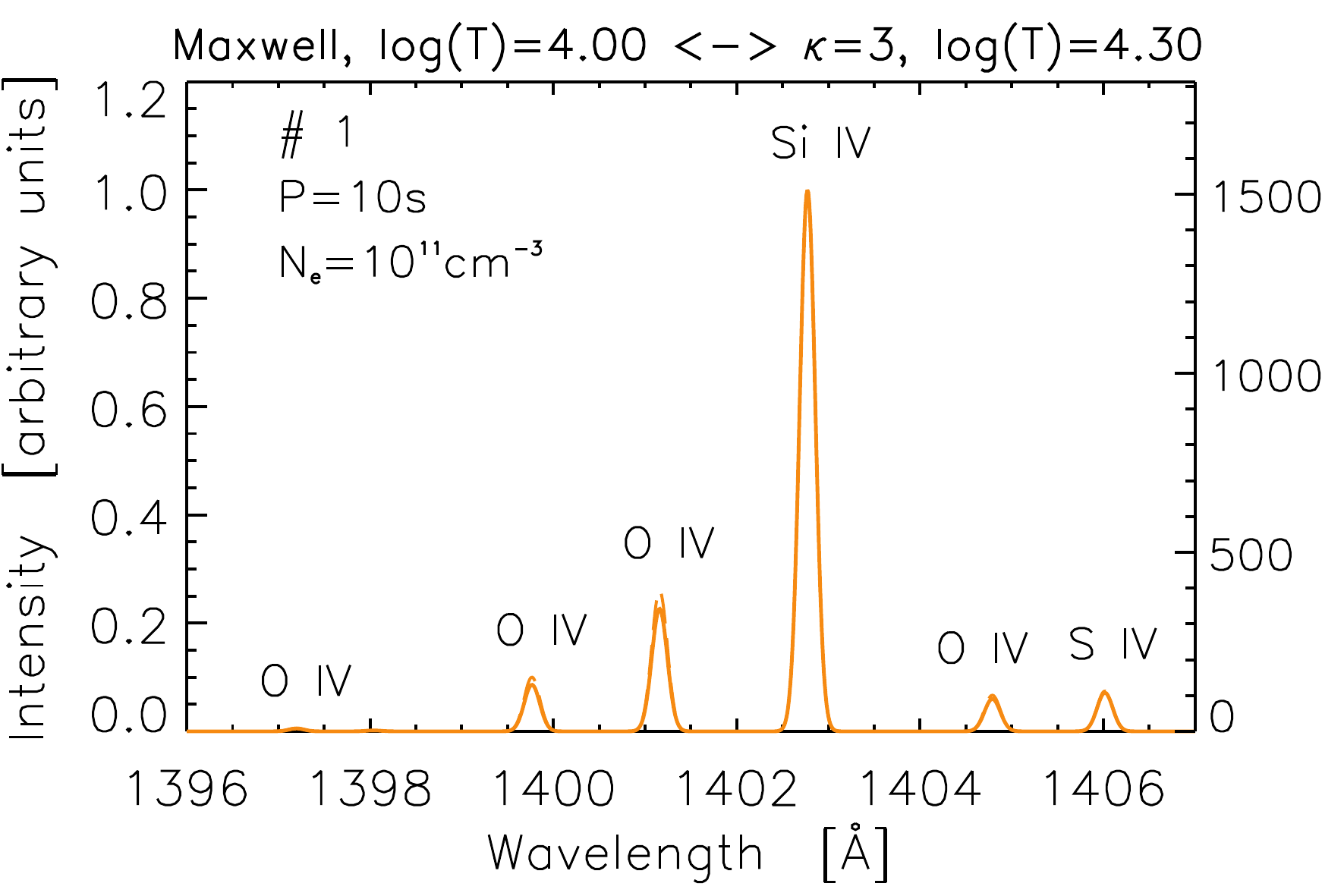}
        \includegraphics[width=6.0cm]{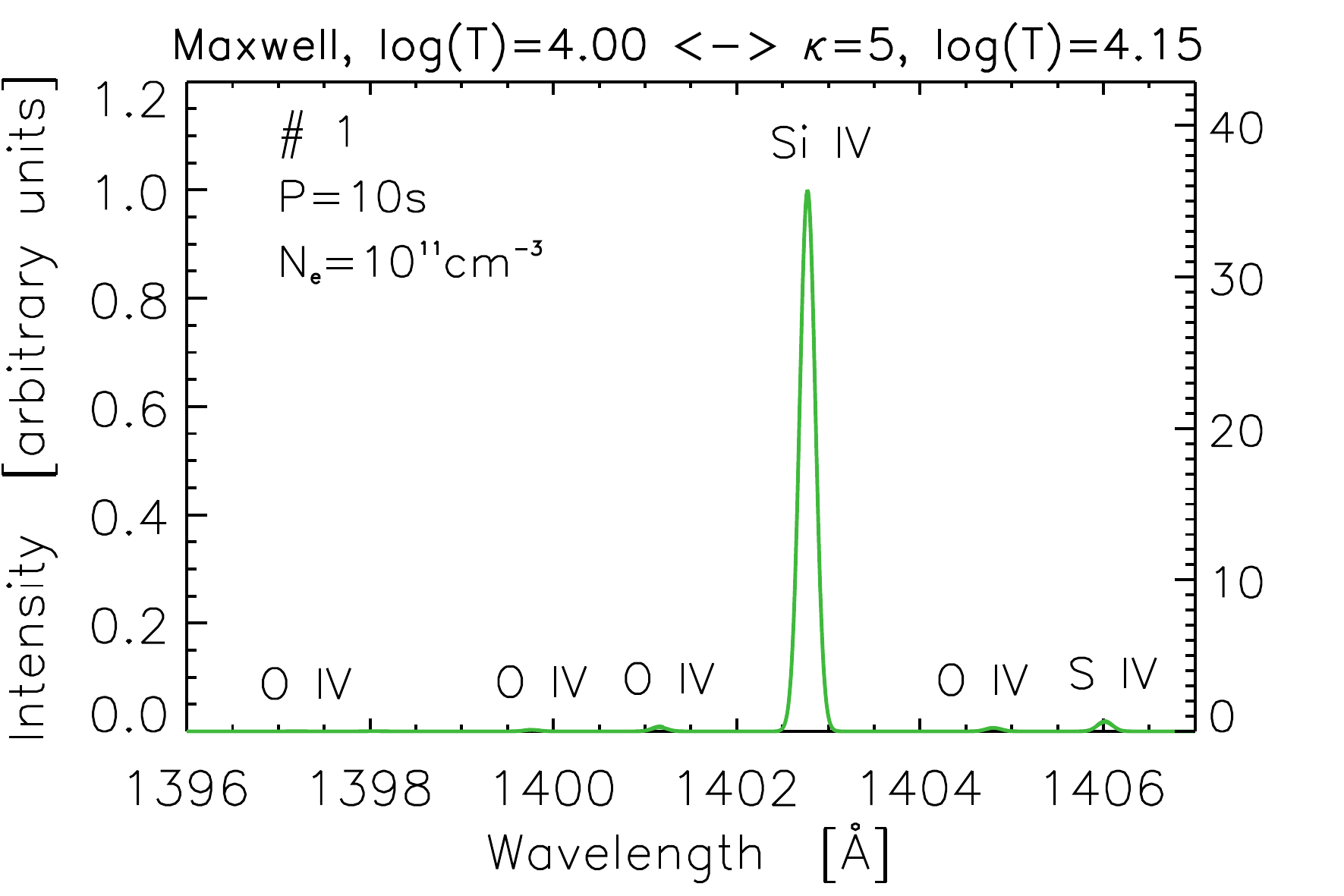}
\caption{Synthetic TR spectra averaged over the first period, calculated for $T=10^4$ K. The layout of this figure corresponds to Fig. \ref{Fig:Teff_evol}. The parameters of each model are given in each panel. Red, orange, and green stand for a model with $\kappa$\,=\,2, 3, and 5 during the second half-period, respectively. Full lines show the spectra with photoexcitation, while the dashed lines correspond to spectra without photoexcitation. Each spectrum is normalized with respect to the \ion{Si}{IV} 1402.8\,\AA~intensity. Original intensities (in arbitrary units) are shown in the right-hand side of each image.}
\label{Fig:sp1_T4}
\end{figure*}
%

%
\begin{figure*}[!ht]
        \centering
        \includegraphics[width=6.0cm]{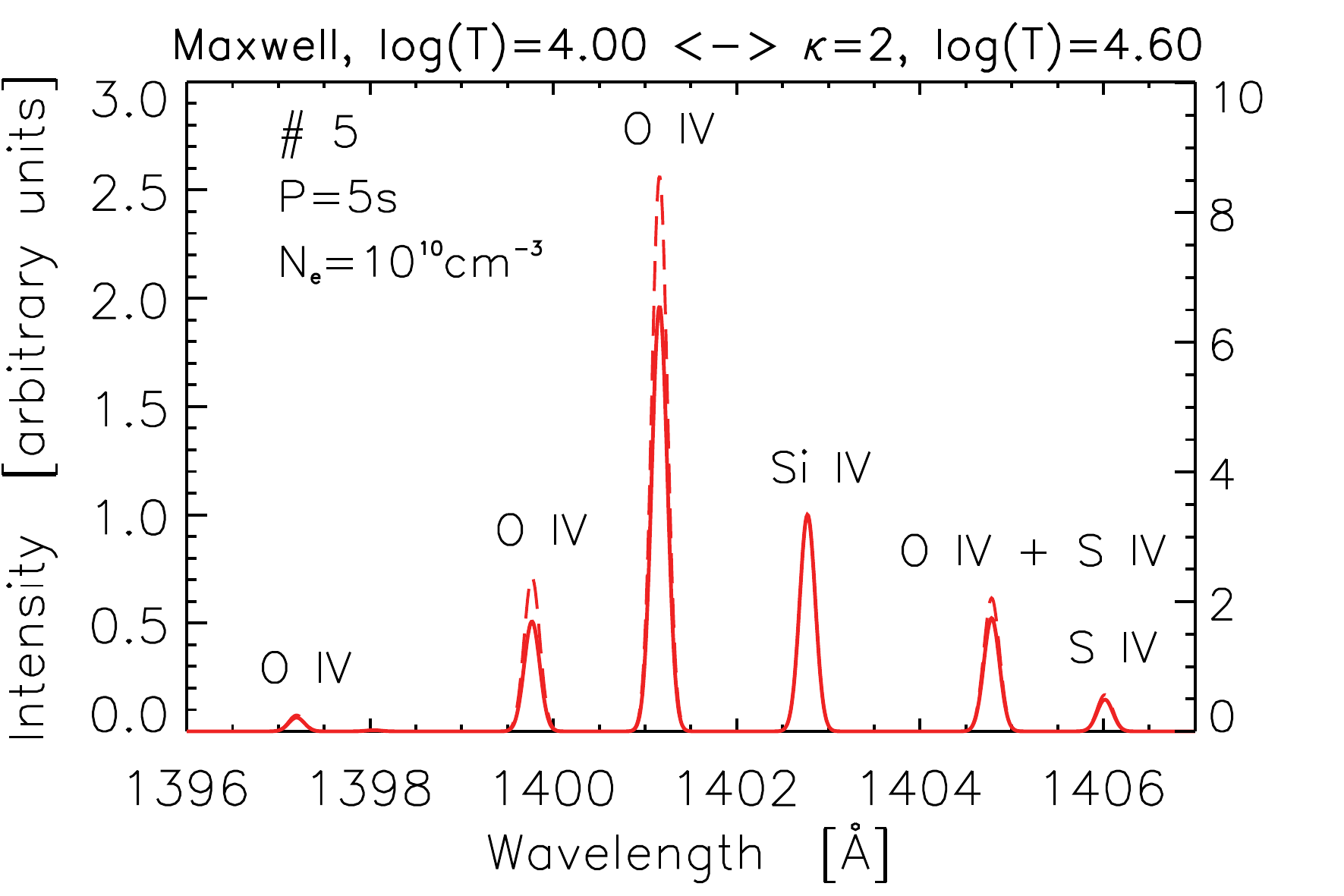}
        \includegraphics[width=6.0cm]{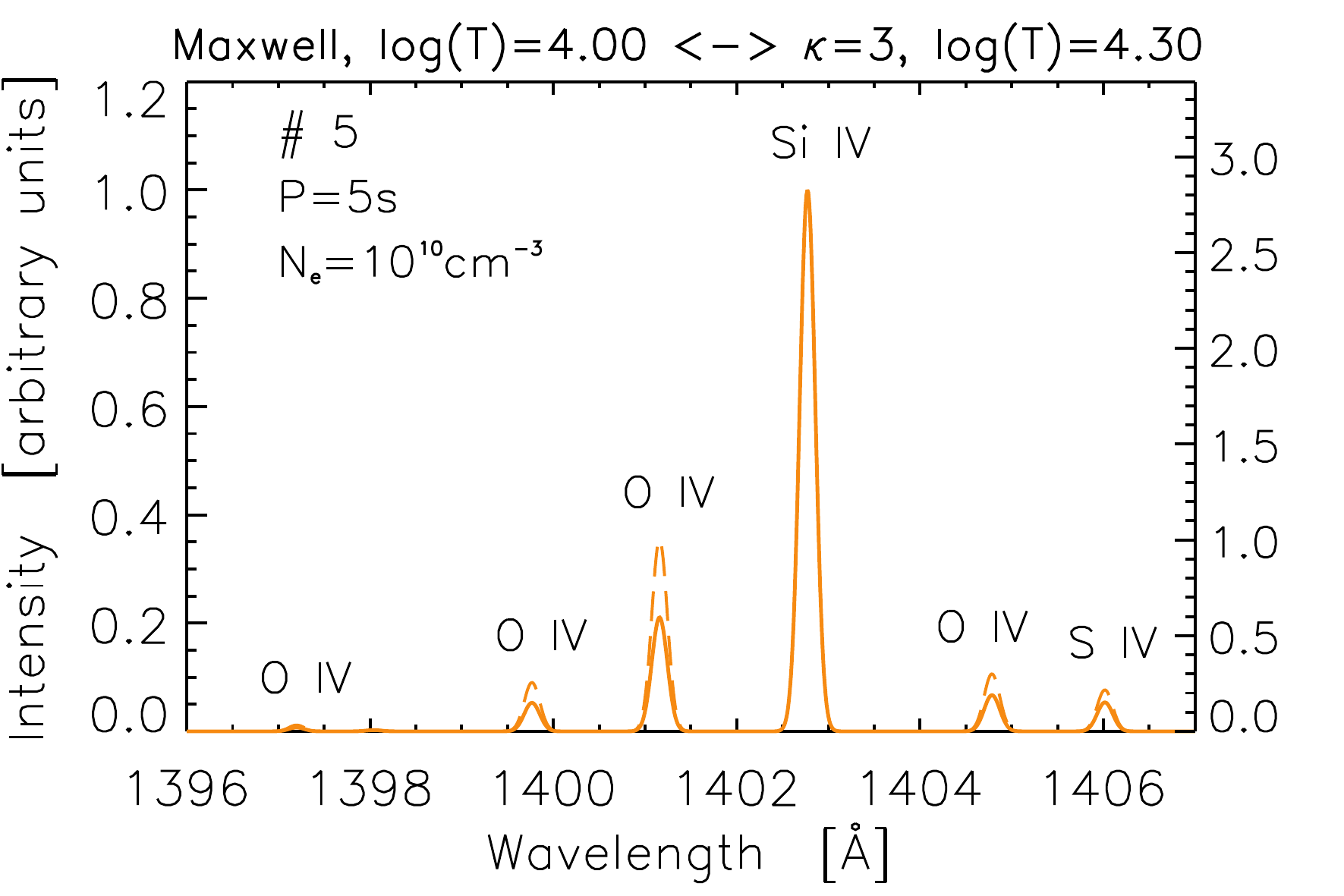}
        \includegraphics[width=6.0cm]{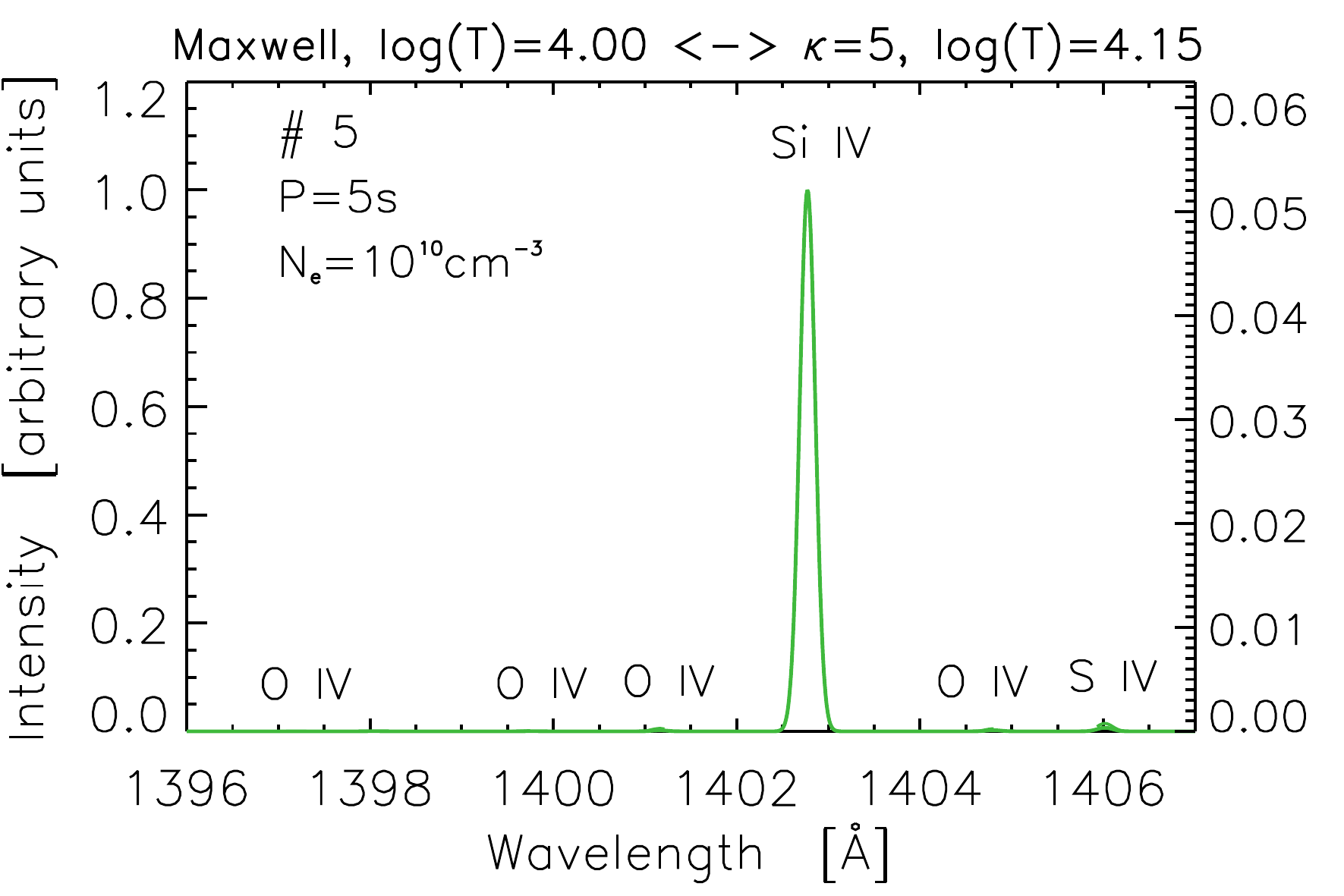}
        \includegraphics[width=6.0cm]{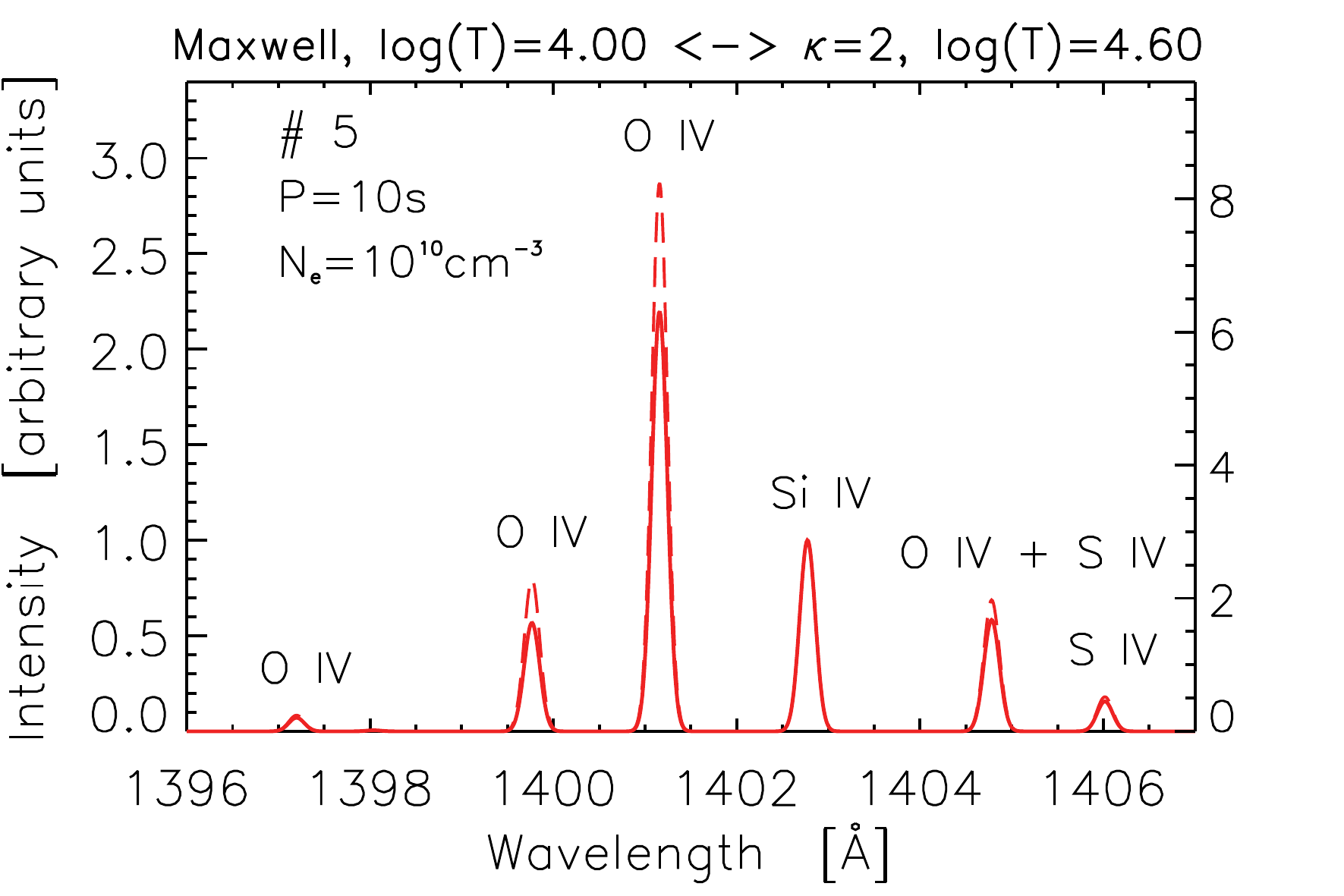}
        \includegraphics[width=6.0cm]{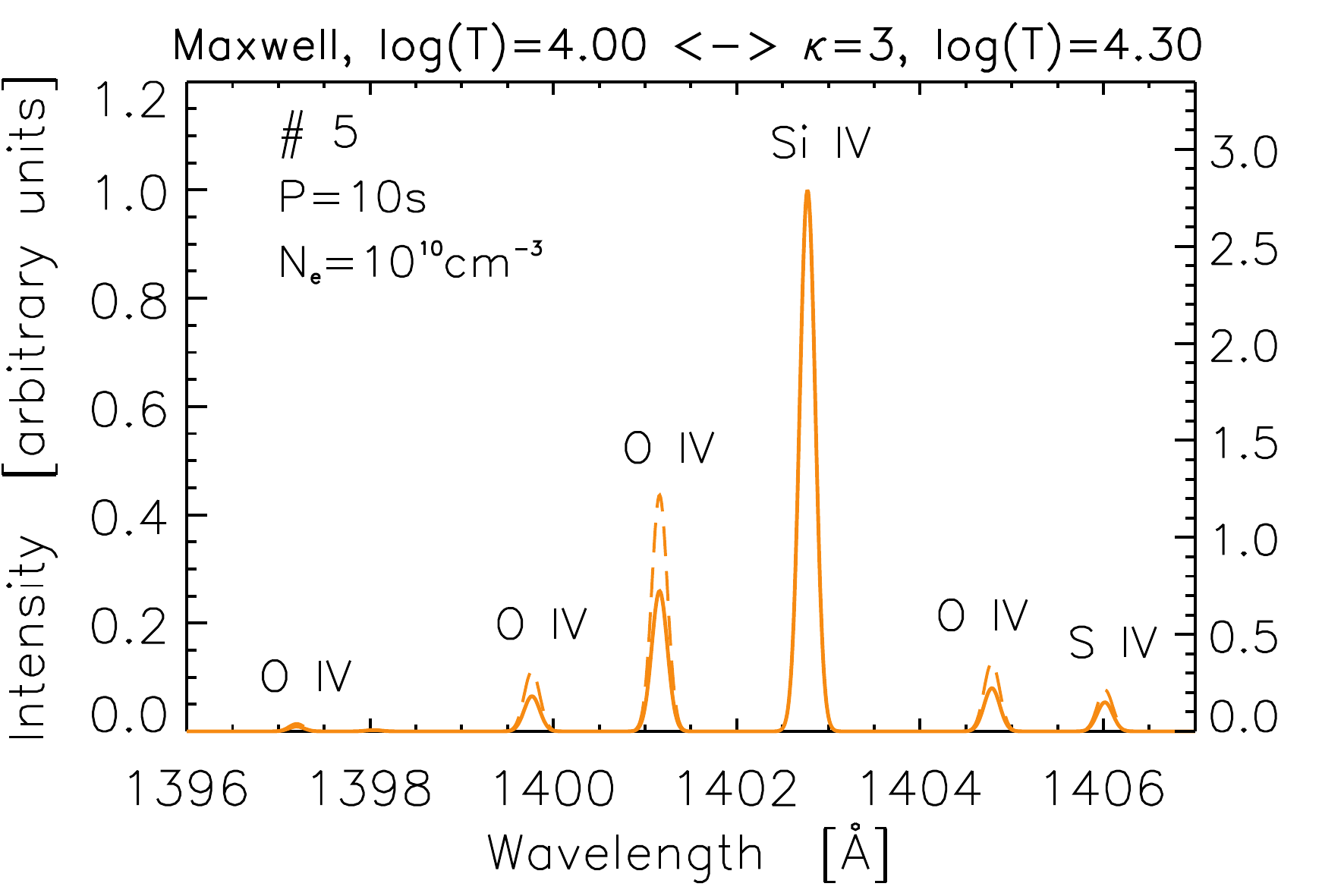}
        \includegraphics[width=6.0cm]{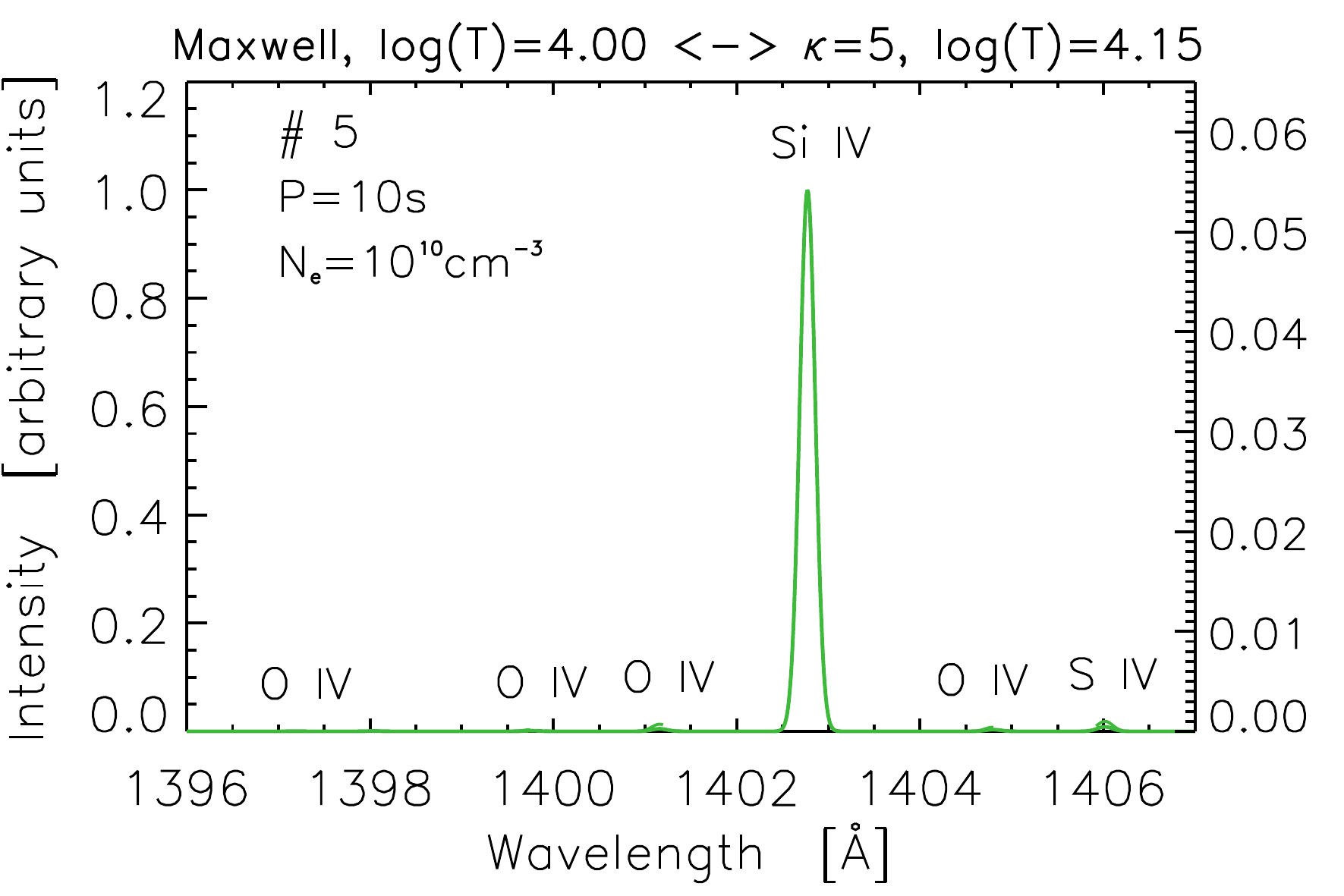}
        \includegraphics[width=6.0cm]{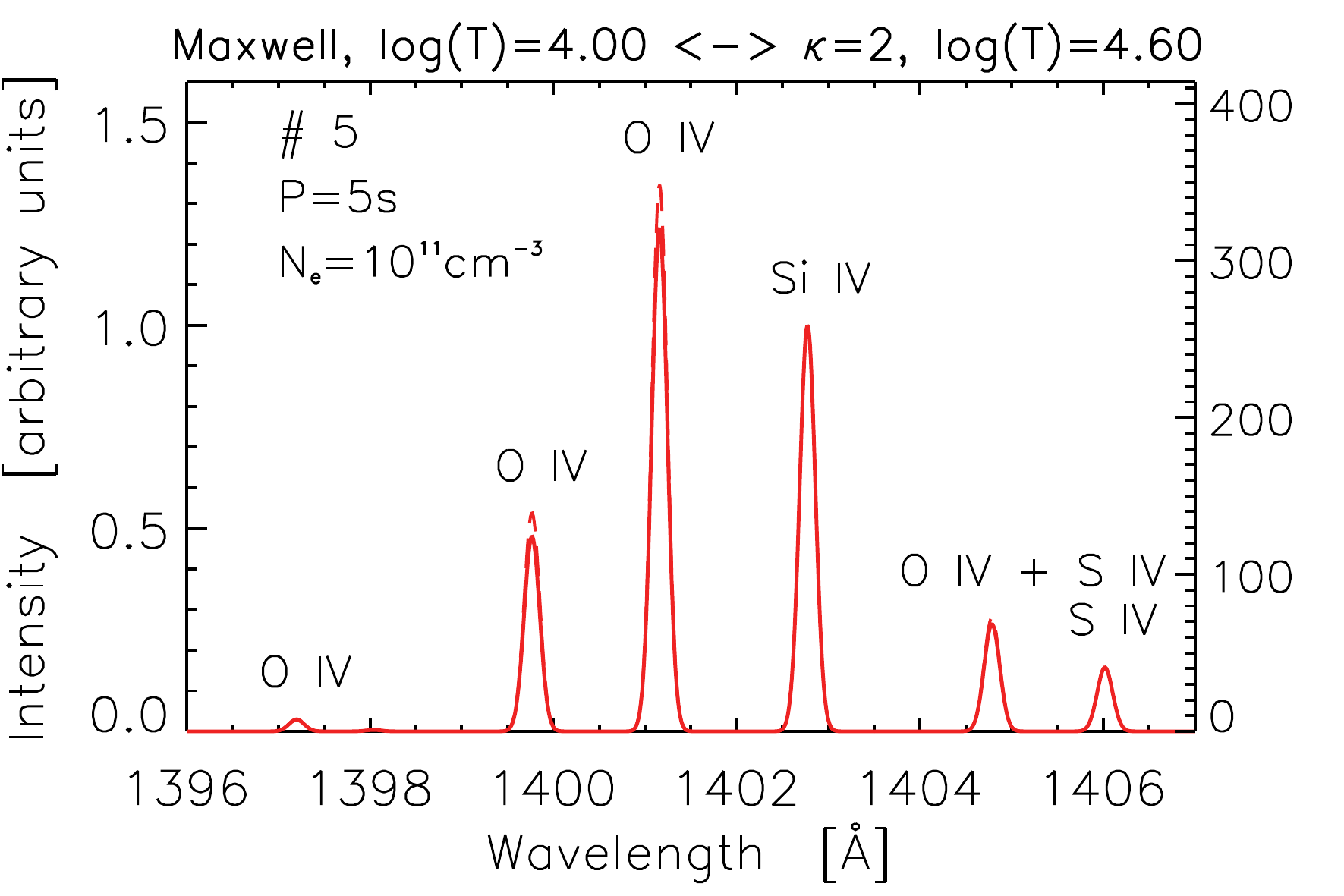}
        \includegraphics[width=6.0cm]{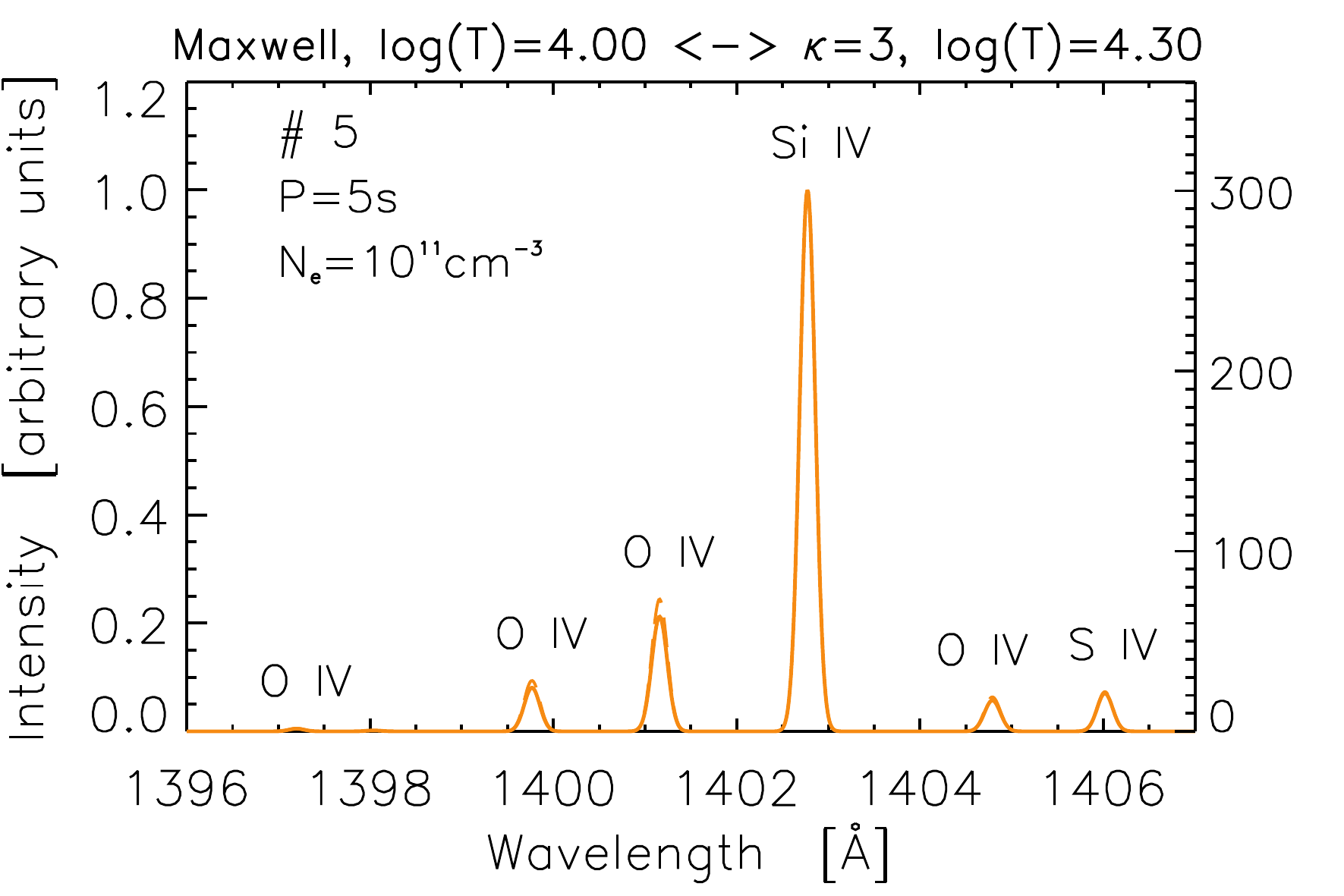}
        \includegraphics[width=6.0cm]{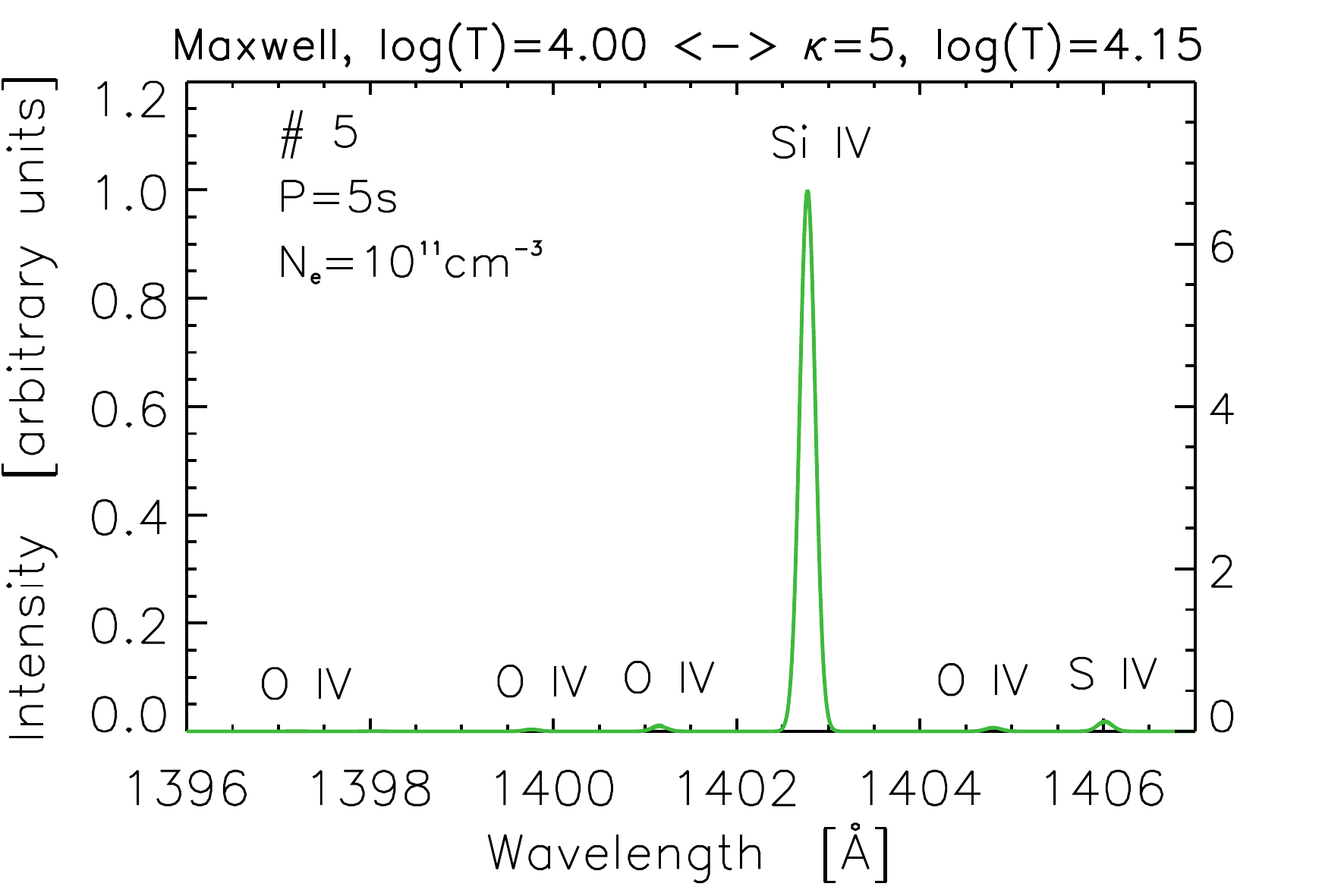}
        \includegraphics[width=6.0cm]{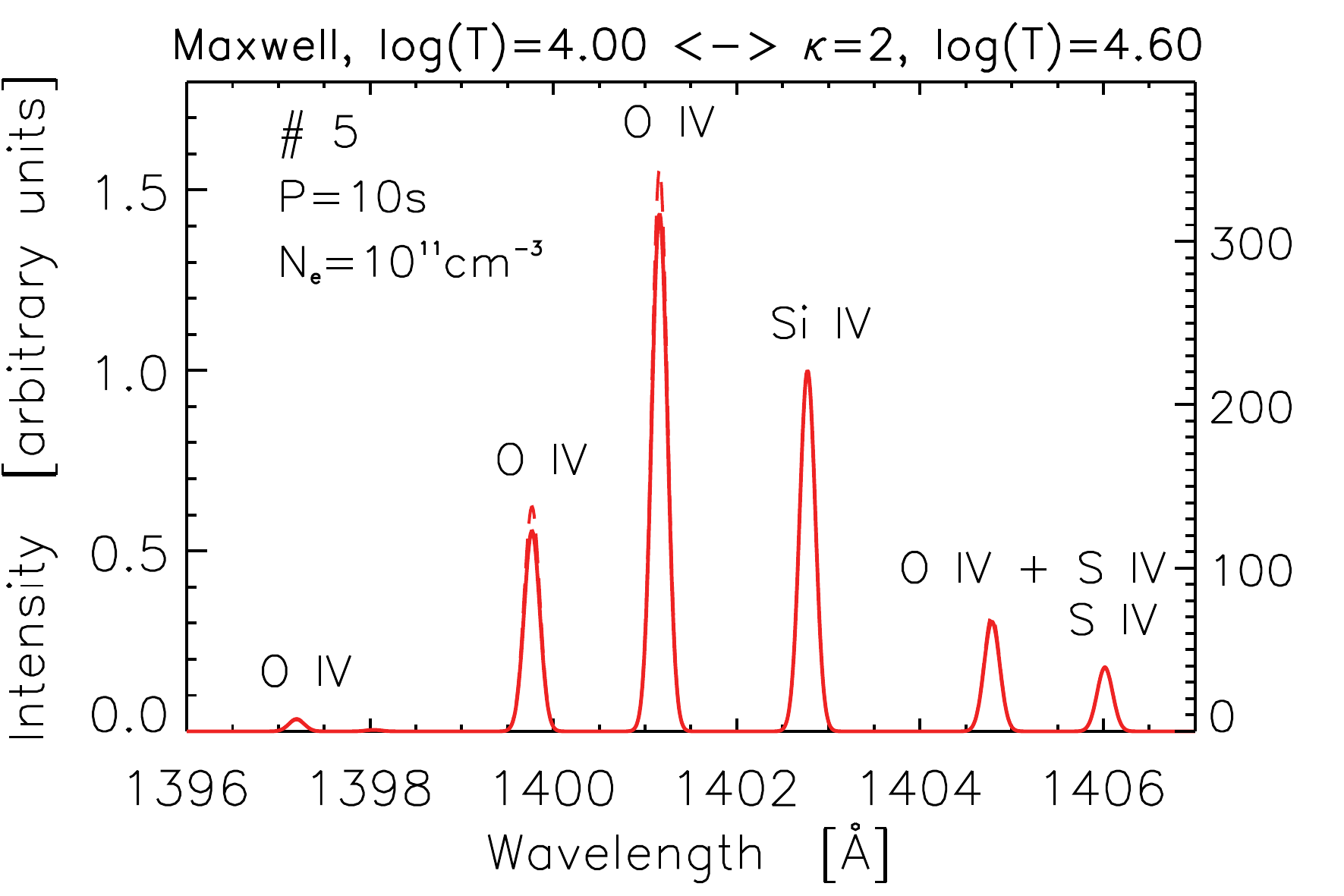}
        \includegraphics[width=6.0cm]{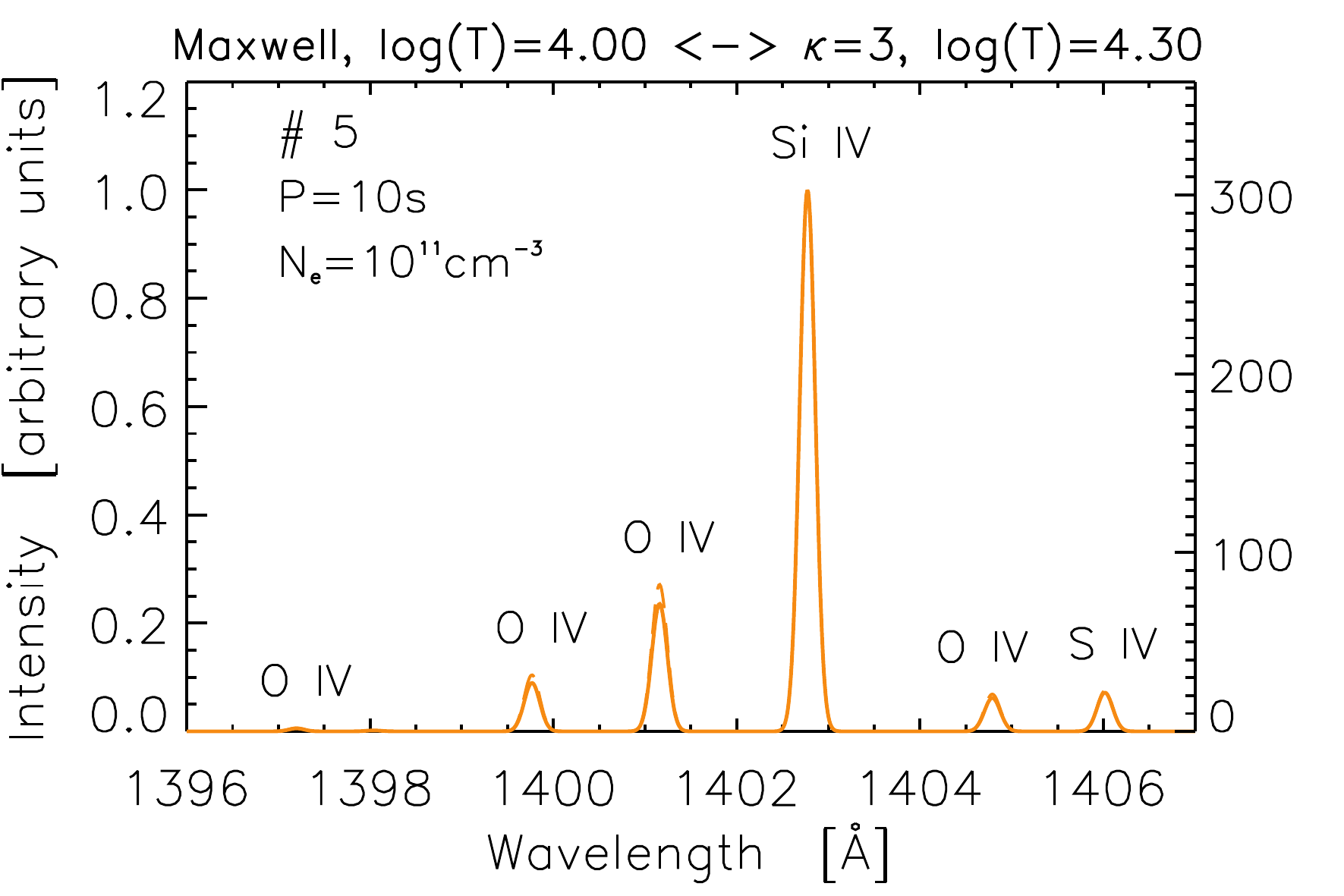}
        \includegraphics[width=6.0cm]{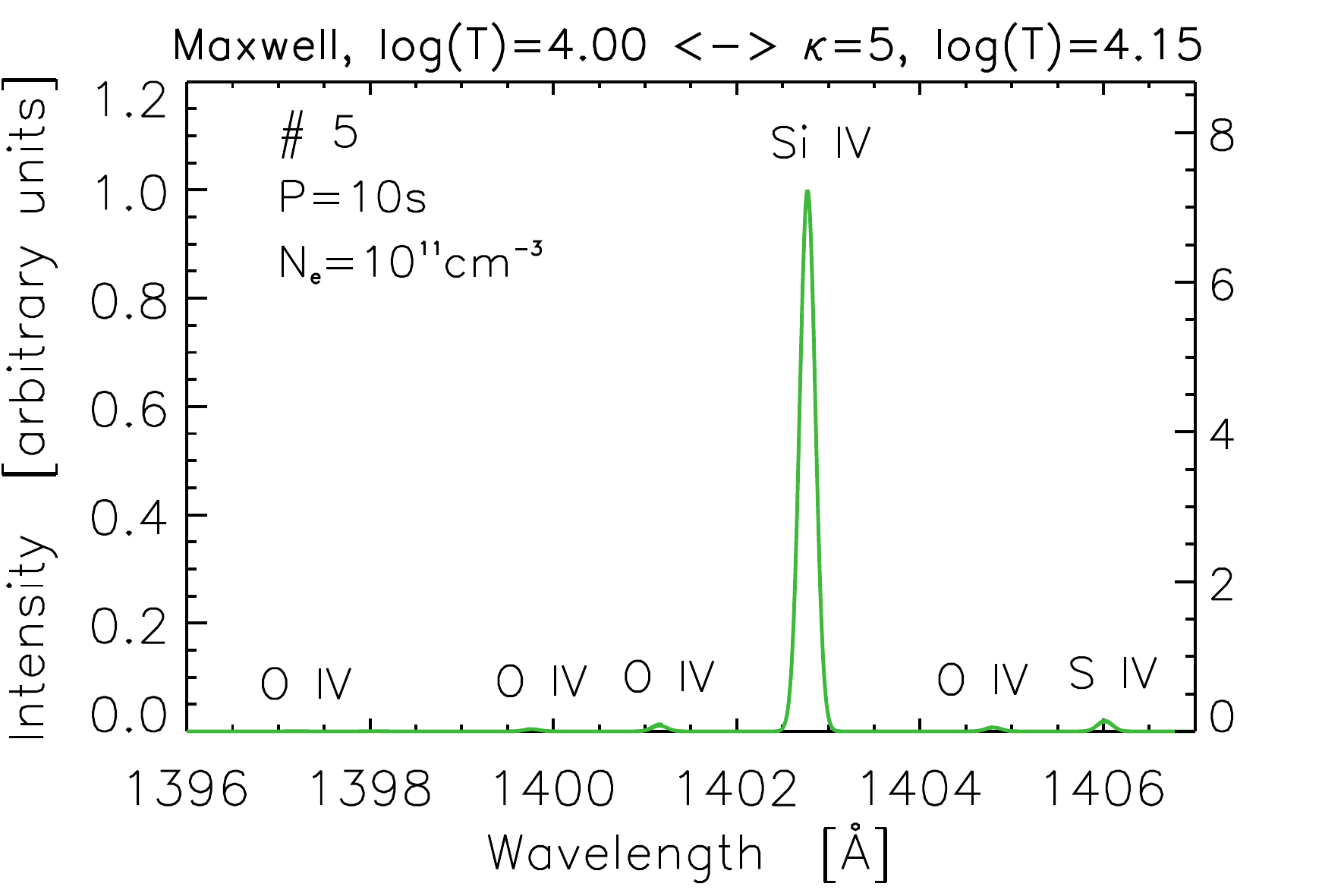}
\caption{Same as in Fig. \ref{Fig:sp1_T4}, but for the fifth period.}
\label{Fig:sp5_T4}
\end{figure*}
%
%
%

\begin{figure*}[!ht]
        \centering 
        \includegraphics[width=6.0cm]{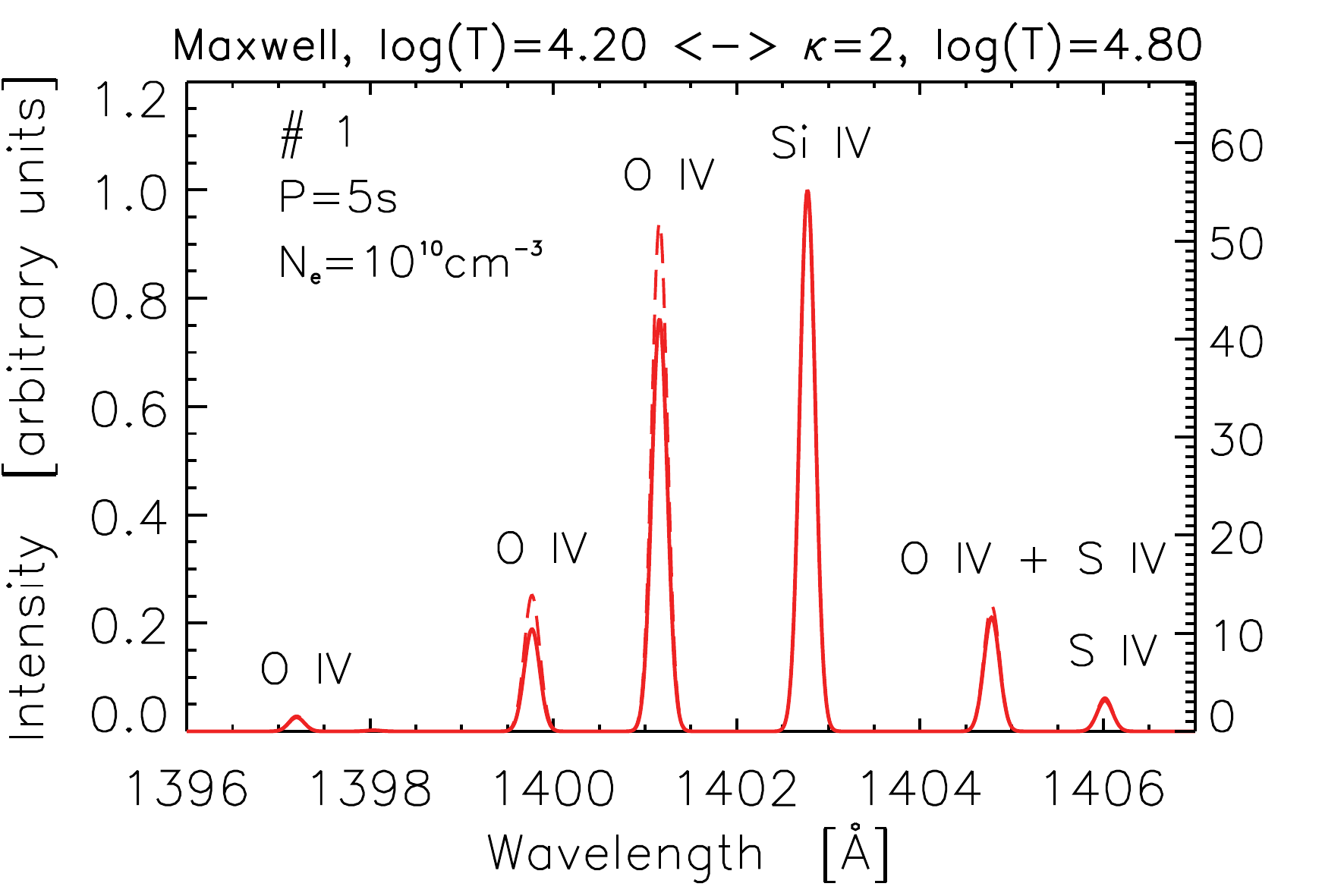}
        \includegraphics[width=6.0cm]{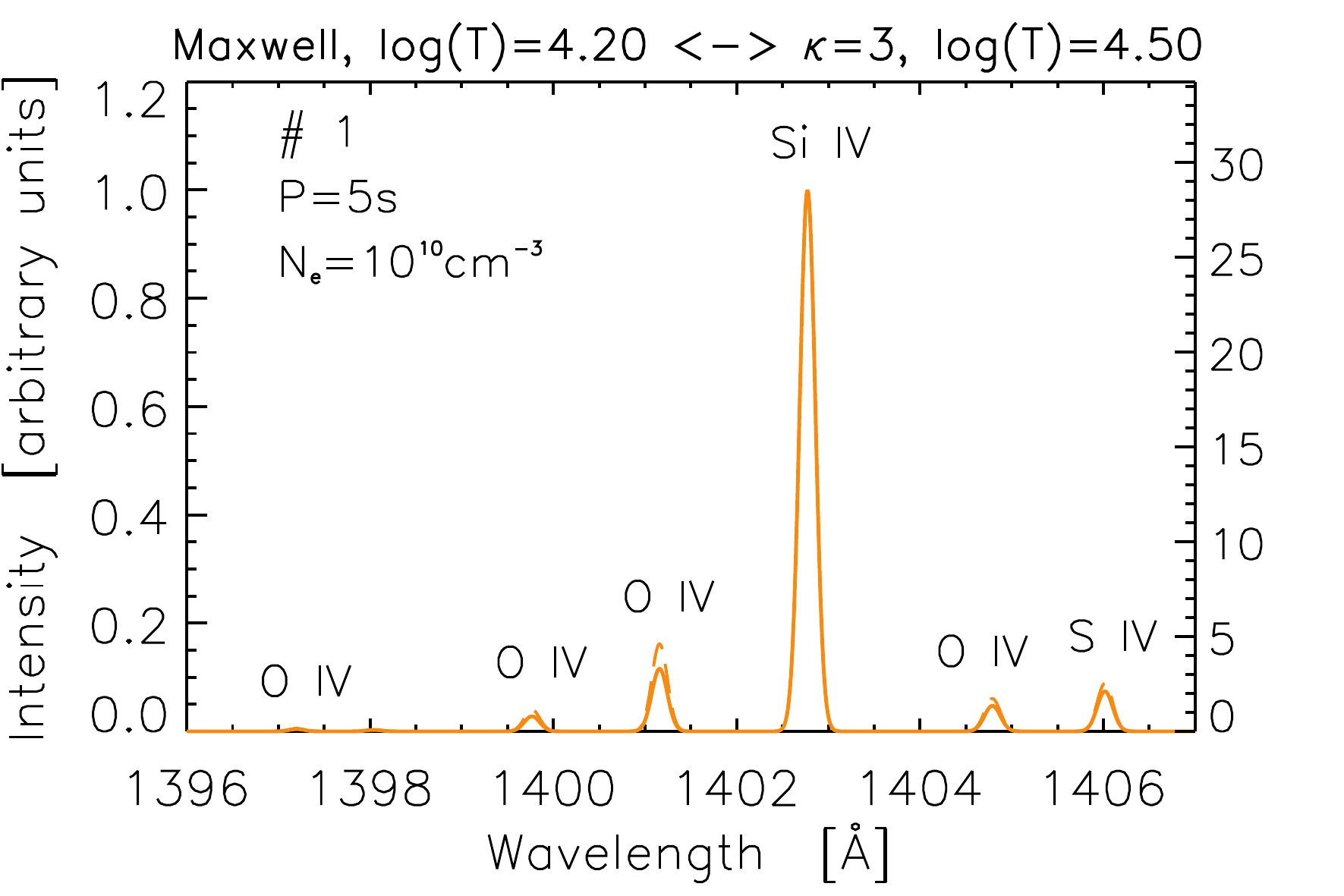}
        \includegraphics[width=6.0cm]{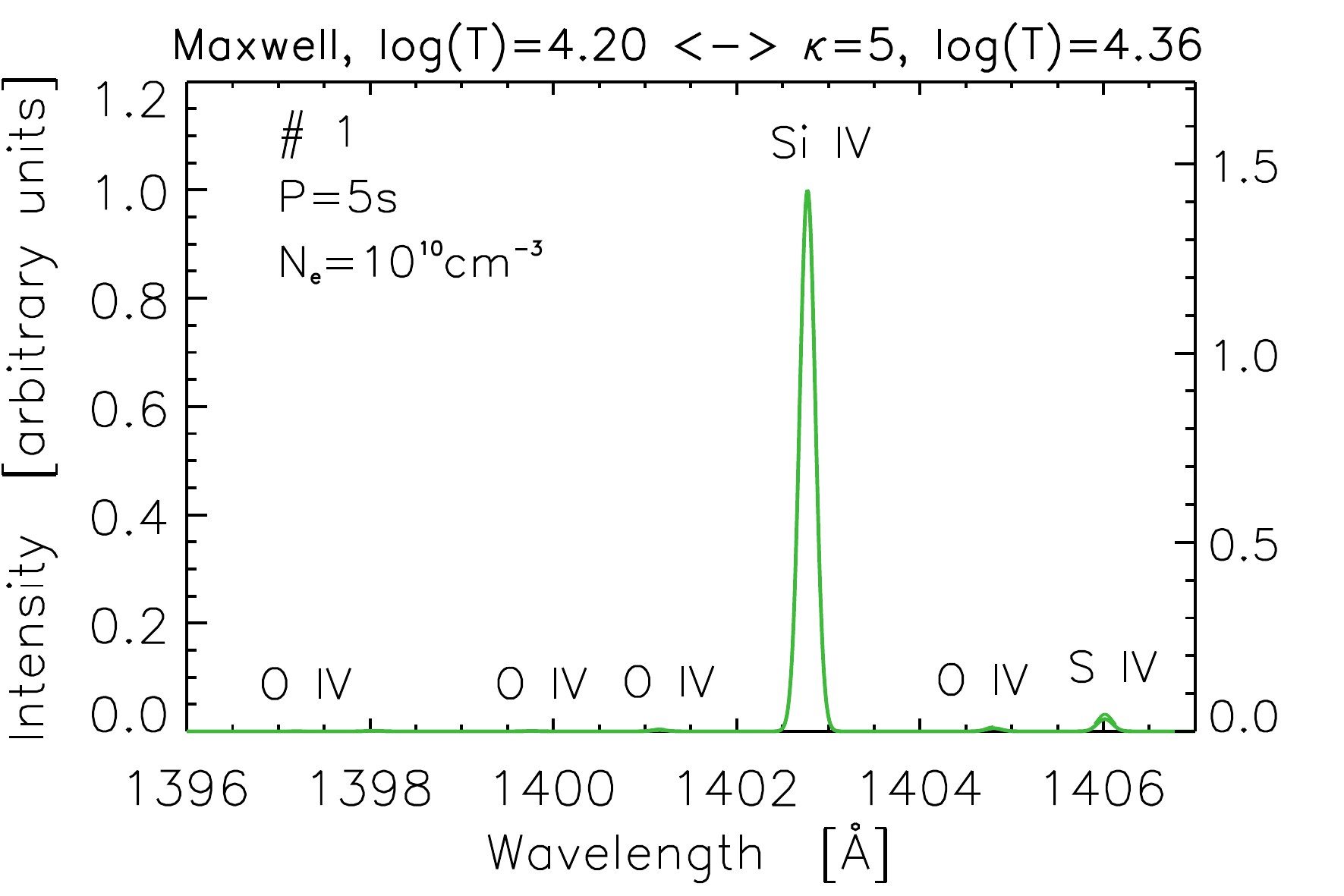}
        \includegraphics[width=6.0cm]{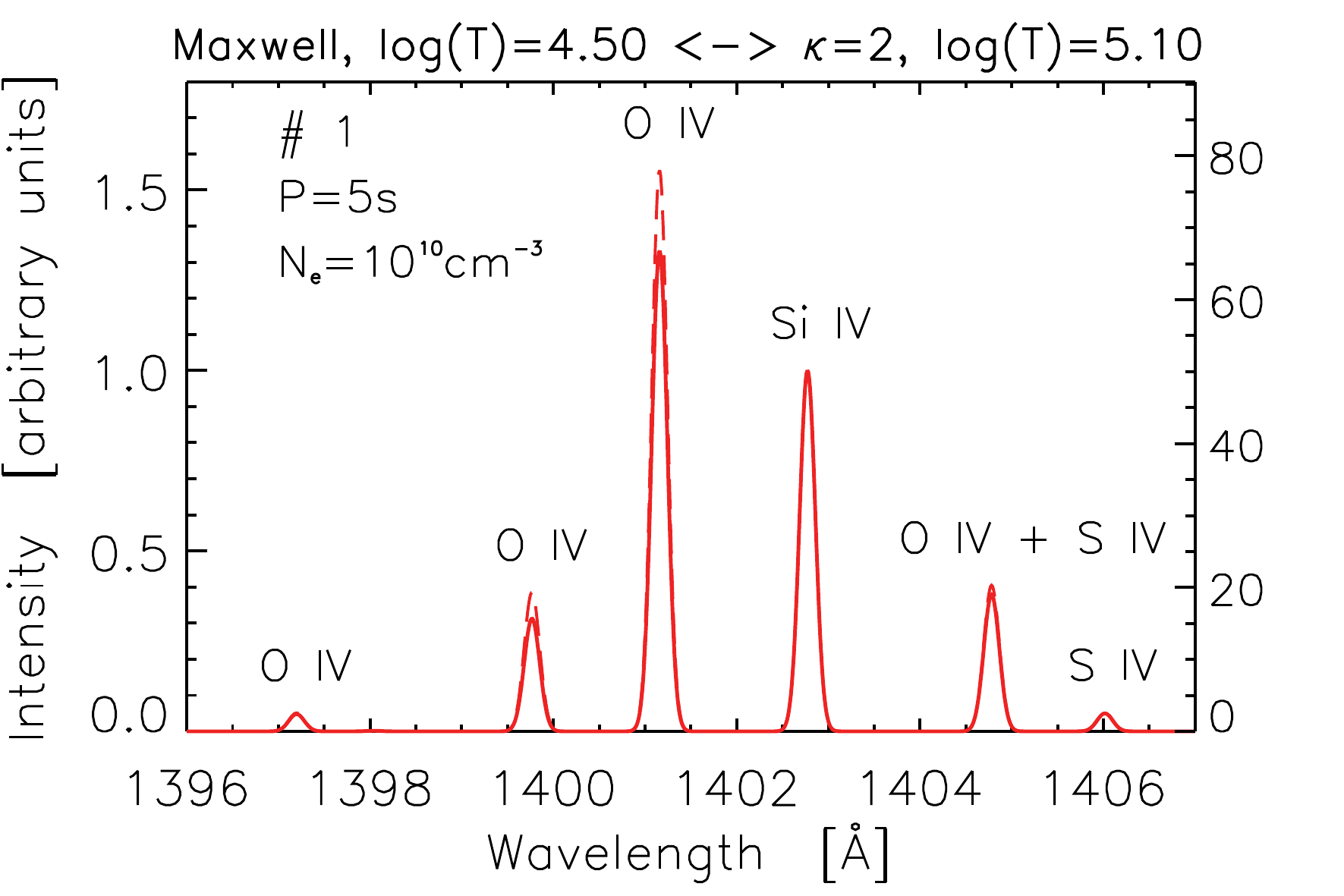}
        \includegraphics[width=6.0cm]{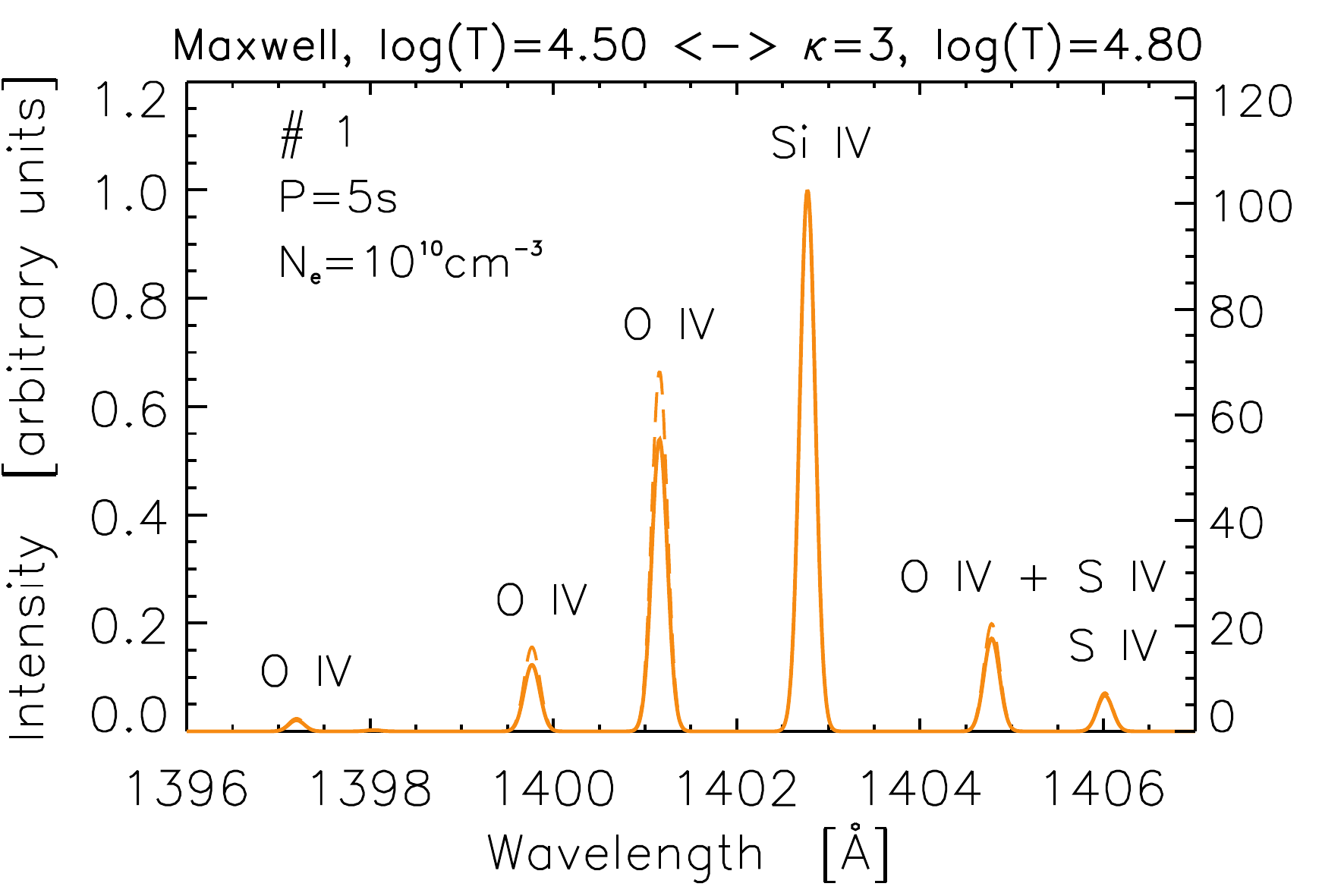}
        \includegraphics[width=6.0cm]{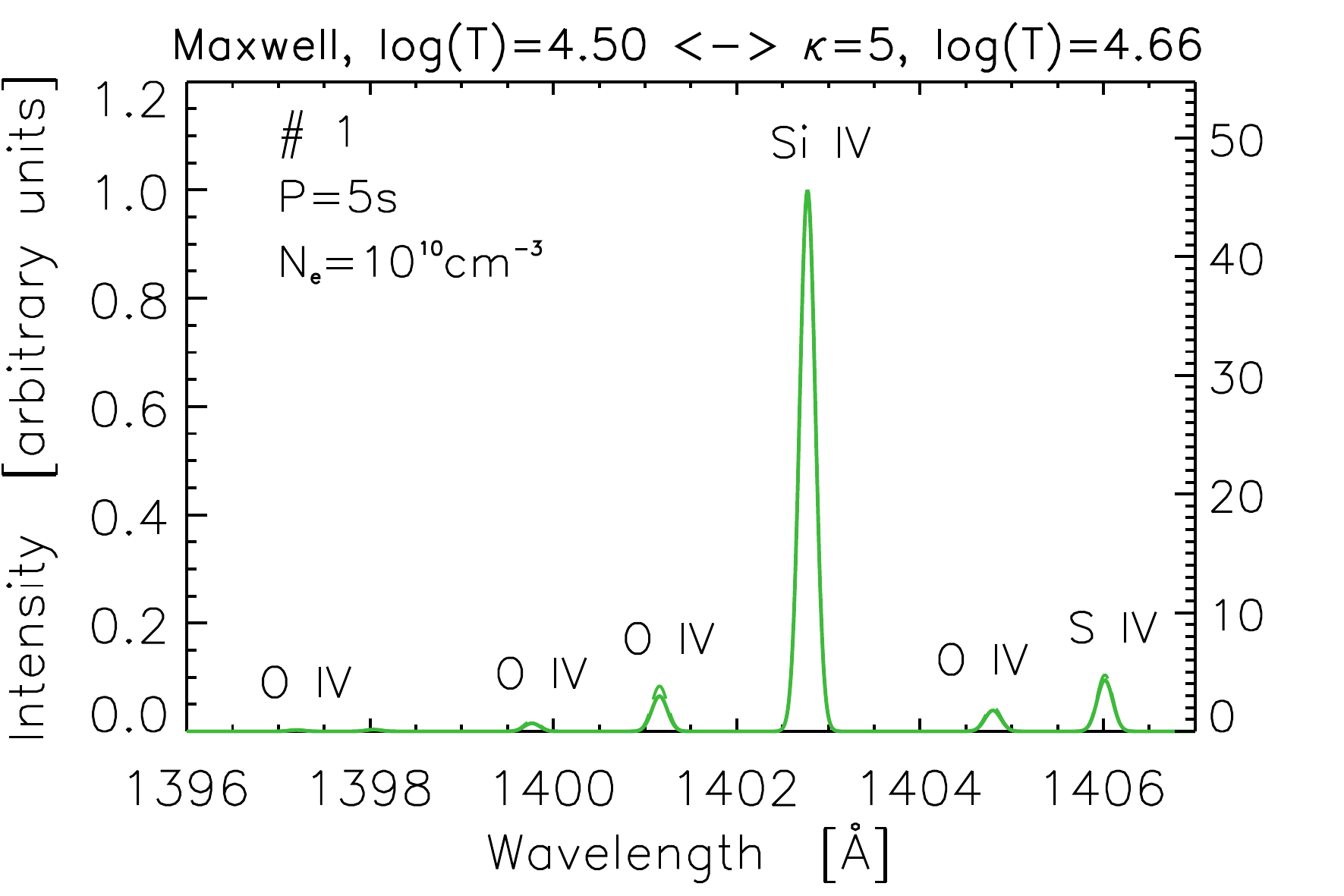}
        \includegraphics[width=6.0cm]{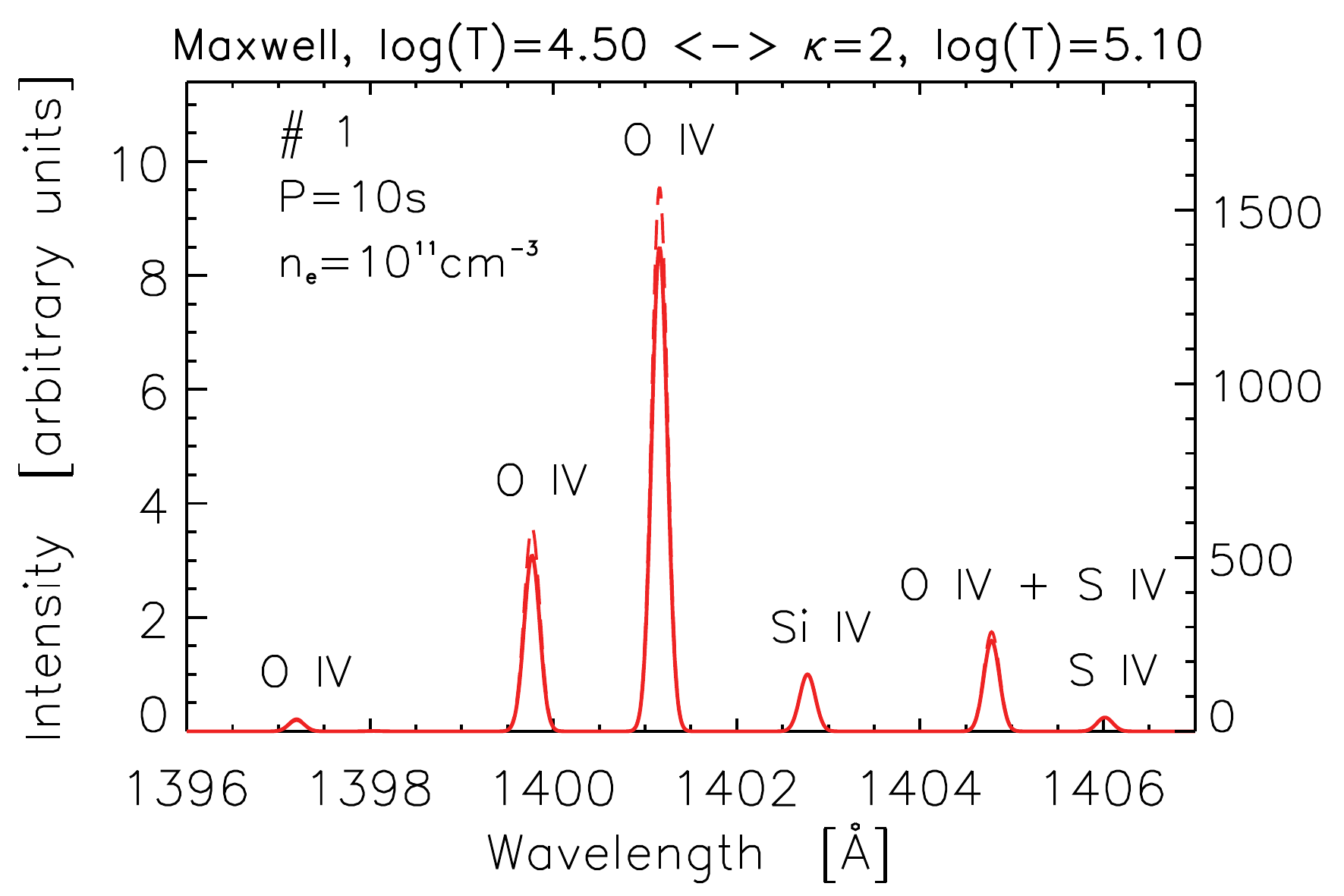}
        \includegraphics[width=6.0cm]{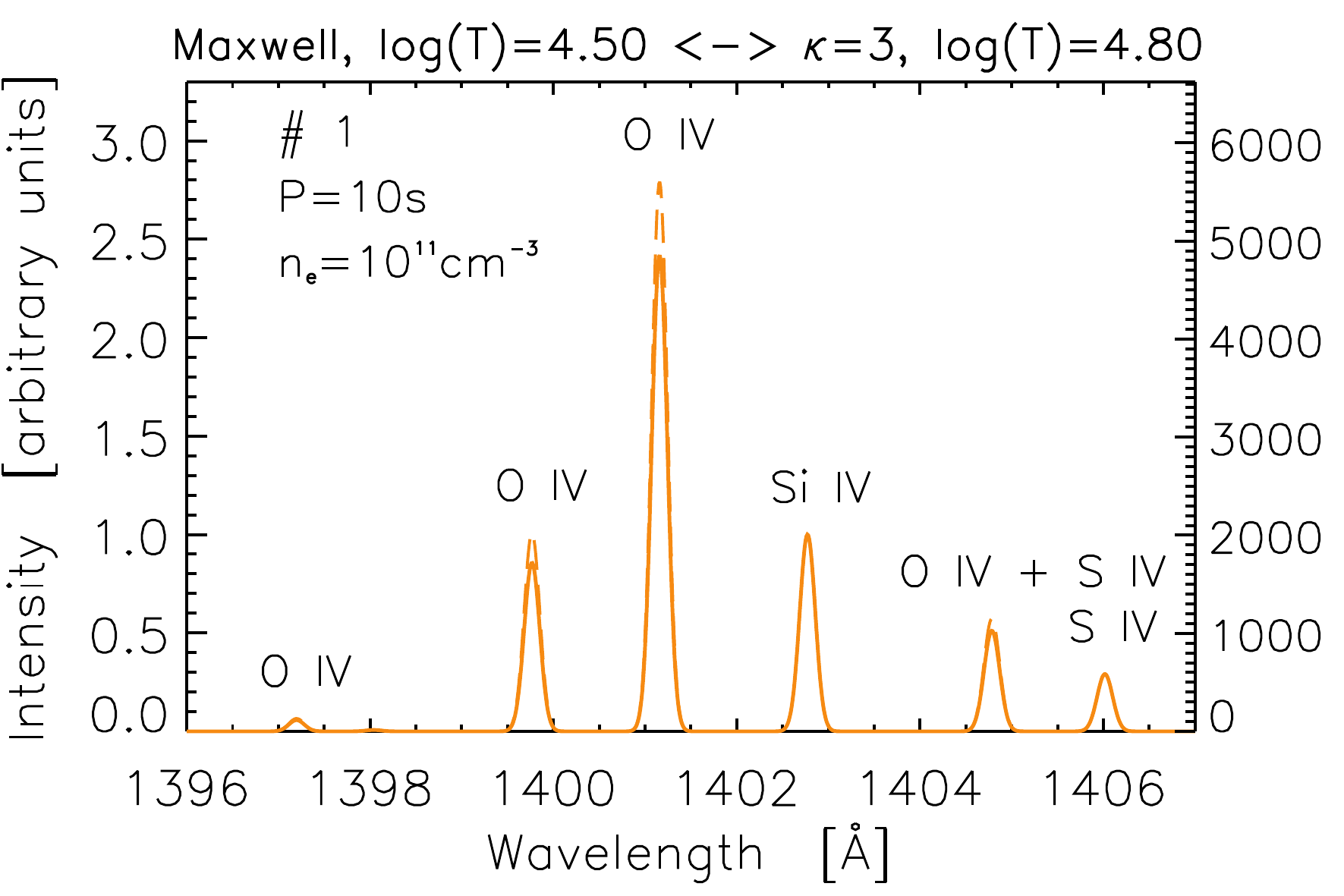}
        \includegraphics[width=6.0cm]{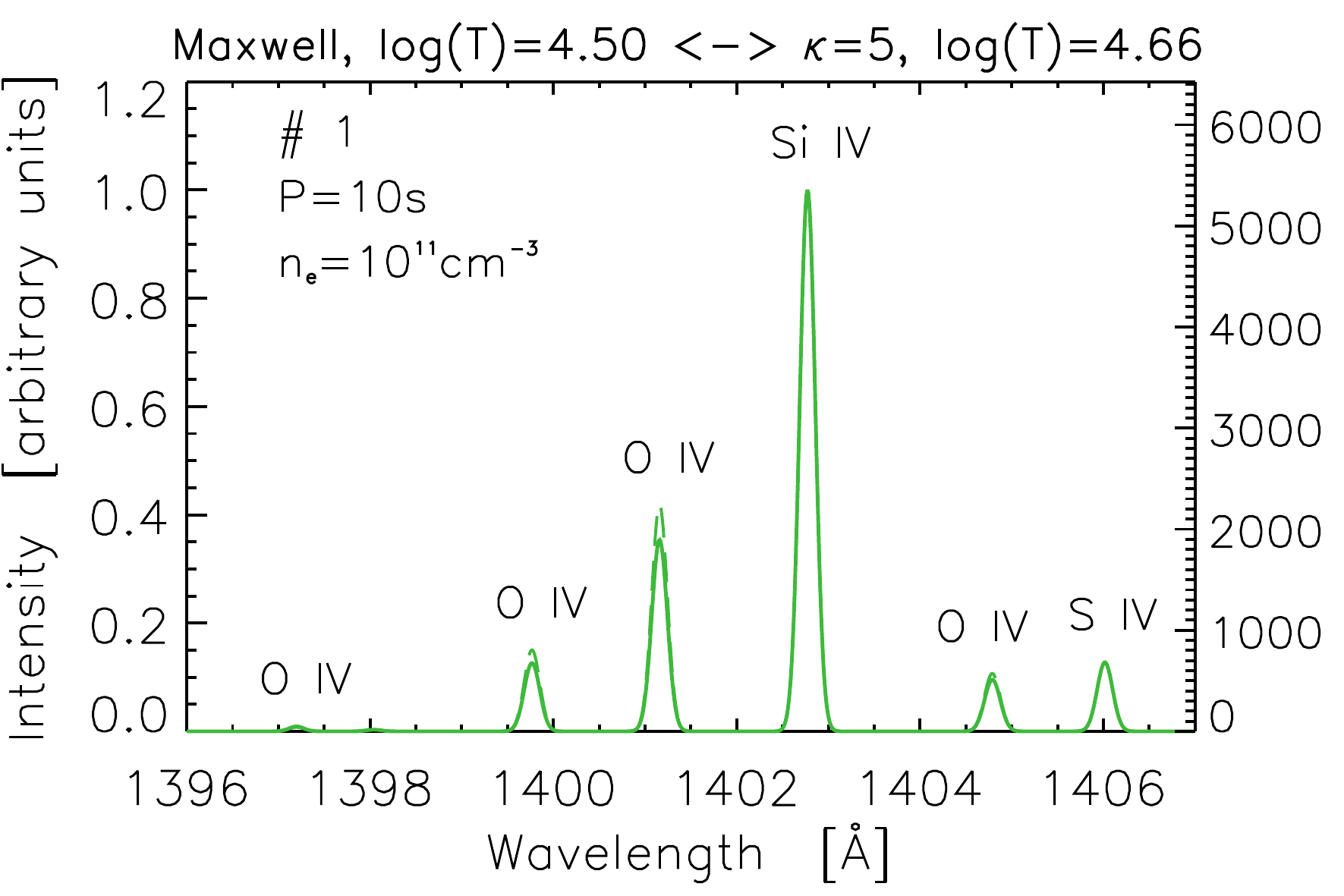}
\caption{Same as in Fig. \ref{Fig:sp1_T4}, but for different initial Maxwellian temperatures, log($T$\,[K])\,=\,4.2 (top), and 4.5 (middle and bottom).}
\label{Fig:sp1_Tdiff}
\end{figure*}

%
%
\subsection{Synthetic IRIS TR spectra}
\label{Sect:3.2}

After calculating the evolution of the ionic composition and the corresponding $T_\mathrm{eff}$, we now discuss the resulting spectra at 1396--1407\,\AA, both instantaneous and period averaged. The spectral range chosen corresponds to the FUV2 channel of the IRIS instrument and contains a strong \ion{Si}{IV} allowed line at 1402.8\,\AA~together with the neighboring \ion{O}{IV} intercombination multiplet and a \ion{S}{IV} intercombination line at 1406\,\AA. We note that the periods $P$ considered here can correspond to typical integration times of spectrometers.

\subsubsection{TR spectra for $\kappa$-distributions and ionization equilibrium}
\label{Sect:3.2.1}

To discuss the non-equilibrium character of the spectra, we compare them to the equilibrium spectra for $\kappa$-distributions at the corresponding temperatures $T_\kappa$. As discussed in Sect. \ref{Sect:2}, photoexcitation was included in the calculation of these spectra alongside the collisional excitation and de-excitation, and radiative de-excitation. These equilibrium spectra are shown in Fig. \ref{Fig:sp_eq}, where the dashed and full lines stand for $N_\mathrm{e}$\,=\,10$^{10}$ and 10$^{11}$\,cm$^{-3}$, respectively. The values of the relative ion abundances of \ion{Si}{IV} and \ion{O}{IV} for the $\kappa$-distributions at temperatures $T$ during the second half-period are given in Table \ref{Table:2}. 

For $\kappa$\,=\,2, the corresponding plasma temperature during the second half-period is log($T_\kappa$ [K])\,=\,4.6 (or $T_\kappa$\,= 4\,$\times$\,10$^4$\,K). At such low $\kappa,$ the maxima of the \ion{Si}{IV} and \ion{O}{IV} abundances occur at lower $T_\mathrm{max}$\ than $T_\kappa$ \citep[Table \ref{Table:1} and red arrows in Fig. \ref{Fig:rates}, see also Fig. 1 in][]{Dudik14b}. For \ion{Si}{IV},  $T_\mathrm{max}$ is much lower than for \ion{O}{IV}, meaning that the equilibrium spectra are expected to show very little \ion{Si}{IV} emission compared to \ion{O}{IV} (cf. Table \ref{Table:2}). This is indeed the case, see Fig. \ref{Fig:sp_eq} left. In these conditions, the \ion{O}{IV} 1401.2\,\AA~line is a factor of 2--4 higher than the neighboring \ion{Si}{IV} 1402.8\,\AA~line, depending on the density.

For $\kappa$\,=\,3, log($T_\kappa$ [K])\,=\,4.3, which is near the peak of the corresponding \ion{Si}{IV} abundance (Table \ref{Table:1}), while the \ion{O}{IV} ion abundance peak occurs at higher temperatures. The equilibrium spectra are thus dominated by \ion{Si}{IV} (middle panel of Fig. \ref{Fig:sp_eq} and Table \ref{Table:2}).

For $\kappa$\,=\,5, log($T_\kappa$ [K])\,$\approx$\,4.15, which is much lower than the corresponding \ion{Si}{IV} and \ion{O}{IV} peak temperatures, which occur at 4.65 and 4.95, respectively (Table \ref{Table:1}). We thus expect very low \ion{O}{IV} intensities, which is indeed the case (Fig. \ref{Fig:sp_eq}, right).

\subsubsection{Non-equilibrium TR spectra: First period}
\label{Sect:3.2.2}

We first discuss the non-equilibrium TR spectra during the first period, calculated with initial $T$\,=\,10$^4$\,K. Although the beam is assumed to be periodic, the behavior of the spectra during the first period can be used as an approximation for situations where the beam occurs only once, in analogy to a single heating pulse or burst of reconnection in plasma at initially chromospheric temperatures.

During the first period, the situation is such that no lines are present for the duration of $P/2$ when the distribution is Maxwellian, followed by a fast increase in $T_\mathrm{eff}$ and line intensities when the beam is switched on. Typically, the timescale of this increase is on the order of seconds or shorter, with the increase being faster for \ion{Si}{IV} lines than for the \ion{O}{IV} lines (Fig. \ref{Fig:Si4_O4}). This behavior can be discerned from the instantaneous spectra and their evolution shown in Fig. 1 and the accompanying online animation presented in Appendix \ref{Appendix:A}. After the initial increase, the evolution during the second $P/2$ slows down, with the \ion{Si}{IV} decreasing toward the end of the first period, as the ionic composition evolves toward higher ionization stages. 

Generally, for densities of 10$^{10}$\,cm$^{-3}$, the ionization state is always out of equilibrium (Sect. \ref{Sect:3.1}) and thus the spectra are fully in non-equilibrium as well (Fig. \ref{Fig:sp1_T4}, rows 1--2). Equilibrium ion abundances of \ion{Si}{IV} and \ion{O}{IV} (Table \ref{Table:2}) are reached only toward the end of the first period and only for some combination of parameters, such as $\kappa$\,=\,2 and at higher density of 10$^{11}$\,cm$^{-3}$. The corresponding spectra are then closer to the equilibrium spectra; compare row 4 of Fig. \ref{Fig:sp1_T4} with Fig. \ref{Fig:sp_eq}. 

For the strong electron beam, $\kappa=2$, and $P=5$\,s, the period-averaged spectra show that the \ion{O}{IV} intensities are lower than the \ion{Si}{IV} 1402.77\,\AA~intensity (Fig. \ref{Fig:sp1_T4} top left). This spectrum is very different from the equilibrium spectrum (Fig. \ref{Fig:sp_eq}, dashed line in the left panel). With longer periods or higher densities, the \ion{O}{IV} lines increase compared to \ion{Si}{IV}. At $N_\mathrm{e}=10^{11}$\,cm$^{-3}$ and $P=10$\,s, \ion{O}{IV} 1401.2\,\AA~is a factor of $\approx$1.3 higher than \ion{Si}{IV} 1402.8\,\AA~(Fig. \ref{Fig:sp1_T4}, bottom left), approaching the equilibrium case. This increase would be even higher if the photoexcitation were not included (dashed lines in Fig. \ref{Fig:sp1_T4}). Photoexcitation notwithstanding, this behavior of the \ion{O}{IV} lines compared to \ion{Si}{IV} is caused by the high log($T_\mathrm{eff}^{\kappa=2}$\,[K])\,=\,4.6 that is reached toward the end of the first period. As explained in Sect. \ref{Sect:3.2.1}, this $T_\mathrm{eff}^{\kappa=2}$ is much higher than the  $T_\mathrm{max}$ for \ion{Si}{IV} at $\kappa$\,=\,2 (Table \ref{Table:1}), but only 0.15 dex higher than the compared $T_\mathrm{max}$ for \ion{O}{IV}. If the charge state is interpreted through $T_\mathrm{eff}^\mathrm{Mxw}$ for the Maxwellian distributions, from Fig. \ref{Fig:Teff_evol} we obtain for this case values of log$(T_\mathrm{eff}^\mathrm{Mxw}$\,[K])\,$\approx$\,5.3--5.45 for O and Si, respectively, which is much higher than the corresponding Maxwellian $T_\mathrm{max}$ for \ion{Si}{IV}, but closer to the $T_\mathrm{max}$ for \ion{O}{IV}. Therefore, the \ion{O}{IV} lines will dominate the \ion{Si}{IV} for these conditions.

A weaker beam with $\kappa=3$ can form plasma with an effective temperature of log($T_\mathrm{eff}^{\kappa=3}$\,[K])\,=\,$2\times10^4$\,K (or log($T_\mathrm{eff}^\mathrm{Mxw}$\,[K])\,$\approx$\,4.8). These temperatures are close to the \ion{Si}{IV} abundance maximum for both $\kappa=3$ and Maxwellian distributions, while \ion{O}{IV} is formed at higher $T_\mathrm{max}$. This significantly reduces
the \ion{O}{IV} line intensities, and the spectra for a beam with $\kappa$\,=\,3 are dominated by \ion{Si}{IV} lines. These synthetic spectra can approximate the typically observed spectra \citep[e.g.,][and references therein]{Dudik17b}. At lower densities, however, where non-equilibrium effects are stronger, \ion{Si}{IV} 1402.8\,\AA~can be about an order of magnitude or more higher than the neighboring \ion{O}{IV} 1401.2\,\AA~line. This spectrum is far from the equilibrium spectrum shown with the
dashed line in the middle panel of Fig. \ref{Fig:sp_eq}. In contrast to this, for $N_\mathrm{e}=10^{11}$\,cm$^{-3}$, the TR spectrum is similar to the equilibrium spectrum, with the \ion{O}{IV} 1401.2\,\AA~intensity being about a factor of 3--5 lower than the \ion{Si}{IV} 1402.8\,\AA~intensity.

A weak electron beam with $\kappa=5$ causes only a small rise in $T_\mathrm{eff}$; log($T_\mathrm{eff}^{\kappa=5}$\,[K])\,$\lesssim$\,4.2, or equivalently log($T_\mathrm{eff}^\mathrm{Mxw}$\,[K])\,=\,4.4--4.6. These low effective temperatures combined with the non-equilibrium effects lead to spectra with very weak or unobserved \ion{O}{IV} lines. Similar relative intensities observed during explosive events were interpreted as being located deep in the solar atmosphere \citep{Peter14Sci}. When interpreted in terms of ionization equilibrium and equilibrium Maxwellian distribution, spectra like this are only possible at electron densities higher by at least one order of magnitude \citep{Judge15}, possibly even three orders of magnitude \citep{Peter14Sci}! Here, we show that it is possible to create such spectra at much lower densities, if energetic particles \citep[e.g., due to the reconnection invoked by][]{Peter14Sci} are present.

\subsubsection{Non-equilibrium TR spectra: Fifth period}
\label{Sect:3.2.3}

At the end of the first period, the electron temperature reverts to 10$^4$\,K for another $P/2$. Since the timescales for excitation are very short, the spectra cease to show any emission lines (see the animation attached to Fig. 1). The ions recombine, and thus both $T_\mathrm{eff}^\mathrm{Mxw}$ and $T_\mathrm{eff}^\kappa$ decrease (Fig. \ref{Fig:Teff_evol}). As shown in Sect. \ref{Sect:3.1}, at  $N_\mathrm{e}$\,=\,10$^{10}$ cm$^{-3}$, ionization pumping can occur; that is, the ionic composition never reaches the initial composition before the beam is switched on again. Following Paper~I, we now describe the main features of the TR spectra during the fifth period, when the averaged $T_\mathrm{eff}$ behavior is not significantly different from near stationary oscillatory behavior.

For $\kappa$\,=\,2 and $N_\mathrm{e}$\,=\,10$^{10}$\,cm$^{-3}$, the $T_\mathrm{eff}^\kappa$ calculated for Si increases more slowly than the corresponding oxygen temperature. This means that during the fifth period, the temperatures are too high for \ion{Si}{IV}, and most of Si exists at higher ionization stages than during the first period. The relative ion abundance of \ion{Si}{IV} decreases (Fig. \ref{Fig:Si4_O4}) along with its intensities. The spectra become dominated by \ion{O}{IV} lines (Fig. \ref{Fig:sp5_T4}, red color in rows 1--2).

For $\kappa$\,=\,3, oxygen catches up with silicon, and the \ion{O}{IV} lines increase relative to the \ion{Si}{IV} lines. The spectra are nevertheless still dominated by the \ion{Si}{IV} 1402.8\,\AA.

Since the $T_\mathrm{eff}^{\kappa=5}$ for oxygen does not depart from 10$^4$\,K at all if $\kappa$\,=\,5, the spectra during the fifth period look very similar to those from the first period. In particular, there are no visible \ion{O}{IV} lines.

For $N_\mathrm{e}$\,=\,10$^{11}$\,cm$^{-3}$, the spectra during the fifth period are not significantly different from those during the first period for all $\kappa$ (compare rows 3--4 of Figs. \ref{Fig:sp1_T4} and \ref{Fig:sp5_T4}). This is because the $T_\mathrm{eff}$ at this density already reached a stationary oscillatory behavior very soon into the simulation (rows 3--4 of Figs. \ref{Fig:Teff_evol} and \ref{Fig:Si4_O4}).

\subsubsection{Higher initial temperatures}
\label{Sect:3.2.4}

To study the effect of higher initial Maxwellian temperatures $T$, we recalculated the model spectra using values of log($T$\,[K])\,=\,4.2 and 4.5. The resulting spectra are shown in Fig. \ref{Fig:sp1_Tdiff}. For initial temperatures of  $T=1.5\times10^4$\,K, a lower electron density, and $P$\,=\,5\,s (row 1 of Fig. \ref{Fig:sp1_Tdiff}), the \ion{O}{IV} intensities are still lower than for  \ion{Si}{IV}. In fact, the spectra look very similar to the case of $T$\,=\,10$^4$\,K, except that the \ion{O}{IV} intensities have increased for $\kappa$\,=\,2 and 3.

At higher initial temperatures, \ion{O}{IV} achieves higher abundances faster, and its line intensities increase with the period and electron density (Fig. \ref{Fig:sp1_Tdiff}). While \ion{O}{IV} intensities increase, the effective temperatures for Si increase as well, resulting in the \ion{Si}{IV} abundances passing over its peak, followed by a decrease. This behavior increases the
relative emissivities of the \ion{O}{IV} lines to the \ion{Si}{IV} lines (Fig. \ref{Fig:sp1_Tdiff}, middle and right). For the highest initial temperature, a strong electron beam ($\kappa=2$), the
longest periods, and high density, the emissivities of the \ion{O}{IV}  lines can strongly overcome the \ion{Si}{IV} line emissivities (Fig. \ref{Fig:sp1_Tdiff}, bottom left). That situations like
this are never observed is perhaps not surprising, as this scenario would require acceleration of a large number of particles in a high-density plasma that is already heated to TR temperatures.

%
%
\section{Summary}
\label{Sect:4}

We have considered the occurrence of a periodic electron beam, described by a $\kappa$-distribution with a high-energy tail, in TR plasma. Our initial plasma temperatures chosen were 10$^4$\,K, which roughly corresponds to the upper chromosphere, while the densities considered were 10$^{10}$ and 10$^{11}$\,cm$^{-3}$, at which non-equilibrium ionization is likely to occur. We considered periods of 5 and 10\,s, similar to the timescale of the fast intensity changes reported from IRIS observations, and included the effect of photoexcitation.

Similarly to Paper~I, we have shown that the periodic electron beam, represented by additional high-energy particles, both increases the plasma temperature and drives the plasma out of equilibrium, and that the synthetic intensities of the TR lines of \ion{Si}{IV} and \ion{O}{IV} observable by IRIS are non-equilibrium as well. Ionization pumping can occur at lower densities of 10$^{10}$\,cm$^{-3}$. We found that the stronger the beam (i.e., the lower $\kappa$), the higher the effective ionization temperatures that are reached toward the end of the period. Combined with the strong shifts of the peaks of the relative ion abundances of \ion{O}{IV} and \ion{Si}{IV} toward lower temperatures in equilibrium, this translated into relatively high \ion{O}{IV} intensities compared to \ion{Si}{IV}. For higher $\kappa$\,=\,3, the effective ionization temperatures reached in the model corresponded to the peak of \ion{Si}{IV} abundances for such $\kappa$, resulting in \ion{O}{IV}\,/\,\ion{Si}{IV} ratios similar to the typically observed ratios. A surprising finding was that a relatively low number of energetic particles, corresponding to $\kappa$\,=\,5, that is, the weakest high-energy tails considered here (with only $\approx$19\% additional particles carrying 40\% extra energy, see Sect. 2.1 of Paper~I), was able to generate spectra with negligible \ion{O}{IV} lines compared to \ion{Si}{IV}. This could might explain the peculiar spectra observed during bursty situations, such as in UV (Ellerman) bursts.

Overall, we showed that a simple combination of non-equilibrium ionization with high-energy particles can produce \ion{Si}{IV} lines that are stronger than \ion{O}{IV}, similar to what is typically observed. Here, such spectra could be generated with fewer energetic particles than in the time-independent non-Maxwellian scenario investigated by \citet{Dudik14b}. These results underline the importance of the non-equilibrium processes, the non-equilibrium ionization, departures from a Maxwellian, and their combinations, for the interpretation of IRIS TR spectra.

\begin{acknowledgements}
We acknowledge support from Grants No. 16-18495S and 17-16447S of the Grant Agency of the Czech Republic, as well as institutional support RVO:67985815 from the Czech Academy of Sciences. IRIS is a NASA small explorer mission developed and operated by LMSAL with mission operations executed at NASA Ames Research center and major contributions to downlink communications funded by the Norwegian Space Center (NSC, Norway) through an ESA PRODEX contract.
CHIANTI is a collaborative project involving the NRL (USA), the University of Cambridge (UK), and George Mason University (USA).
\end{acknowledgements}

\bibliographystyle{aa}
\bibliography{si4_noneq}

\begin{appendix}

%
\begin{figure}[!ht]
        \centering
        \includegraphics[width=8.8cm]{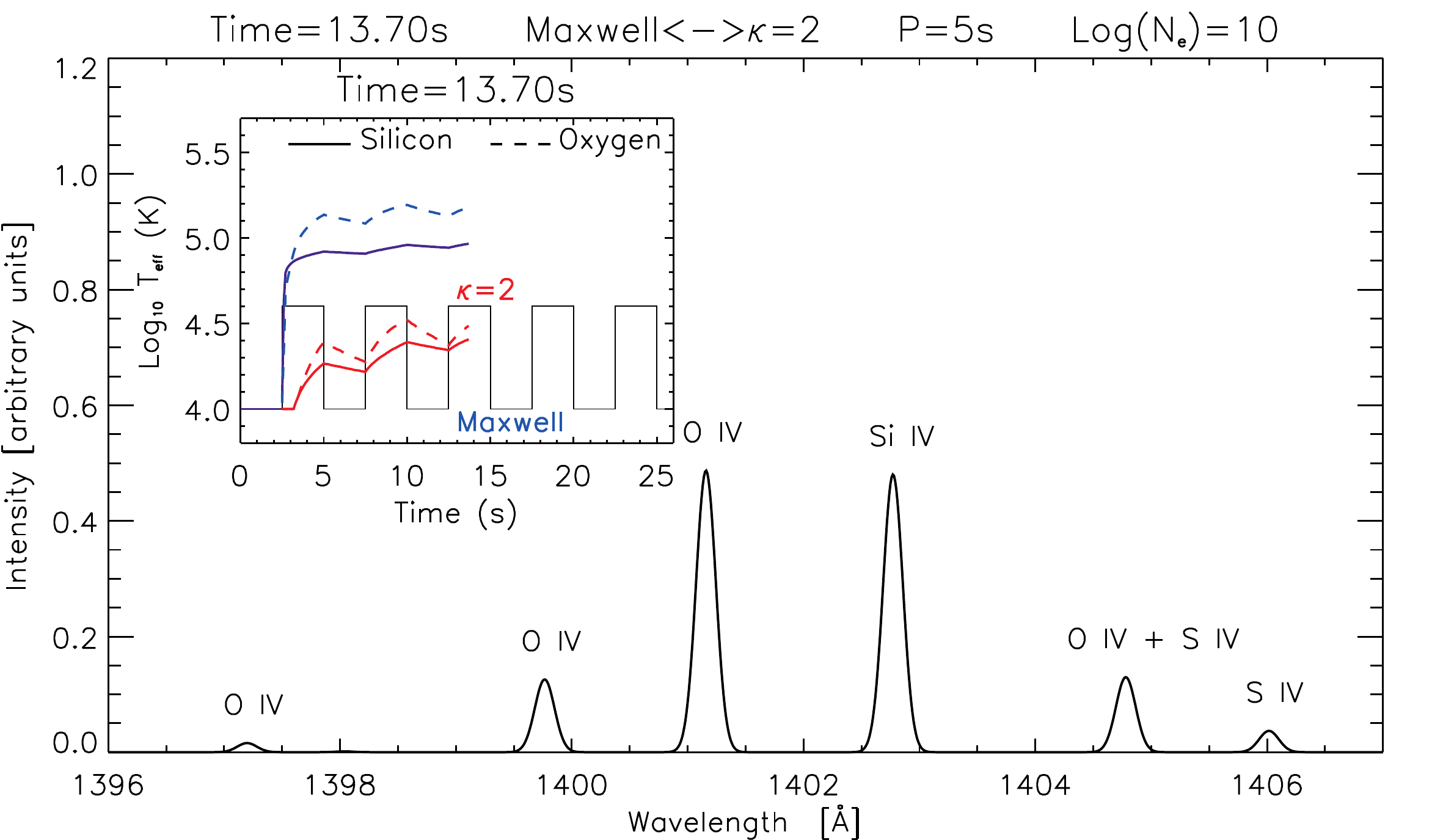}
        \includegraphics[width=8.8cm]{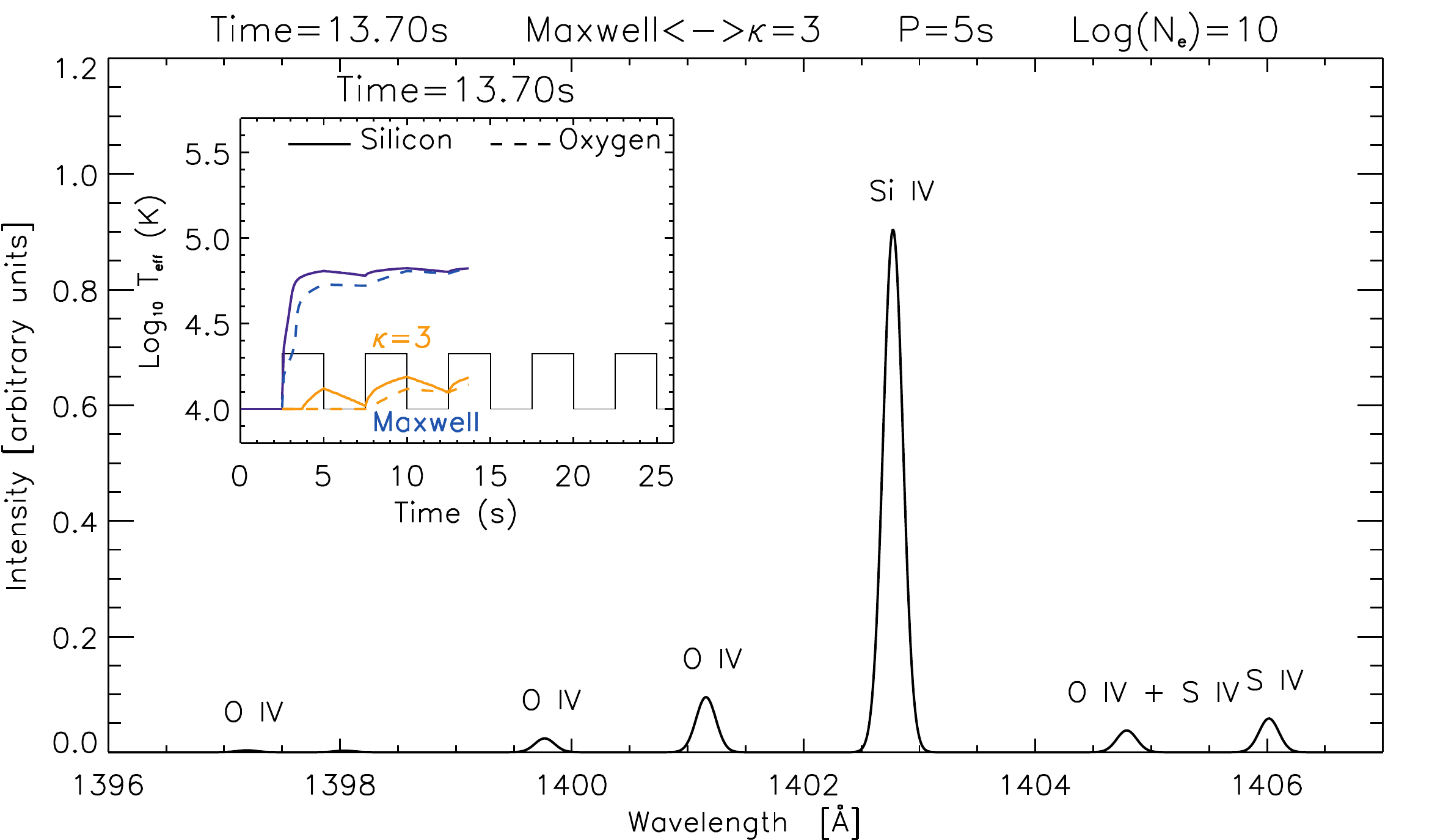}
        \includegraphics[width=8.8cm]{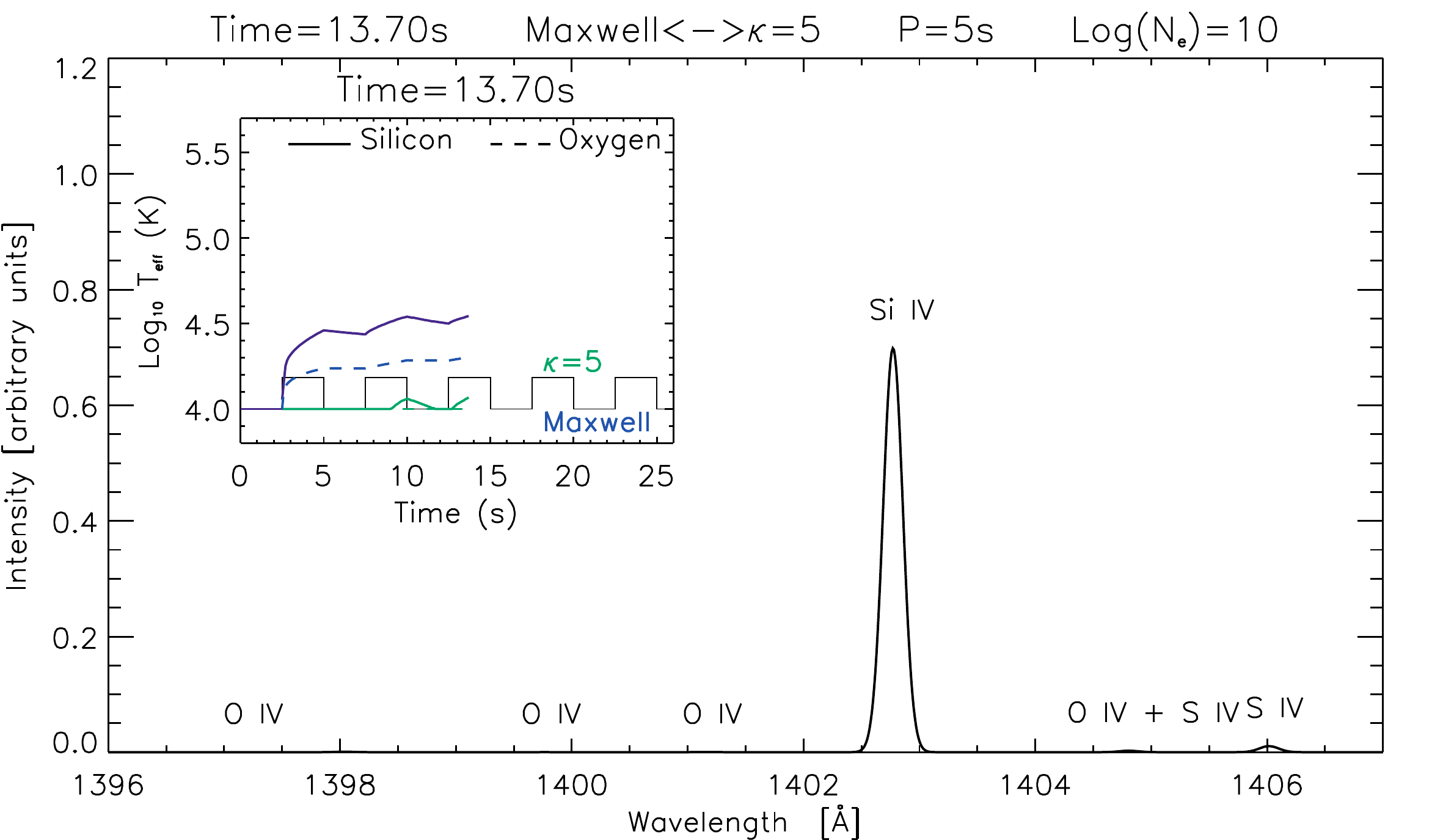}
\caption{Example of instantaneous spectra obtained for the case of $P$\,=\,5\,s and \logne\,=\,10 during the time $t$\,=\,13.7\,s; i.e., during the third period. The inset image shows the instantaneous $T_\mathrm{eff}^\mathrm{Mxw}$ and $T_\mathrm{eff}^\kappa$ for \ion{Si}{IV} (full lines) and \ion{O}{IV} (dashed lines). The evolution of these instantaneous spectra is shown in the animations attached to this figure.}
\label{Fig:Evolution}
\end{figure}
%

%
%
\section{Evolution of the line intensities}
\label{Appendix:A}

Here, we show the example instantaneous spectra and their evolution obtained for the case of $P$\,=\,5\,s, $N_\mathrm{e}$\,=\,10$^{10}$ cm$^{-3}$, and different values of $\kappa$. The instantaneous spectra are chosen during the first quarter of the third period, that is, at $t$\,=\,13.7\,s. The evolution of these instantaneous spectra is shown in animations attached to Fig. 1. There, the entire evolution of the spectra is shown between $t$\,=\,0 and 25\,s, that is, during the first five periods. We note that during the Maxwellian half-periods with $T$\,=\,$10^4$\,K, there are no lines, since these conditions are not sufficient to excite any transitions in \ion{Si}{IV} or \ion{O}{IV}.

\end{appendix}
\end{document}